\documentclass[10pt,twocolumn]{article}
\pdfoutput=1
\usepackage{amsmath}
\usepackage{amsfonts}
\usepackage{amssymb}
\usepackage{graphicx}
\usepackage{longtable}
\usepackage{booktabs}
\usepackage[usenames,dvipsnames]{xcolor}
\newcommand{\bX}{{\bf X}}
\newcommand{\bZ}{{\bf Z}}
\usepackage{multirow}
\usepackage{url}
\usepackage{natbib}
\usepackage{authblk}
\newcommand{\beginsupplement}{%
        \setcounter{table}{0}
        \renewcommand{\thetable}{S\arabic{table}}%
        \setcounter{figure}{0}
        \renewcommand{\thefigure}{S\arabic{figure}}%
        \setcounter{section}{0}
        \renewcommand{\thesection}{S\arabic{section}}%
     }
     
\begin{document}
\date{}
\title {Inferring the perturbation time from biological time course data}
\author[1] {Jing Yang}
\author[2] {Christopher A. Penfold}
\author[3] {Murray R. Grant}
\author[1] {{Magnus Rattray}\thanks{Correspondence: Magnus.Rattray@manchester.ac.uk}}
\affil[1] {Faculty of Life Sciences, University of Manchester, Manchester, UK.}
\affil[2] {Warwick Systems Biology Centre, University of Warwick, Coventry, UK.}
\affil[3] {School of Biosciences, University of Exeter, Exeter, UK.}

\maketitle

\begin{abstract}

Time course data are often used to study the changes to a biological process after perturbation. Statistical methods have been developed to determine whether such a perturbation induces changes over time, e.g. comparing a perturbed and unperturbed time course dataset to uncover differences. However, existing methods do not provide a principled statistical approach to identify the specific time when the two time course datasets first begin to diverge after a perturbation; we call this the perturbation time. Estimation of the perturbation time for different variables in a biological process allows us to identify the sequence of events following a perturbation  and therefore provides valuable insights into likely causal relationships. 

In this paper, we propose a Bayesian method to infer the perturbation time given time course data from a wild-type and perturbed system. We use a non-parametric approach based on Gaussian Process regression. We derive a probabilistic model of noise-corrupted and replicated time course data coming from the same profile before the perturbation time and diverging after the perturbation time. The likelihood function can be worked out exactly for this model and the posterior distribution of the perturbation time is obtained by a simple histogram approach, without recourse to complex approximate inference algorithms. We validate the method on simulated data and apply it to study the transcriptional change occurring in Arabidopsis following inoculation with \emph{P. syringae} pv. tomato DC3000 versus the disarmed strain DC3000\emph{hrpA}. 

An R package, DEtime, implementing the method is available at \url{https://github.com/ManchesterBioinference/DEtime} along with the data and code required to reproduce all the results.

\end{abstract}

\section{Introduction}

Gene expression time profiles can reveal important information about cellular function and gene regulation~\citep{bar2004analyzing}. A common experimental design is to perturb a biological system either before or during a time course experiment. In this case, a fundamental problem is to identify the precise {\em perturbation time} when a gene's time profile is first altered. In this paper we present an exactly tractable Bayesian inference procedure to infer the perturbation time by comparing perturbed and wild-type gene expression profiles. Ordering genes by their perturbation time gives valuable insight into the likely causal sequence of events following a perturbation. We demonstrate the applicability of our method by studying the timing of transcriptional changes in \emph{Arabidopsis thaliana} leaves following inoculation with the hemibiotrophic bacteria \emph{Pseudomonas syringae} pv. tomato DC3000 versus the disarmed strain DC3000\emph{hrpA}.

Most methods for the analysis of differentially expressed genes are based upon snapshots of gene expression \citep{kerr2000analysis,dudoit2002statistical} and there are many well-established software packages for that purpose targeted at microarray and RNA-Seq data~\citep{robinson2010edger, hardcastle2010bayseq,anders2010differential}. However, most of these methods cannot easily be extended to time course gene expression data and ignoring the temporal nature of the data is statistically inefficient.  Methods have therefore been developed specifically for time-series applications. In the case of gene expression profiles under a single condition, one-sample methods have been developed to discriminate differentially expressed genes from constitutively expressed genes. For example, probabilistic models have been designed for this purpose which use a likelihood-ratio test to rank genes based on a comparison between a dynamic and a constant profile~\citep{angelini2008bats,kalaitzis2011simple}. 

When expression profiles are available from two or more conditions then a two-sample test is more appropriate~\citep{storey2005significance,conesa2006masigpro,kim2013method, Stegle2010robust}. \citet{storey2005significance} apply a polynomial regression model to simulate the temporal behaviour of genes and a statistical test to identify differentially expressed genes. \citet{conesa2006masigpro} adopt a two-step regression model in analysing temporal profiles of genes with time treated as an extra experimental factor. \citet{kim2013method} apply Fourier analysis to time course gene expression data and identify differentially expressed genes in the Fourier domain. \citet{Stegle2010robust} apply a  model based on Gaussian Process (GP) regression which is closely related to our proposed approach. In this model, when two time series are the same they are represented by a shared GP function but where they differ they are better represented by two independent GP functions. Binary latent variables are used to model whether a particular time interval is better represented by two independent GPs or one combined GP. More recently, the GP regression framework has been refined through use of a non-stationary covariance function and a simplified scoring approach to detect time periods of differential gene expression~\citep{Heinonen2014}. Similar to the work of \citet{Stegle2010robust}, a log-likelihood ratio is used to identify time periods of differential expression. In order to better adapt to the case where unevenly or sparsely distributed times are used, they introduce a non-stationary covariance function and proposed two novel likelihood ratio tests to evaluate the likelihood at arbitrary time points. All these approaches can be used to find differentially expressed genes and some can be used to identify temporal domains where there is support for profiles being different. However, these methods do not directly score the probability of the perturbation time where two profiles first diverge, which is the aim of our approach. Although the methods of \citet{Stegle2010robust} and \citet{Heinonen2014} can be adapted to provide an estimate of the perturbation time, e.g. by applying a thresholding procedure to their differential expression scores, we show here that direct inference of the perturbation time is a more powerful approach when that is the object of interest.

In this paper, we propose a method to identify the perturbation point given data from two time course experiments. We use a non-parametric GP to describe the joint posterior distribution of two time profiles which are equal up to a proposed perturbation time. The perturbation time is then a model parameter which can be inferred. We derive the covariance function of the GP model and show that the likelihood function is exactly tractable. The posterior distribution of the perturbation time can be computed through a simple one-dimensional histogram approach, with no assumptions over the shape of the posterior distribution and no need to resort to complex approximate inference schemes. This differs from \citet{Stegle2010robust} and \citet{Heinonen2014} in that we focus specifically on inferring the perturbation time and derive an exact approach to this problem.  \citet{Stegle2010robust} creates a mixed model in pre-specified time intervals with the transition between independent GPs and shared GPs. The likelihood in that case must be approximated using Expectation Propagation (EP) due to its non-Gaussian nature.  \citet{Heinonen2014} provide a simpler approach by adopting the expected marginal log-likelihood ratio or the noisy posterior concentration ratio to construct a smooth curve indicating time periods of differential expression. However, their approach does not allow direct inference of the perturbation time.

The paper is organised as follows. In Section 2, we present background on GP regression  and derive the covariance function, likelihood function and posterior inference procedure for our new model. In Section 3, the algorithm is demonstrated on simulated data and subsequently applied to identify the perturbation times for Arabidopsis genes in a microarray time series dataset detailing the transcriptional changes that occur in Arabidopsis following inoculation with DC3000 versus the disarmed strain DC3000\emph{hrpA} \citep{Lewis:15} and with a brief conclusion presented in Section 4.        

\section{METHODS}
\subsection{Gaussian Process regression}

Gaussian Processes (GPs) \citep{Williams2006} extend multivariate Gaussian distributions to infinite dimensionality and can be used as probabilistic models that specify a distribution over functions \citep{Lawrence2005}. GPs have been used in a range of gene expression applications, e.g. to model the dynamics of transcriptional regulation \citep{Gao2008, honkela2010model} and in temporal differential expression scoring~\citep{yuan2006flexible,kalaitzis2011simple,Stegle2010robust, Heinonen2014}. 



We have a data set $\mathcal{D}$ with $N$ inputs ${\bf X} = \{x_n\}_{n=1}^N$ and corresponding real valued targets ${\bf Y} = \{y_n\}_{n=1}^N$. In the case of time course data the data are ordered such that $x_n \geq x_{n-1}$ but there is no restriction on the spacing since GPs operate over a continuous domain. We allow the case $x_n = x_{n-1}$ since that provides a simple way to incorporate replicates. We assume that measurement noise in $\bf Y$, denoted by $\epsilon$, is i.i.d Gaussian distributed $\epsilon \sim \mathcal{N}(0,\sigma^2 \mathbf{I})$ and the underlying model for $\bf Y$ as a function of $\bf X$ is $f(\cdot)$, so that
$$ {\bf Y} = f({\bf X})+\epsilon, $$

\noindent and $f({\bf X})$ represents the mean of the data generating process. Our prior modelling assumption is that the function $f$ is drawn from a GP prior with mean function $\mu(\mathbf{X})$, covariance function $K(\mathbf{X},\mathbf{X})$ and hyperparameters $\theta$. We write,
$$ f (\mathbf{X}) \sim \mathcal{GP}(\mu(\mathbf{X}),K(\mathbf{X},\mathbf{X})), $$ 
\noindent and the likelihood of $\mathbf{Y}$ becomes
$$ p(\mathbf{Y}| \mathbf{X},\theta ) \sim \mathcal{N}(\mu(\mathbf{X}),K(\mathbf{X},\mathbf{X}) + \sigma^2\mathbf{I}),$$ 
where $K({\bf X},{\bf X})$ is the $N\times N$ covariance matrix with elements $K(x_n,x_m)$. The covariance function describes typical properties of the function $f$, e.g. whether it is rough or smooth, stationary or non-stationary etc. We choose the squared exponential function,
\begin{equation}
K(x_n,x_m) = \alpha\exp\left(\frac{-(x_n-x_m)^2}{l}\right),
\label{eqn:sqr}
\end{equation}
with hyper-parameters $\theta = (\alpha, l)$ specifying the amplitude and lengthscale of samples drawn from the prior. This choice corresponds to a prior assumption of smooth and stationary functions. However, our model can be applied with any other choice of covariance function, e.g. the non-stationary covariance introduced by \citet{Heinonen2014}. The hyper-parameters can be estimated from the data by maximum likelihood or through a Bayesian procedure~\citep{Williams2006}. We can also consider the noise variance, $\sigma^2$, as an additional hyper-parameter to be estimated similarly. 

 A typical regression analysis will be focused on a new input $x_*$ and its prediction $f_*$. Based upon Gaussian properties \citep{Williams2006} the posterior distribution of $f_*$ given data $\bf Y$ is $ p(f_*|{\bf Y})  \sim  \mathcal{N}(\mu_*,{C}_*) $ with
\begin{eqnarray*}
\mu_* & = & K({\bf X},x_*)^{\top} (K({\bf X,X})+\sigma^2{\bf {I}})^{-1} {\bf Y}, \\
C_* &=&  K(x_*,x_*) \\
& & - K({\bf X},x_*)^{\top}(K({\bf X,X})+\sigma^2{\bf I})^{-1}{K}({\bf X}, x_*) \ .
\end{eqnarray*}    
We see then that the posterior distribution is also a GP but it is adapted to the data. The mean prediction is a weighted sum over data with weights larger for nearby points in a manner determined by the covariance function. The posterior covariance captures our uncertainty in the inference of $f*$ and will typically be reduced as we incorporate more data.

A special case of GP regression, which is useful in deriving our model below, is the case where $\bf (X,Y)$ is a single point $(x_p, u)$ measured with zero noise. In this case the GP regression of all new points $\bf \bar{X}$ given $(x_p, u)$ is then 
\begin{eqnarray} \label{GPprior_f}
p(f(\bf \bar{X})|{\bf Y}) & \sim & \mathcal{N}(\mu(\bf \bar{X}),C({\bf \bar{X},\bar{X}})),
\end{eqnarray}
 
\noindent with
\begin{eqnarray}
\mu({\bf \bar{X}}) & = & \frac{K({\bf \bar{X}},x_p)u}{K(x_p,x_p)}, \\
C({\bf \bar{X},\bar{X}}) & = & K({\bf \bar{X},\bar{X}}) - \frac{K({\bf \bar{X}},x_p)K({\bf \bar{X}},x_p)^{\top}}{K(x_p,x_p)}. 
\end{eqnarray}

\subsection{Joint distribution of two functions constrained to cross at one point}

Consider the case where two time profiles, $f({\bf X})$ and $g({\bf Z})$, evaluated at specified sets of time points $\bf X$ and $\bf Z$, respectively, cross at the point $x_p$ with $f(x_p)=g(x_p)=u$ at the crossing point. Before considering the constraint we use the same GP prior for each function with hyperpaprameters $\theta$,
\begin{eqnarray*}
f({\bf X}) &\sim& \mathcal{GP}(\mu(\mathbf{X}),K(\mathbf{X},\mathbf{X})) \ , \\
g({\bf Z}) &\sim& \mathcal{GP}(\mu(\mathbf{Z}),K(\mathbf{Z},\mathbf{Z})) \ .
\end{eqnarray*}
\noindent Imposing the constraint that the functions cross at $x_p$  is equivalent to observing a data point $(x_p,u)$ with zero noise. Then $p(f|{\bf X},u)$ and $p(g|{\bf Z},u)$ are as in Eqn. (\ref{GPprior_f}), 
\begin{eqnarray*}
p(f({\bf X})|u) &\sim& \mathcal{N}(\mu_{\bf X},C_{\bf X}) \ , \\
p(g({\bf Z})|u) &\sim& \mathcal{N}(\mu_{\bf Z},C_{\bf Z}),
\end{eqnarray*}
with  
\begin{eqnarray*}
\mu_{\bf X} \!\! & = & \!\! \frac{K({\bf X},x_p)u}{K(x_p,x_p)} \  \\
C_{\bf X} \: &=& \: K({\bf X,X}) - \frac{K({\bf X},x_p)K({\bf X},x_p)^\top}{K(x_p,x_p)}, \\
\mu_{\bf Z}  \!\! & = &  \!\! \frac{K({\bf Z},x_p)u}{K(x_p,x_p)} \ , \\
\quad C_{\bf Z} \: & = & \: K({\bf Z,Z}) - \frac{K({\bf Z},x_p)K({\bf Z},x_p)^\top}{K(x_p,x_p)},
\end{eqnarray*}

\begin{figure}[!t]
\centering
\begin{minipage}[c]{0.4\textwidth}
  \includegraphics[width=\textwidth]{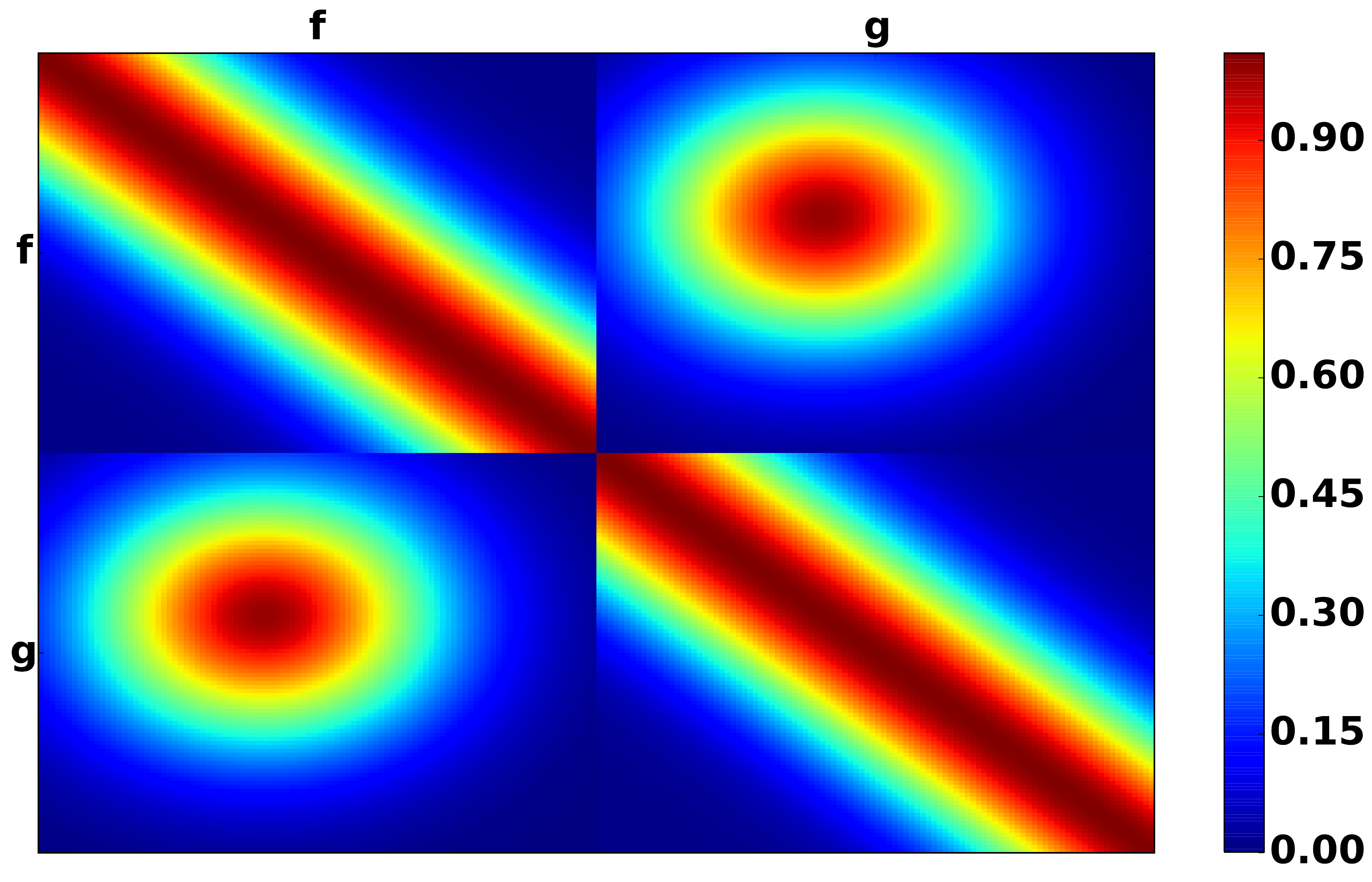}
 \end{minipage}
\begin{minipage}[c]{0.4\textwidth}
 \includegraphics[width=\textwidth]{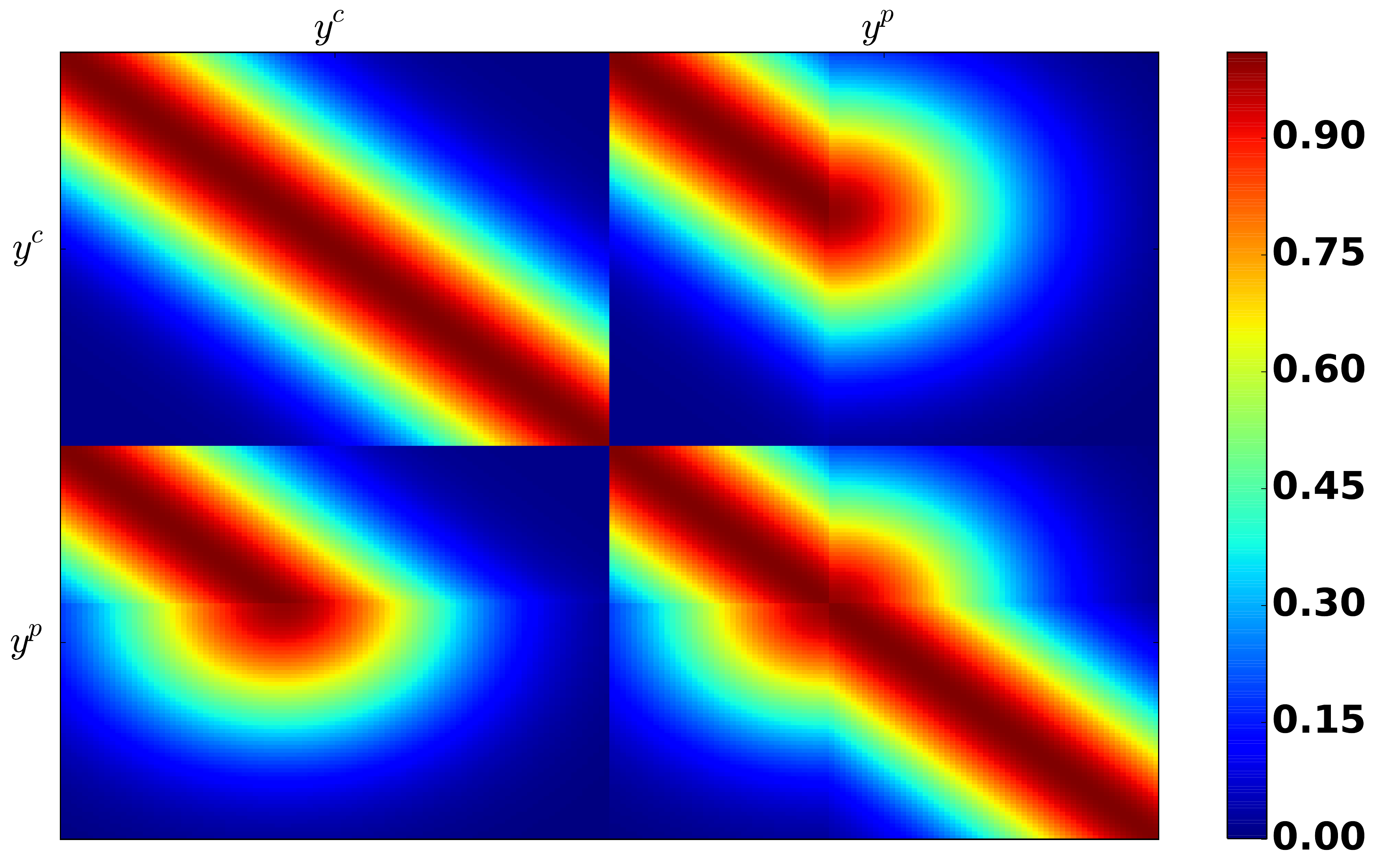}
 \end{minipage}
 \caption{Illustration of the covariance matrix, $\Sigma$, for two functions $f$ and $g$ evaluated at points evenly distributed in [0,100] and crossing at $x_p=40$ (upper) and the resulting data covariance matrix, $\hat{\Sigma}$, for time course data $y^c$ and $y^p$ from a wild-type and perturbed system respectively (lower).}
 \label{cov_illustration}
 \end{figure}

\noindent In practice, the time profiles $f(\bf X)$ and $g(\bf Z)$ are typically measured at the same time points, so that $\bf Z$ can be replaced by $\bf X$. The value of the functions at the crossing point, $u$, is not known and we marginalise it out using the prior Gaussian distribution $u \sim {\mathcal N}(0,K(x_p,x_p))$. The joint probably distribution of $f$ and $g$ is then given by Eqn. (\ref{joint_prob}) below,
\begin{align}\label{joint_prob}
p \left( f({\bf X}),g({\bf X}) \right)  & =   \int p(f|{\bf X},u)p(g|{\bf X},u)p(u) du, \nonumber\\
 & \propto  \exp \left( -\frac{1}{2} \begin{pmatrix} f & g \end{pmatrix} \Sigma^{-1} \begin{pmatrix} f & g \end{pmatrix}^\top \right),  
\end{align}
so that the two functions are jointly Gaussian distributed as ${\mathcal N}(0,\Sigma)$ with covariance given by,
\begin{eqnarray}\label{original_C}
\Sigma &= \begin{pmatrix}  K_{ff} &  K_{fg} \\ K_{gf} & K_{gg} \end{pmatrix} = \begin{pmatrix}  K_{\bf X} &  \frac{k_{\bf X}k_{\bf X}^\top}{k_{x_p}} \\ \frac{k_{\bf X}k_{\bf X}^\top}{k_{x_p}} & K_{\bf X}, \end{pmatrix},  
\end{eqnarray}

\noindent where $K_{\bf X}$, $k_{x_p}$ and $k_{\bf X}$ are abbreviations for $K(\bf X, X)$,  $K(x_p,x_p)$ and $K({\bf X},x_p)$, respectively. We show an example of this covariance function in Fig.~\ref{cov_illustration} (upper panel) for $\bf X$ in the range [0,100] and $x_p=40$. The detailed derivations of Eqn. (\ref{joint_prob}) and Eqn. (\ref{original_C}) are illustrated in the Supplementary Section \ref{section_s1} and \ref{section_s2}. 

\subsection{The data likelihood under the model}

We define the perturbation time $x_p$ as the point where two time profiles first begin to diverge. If the time profiles are measured without noise then it would be trivial to identify this point. However, biological time course data from high-throughput experiments are often corrupted by significant biological and technical sources of noise and our task is to {\em infer} the perturbation time given noisy time course data. In order to do that we must first derive the likelihood function under the new model. 

Let two sets of gene expression time course data, $y^c(\bf X)$ and $y^p(\bf X)$, represent noisy measurements with i.i.d Gaussian measurement noise, $\mathcal{N}(0,\sigma^2 \mathbf{I})$, from the control condition and perturbed condition, respectively. A GP prior is placed on the mean functions underlying $y^c$ and $y^p$ and a time point $x_p$ is defined as the perturbation time point. The data model is defined as:
\begin{enumerate}
\item The two datasets $y^c$ and $y^p$ before $x_p$ are noise-corrupted versions of the same underlying mean function $f$ which has a GP prior,
\begin{eqnarray*}
y^c(x_n) & \sim & \mathcal{N}(f(x_n),\sigma^2), \\
y^p(x_n) & \sim & \mathcal{N}(f(x_n),\sigma^2) \quad \mbox{for } x_n \leq x_p \ .
\end{eqnarray*}
\item The mean function for $y^c$ stays intact after $x_p$ while the mean function for $y^p$  changes to follow $g$,
\begin{eqnarray*}
y^c(x_n) & \sim & \mathcal{N}(f(x_n),\sigma^2), \\
y^p(x_n) & \sim & \mathcal{N}(g(x_n),\sigma^2) \quad \mbox{for } x_n > x_p \ ,
\end{eqnarray*}
where $f$ and $g$ are constrained to cross at $x_p$ and follow the GP described in Eqn. (\ref{joint_prob}).
\end{enumerate}

\noindent The joint distribution of $y^c$ and $y^p$ is then 
\begin{align}
p \left( y^c({\bf X}),y^p({\bf X})|x_p \right) &= \exp \left( -\frac{1}{2} \begin{pmatrix} y^c \\ y^p \end{pmatrix}^\top \hat{\Sigma}^{-1} \begin{pmatrix} y^c \\ y^p \end{pmatrix} \right), 
\end{align}
where the covariance matrix $\hat{\Sigma}$ can be worked out in terms of the covariance matrix $\Sigma$ for the joint distribution of $f$ and $g$ defined by Eqn. (\ref{original_C}),
\begin{equation}\label{modified_C} 
\hat{\Sigma} = \begin{pmatrix}  \hat{K}_{y^cy^c} &  \hat{K}_{y^cy^p} \\ \hat{K}_{y^py^c} & \hat{K}_{y^py^p} \end{pmatrix} ,
\end{equation} with

\begin{align*}\scriptsize
\hat{K}_{y^c({\bf X_1})y^c({\bf X_2})} &= K_{f({\bf X}_1)f({\bf X}_2)} + \sigma^2 \bf I \! \begin{array}{c} {\bf X}_1\in {\bf X},{\bf X}_2 \in {\bf X} \end{array} \\ 
\hat{K}_{y^c({\bf X_1})y^p({\bf X_2})} & =   \left\{ \begin{array}{c} K_{f({\bf X}_1)f({\bf X}_2)} \\
 K_{f({\bf X}_1)g({\bf X}_2)} \end{array} 
\right. \! \begin{array}{l}{\bf X}_1 \in {\bf X}, {\bf X}_2 \leq x_p \\ 
 {\bf X}_1 \in {\bf X}, {\bf X}_2>x_p \end{array}  \\
\hat{K}_{y^p({\bf X_1})y^c({\bf X_2})} &= \left\{ \begin{array}{c} K_{f({\bf X}_1)f({\bf X}_2)} \\ K_{g({\bf X}_1)f({\bf X}_2)} \end{array} \right. \! \begin{array}{l} {\bf X}_1 \leq x_p, {\bf X}_2 \in {\bf X} \\ {\bf X}_1>x_p, {\bf X}_2 \in {\bf X} \end{array}  \\
\hat{K}_{y^p({\bf X}_1)y^p({\bf X}_2)} &=\left\{\begin{array}{c} K_{f({\bf X}_1)f({\bf X}_2)}  + \sigma^2 \bf I \\ K_{g({\bf X}_1)f({\bf X}_2)} \\ K_{f({\bf X}_1)g({\bf X}_2)} \\ K_{g({\bf X}_1)g({\bf X}_2)} + \sigma^2 \bf I \end{array} \right. \! \begin{array}{l} {\bf X}_1 \leq x_p, {\bf X}_2 \leq x_p \\  {\bf X}_1>x_p, {\bf X}_2 \leq x_p \\   {\bf X}_2>x_p, {\bf X}_1 \leq x_p \\   {\bf X}_1>x_p,{\bf X}_2>x_p \end{array}   
\end{align*}

\noindent The lower panel in Fig. \ref{cov_illustration} shows the data covariance matrix $\hat{\Sigma}$ for $\bf X$ evenly spread in the range $[0,100]$ and with a perturbation occurring at $x_p=40$. 

\subsection{Posterior distribution of the perturbation point} 

According to Bayes' rule the posterior distribution of $x_p$ is, 
\begin{eqnarray*}
p(x_p|y^c({\bf X}),y^p({\bf X})) & = \frac{p({y^c({\bf X}),y^p({\bf X})}|x_p)p(x_p)}
{\int p({y^c({\bf X}),y^p({\bf X})}|x_{p})p(x_p)d x_p} .
\end{eqnarray*}
We assume a uniform prior on $x_p$ within the range $[x_{\mathrm{min}},x_{\mathrm{max}}]$ of the observed data. We use a simple discretization $x_p \in [x_{\mathrm{min}}, x_{\mathrm{min}}+\delta, x_{\mathrm{min}}+2\delta,\ldots,x_{\mathrm{max}}]$ in this range. Then the posterior can be approximated as a simple summation over this grid,
\[
p(x_p|y^c({\bf X}),y^p({\bf X})) \simeq \frac{ p(y^c({\bf X}),y^p({\bf X})|x_p) }{ \sum_{x = x_{\mathrm{min}}}^{x = x_{\mathrm{max}}}p(y^c({\bf X}),y^p({\bf X})|x)},
\]
only requiring that we evaluate the likelihood at each grid point. There are hyper-parameters $\theta$ also involved in the posterior distribution of $x_p$ which would potentially complicate matters. We choose to estimate these hyper-parameters prior to inferring $x_p$. To do this we use maximum likelihood optimisation for the case where $x_p$ approaches -$\infty$ which corresponds to the two GPs for the control and perturbed conditions being independent,
\[
\hat{\theta} = \arg\!\max_{\theta} \left(\lim_{x_p \to -\infty} p_{\theta}(y^c({\bf X}),y^p({\bf X})| x_p, \theta) \right) \ . 
\]
Since we have a simple histogram representation for the posterior distribution of the perturbation time point $x_p$ then we can easily estimate the mean, median or mode (MAP) of the posterior distribution to provide a point estimate. 

\subsection{Pre-filtering to remove non-DE genes}

In many applications a large number of genes will show no strong evidence for DE at any time or will have a low signal-to-noise due to being weakly expressed. We therefore filter genes prior to using our model. A DE gene will be better represented by two independent GPs rather than a shared GP under control and perturbed conditions. We therefore filter genes using the log-likelihood ratio $r$ between the independent GP model (equivalent to $x_p$ approaching $-\infty$ in the perturbation model) and the integrated GP (with $x_p$ approaching $+\infty$):
\begin{align}
 r &= \log\mathcal{L}(y_c({\bf X}),y_p({\bf X})|x_p \to -\infty) \nonumber \\
 &-\log\mathcal{L}(y_c({\bf X}),y_p({\bf X})|x_p \to +\infty)
\end{align}
We note that it is difficult to distinguish genes with a late perturbation time from those that are non-DE and our filtering approach may remove some genuine late perturbation genes. In many applications we are primarily interested on relatively early perturbations (e.g. in the application considered here) in which case this will not significantly impact the results. In the Supplementary Section \ref{section_s4.2} we consider an alternative filtering approach which is based on detecting genes with time-varying profile in either the control or perturbed condition and is therefore less likely to filter out late $x_p$ genes. 

The method has been implemented in the DEtime R-package (\url{github.com/ManchesterBioinference/DEtime}) and also as the DEtime kernel in the GPy Python package (\url{github.com/SheffieldML/GPy}). The running time for the whole genome (32578 genes) for the example in Section \ref{realdata} on a Intel(R) Core(TM) i7-3770 CPU of 3.40GHz is around 11 hrs using the DEtime R-package.

\begin{figure}
\centering
\includegraphics[width=0.5\textwidth]{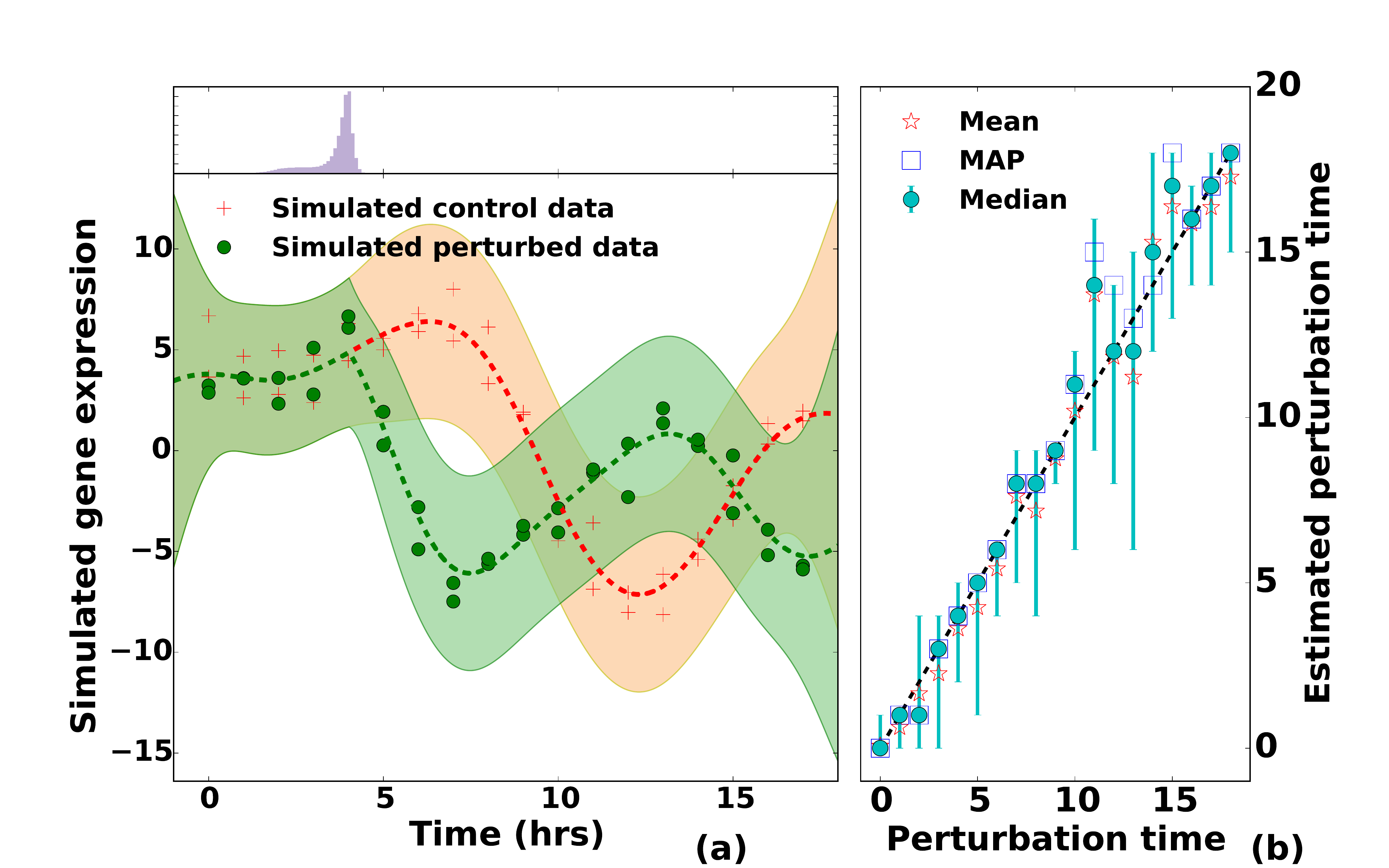}
\caption{ (a) The shaded area in the lower panel represents the 95\% credible region of the GP regression result. In the top panel we show the inferred posterior distribution for the perturbation time $x_p$. (b) The mean, mode and median of the posterior distribution of $x_p$ with the 5-95 percentile coverage of the posterior distribution for 19 simulated dataset at different perturbation time points (dashed line shows the ground truth).} 

 \label{sampleGP}
 \end{figure} 

\section{Results and Discussion}

\subsection{Generating simulated data}

We generated data under a range of different scenarios to explore performance and robustness to deviations from the model. We generated expression profiles from three different covariance models, one matching the one used for inference and the other two generating rougher profiles. We then add noise using three different noise models, one matching the Gaussian model used for inference and two from heavier-tailed distributions. 

\begin{enumerate}

 \item \emph{profile$_1$}: simulated noise-free profile generated from the model $\mathcal{GP}_{\theta}(\mathbf{0},\hat{\Sigma}_{\theta})$ with $\hat{\Sigma}_{\theta}$ given in Eqn. (\ref{modified_C}) assuming a squared exponential covariance function (recall Eqn. (\ref{eqn:sqr})) with the hyperparameters $\theta=\{\alpha=30.0, l=8.0\}$. 
 \item \emph{profile$_2$}: simulated noise-free profile generated from above model with the covariance function in the form of  a matern32 covariance function \citep{Williams2006} with the same hyperparameters as above.
 \item \emph{profile$_3$}: simulated noise-free profile generated from above model with the covariance function in the form of  a matern12 covariance function (an Ornstein-Uhlenbeck (OU) process) with the same hyperparameters as above.
 \end{enumerate}

 Nine simulated dataset are induced with different kinds of i.i.d noise on top of \emph{profile$_1$}, \emph{profile$_2$} and \emph{profile$_3$}, respectively: Gaussian $\mathcal{N}(1.5)$, Student-t distributed with 3 $(T(3))$ and 6 $(T(6))$ degrees of freedom. The simulated data are sampled every hour from 0 hrs until 18 hrs. We simulate data with a range of perturbation times ${\mathbf x_p} \in \{0,1,\cdots, 17,18\}$ and 100 different sets of data are produced for each $x_p$ value.

Fig. \ref{sampleGP} (a) shows an example of simulated data (using the \emph{profile$_1$}$+\mathcal{N}(1.5)$ scenario) with two replicates and a perturbation at 4 hrs. The estimated posterior distribution of $x_p$ is shown in the upper panel and in the lower panel we show the GP regression function after fixing $x_p$ at the MAP value. In this case the MAP estimate for $x_p$ is very close to the ground truth. The mean, mode and median of the posterior distribution of $x_p$ for 19 simulated datasets are illustrated in Fig. \ref{sampleGP} (b) together with the 5-95 percentile coverage of the posterior distribution. It is clear that the posterior distributions of the perturbation time cover the actual perturbation time to a great extent and that the three different point estimates are typically close to the ground truth values. 

\subsection{Comparison with a thresholding approach}
 
Related methods have been introduced to identify regions of differential expression from time course data~\citep{Stegle2010robust, Heinonen2014}. Such methods can in principle also be used to identify the perturbation time by locating the first time point where the DE score passes some threshold value. Here we compare our approach to the most recently published package of this type, developed by \citet{Heinonen2014} implemented in the nsgp R-package. The nsgp package infers the differentially expressed time periods and uses four likelihood ratios: marginal log-likelihood ratio (MLL), expected marginal log-likelihood ratio (EMLL), the posterior concentration (PC) and the noisy posterior concentration (NPC) to quantify these regions. We adopt thresholds of 0.5 and 1.0 to define the initial perturbation points, respectively. The mean, median and mode of the posterior distribution of the inferred perturbation points from our method are also computed. The performance of ranking $x_p$ using each method is measured by Spearman's rank correlation coefficient with the known ground truth and the mean and standard deviation of the rank correlation coefficients across 100 dataset are illustrated in Table \ref{comparison_results}. 

\begin{table*}\scriptsize
\setlength{\tabcolsep}{2pt}
\renewcommand{\arraystretch}{1.1}
\caption{Comparison of the means and stds of the Spearman's rank correlation coefficients of the mean, median and MAP estimation of the perturbation times from our DEtime package and different likelihood ratios with various thresholds from the nsgp package. $M_n$ represents the $M$ ratio with a threshold of $n$. }\label{comparison_results}
\begin{tabular}{c| c| c| c| c| c| c| c| c| c|  c| c}
\hline
  R Package & \multicolumn{3}{c}{DEtime} & \multicolumn{8}{|c}{ nsgp}\\
\hline
 Data & Mean & Median & MAP & MLL$_{0.5}$& EMLL$_{0.5}$ & PC$_{0.5}$ & NPC$_{0.5}$ & MLL$_{1.0}$& EMLL$_{1.0}$ & PC$_{1.0}$ & NPC$_{1.0}$ \\ 
\hline
\emph{profile$_1$}$+\mathcal{N}(1.5)$ & 0.94$\pm$0.04& 0.94$\pm$0.04& 0.93$\pm$0.05 & -0.02$\pm$0.23 & 0.26$\pm$0.22 &0.36$\pm$0.22 &0.26$\pm$0.23 & -0.02$\pm$0.23 & 0.67$\pm$0.16 & 0.29$\pm$0.22 &0.48$\pm$0.21  \\
\emph{profile$_2$}$+\mathcal{N}(1.5)$ & 0.90$\pm$0.06& 0.90$\pm$0.06& 0.88$\pm$0.08 & -0.05$\pm$0.23 & 0.17$\pm$0.23 &0.21$\pm$0.25 &0.24$\pm$0.22 & -0.06$\pm$0.23 & 0.57$\pm$0.20 & 0.18$\pm$0.24 &0.42$\pm$0.21  \\
\emph{profile$_3$}$+\mathcal{N}(1.5)$ & 0.93$\pm$0.04& 0.93$\pm$0.04& 0.92$\pm$0.05 & -0.04$\pm$0.26 & 0.31$\pm$0.22 &0.21$\pm$0.25 &0.28$\pm$0.25 & -0.05$\pm$0.26 & 0.75$\pm$0.16 & 0.20$\pm$0.23 &0.47$\pm$0.23  \\
\emph{profile$_1$}$+T(6)$ & 0.93$\pm$0.05& 0.92$\pm$0.06& 0.91$\pm$0.07 & 0.01$\pm$0.24 & 0.22$\pm$0.24 &0.24$\pm$0.27 &0.23$\pm$0.24 & 0.02$\pm$0.23 & 0.59$\pm$0.18 & 0.17$\pm$0.23 &0.40$\pm$0.23  \\
\emph{profile$_2$}$+T(6)$ & 0.87$\pm$0.07& 0.86$\pm$0.07& 0.83$\pm$0.09 & 0.04$\pm$0.23 & 0.27$\pm$0.25 &0.08$\pm$0.24 &0.15$\pm$0.22 & 0.03$\pm$0.23 & 0.42$\pm$0.24 & 0.02$\pm$0.24 &0.28$\pm$0.24  \\
\emph{profile$_3$}$+T(6)$ & 0.89$\pm$0.06& 0.89$\pm$0.06& 0.87$\pm$0.08 & 0.01$\pm$0.26 & 0.24$\pm$0.24 &0.13$\pm$0.25 &0.26$\pm$0.26 & -0.00$\pm$0.27 & 0.64$\pm$0.20 & 0.07$\pm$0.25 &0.41$\pm$0.23  \\
\emph{profile$_1$}$+T(3)$ & 0.91$\pm$0.05& 0.90$\pm$0.06& 0.89$\pm$0.06 & -0.02$\pm$0.24 & 0.14$\pm$0.22 &0.19$\pm$0.22 &0.20$\pm$0.21 & -0.03$\pm$0.24 & 0.48$\pm$0.21 & 0.15$\pm$0.23 &0.32$\pm$0.21  \\
\emph{profile$_2$}$+T(3)$ & 0.83$\pm$0.09& 0.83$\pm$0.10& 0.80$\pm$0.11 & -0.03$\pm$0.26 & 0.08$\pm$0.23 &0.12$\pm$0.23 &0.16$\pm$0.21 & -0.04$\pm$0.26 & 0.36$\pm$0.23 & 0.05$\pm$0.22 &0.24$\pm$0.24  \\
\emph{profile$_3$}$+T(3)$ & 0.87$\pm$0.07& 0.87$\pm$0.07& 0.84$\pm$0.09 & -0.01$\pm$0.25 & 0.20$\pm$0.21 &0.09$\pm$0.23 &0.20$\pm$0.18 & 0.00$\pm$0.25 & 0.54$\pm$0.22 & 0.09$\pm$0.23 &0.33$\pm$0.23  \\
\hline
\end{tabular}
\end{table*}
 
From the table, it is clear that the mean, median and MAP estimates from the DEtime package provide better ranking performance. The results from the nsgp package vary significantly through different ratios and thresholds, among which, EMLL with threshold 1.0 performs the best in this task, giving rank correlation coefficient of 0.67$\pm$0.16 when tested on the simulated \emph{profile$_1$} contaminated with Gaussian noise $\mathcal{N}(0,1.5)$, which is still considerably lower than the rank correlation coefficients from mean, median or mode of the DEtime package.  In order to compare the performance of the algorithm on data with varied signal-to-noise ratios, we adjusted the signal amplitude hyperparameter $\alpha$ and compared the results from DEtime and nsgp with $\alpha=1.5, 10.0, 20.0, 30.0$. Supplementary Table \ref{table_s1} illustrates the results which shows the robustness of the proposed model. Supplementary Fig. \ref{mean_median_MAP} shows the errorbar of the mean, median, mode from DEtime package and EMLL with thresholds of 0.5 and 1.0 from nsgp package across 100 replicates along all perturbation times for all simulated datasets. We observe that the DEtime package provides reasonable estimation of the initial perturbation time under various noise distributions whereas the performance of the EMLL ratio from nsgp package varies substantially and its performance seems to be deteriorating with later initial perturbations.

We note that methods in the nsgp package are not designed specifically for the task of inferring the initial perturbation point as they were proposed for the more general problem of identifying DE regions. Nevertheless, a common application of time-series DE studies is to distinguish early and late DE events. We have demonstrated that one can obtain greater accuracy by focusing on this specific task rather than adapting a more general DE method. 
    
\subsection{Bacterial infection response in \emph{Arabidopsis thaliana}}\label{realdata}

\begin{figure}
\centering
\includegraphics[width=0.5\textwidth]{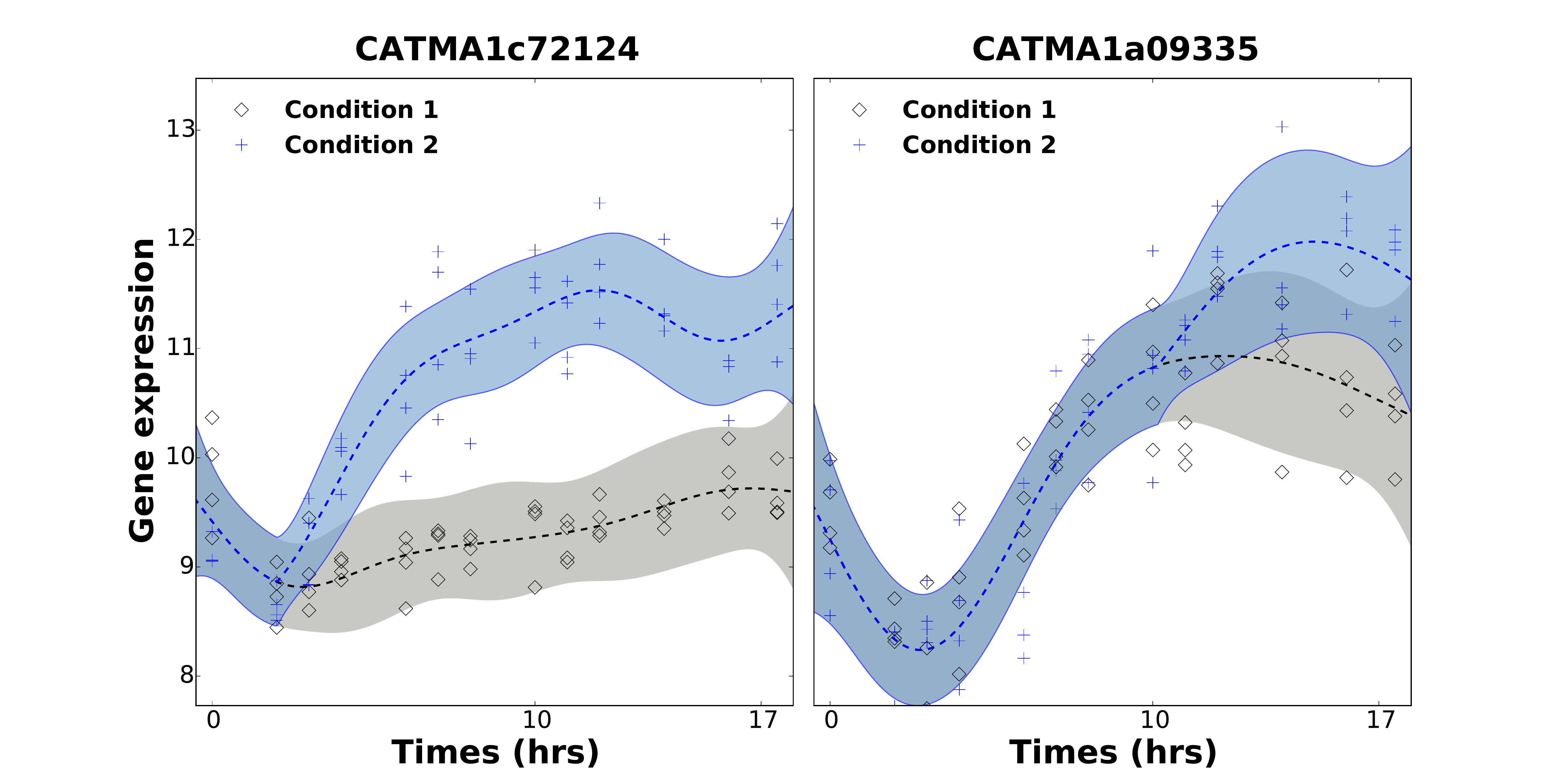}
\caption{Examples of fitting the DEtime model to an early-perturbed (left) and late-perturbed gene from an experiment comparing arabadopsis leaves collected from plants infected with DC3000 (condition 1) and the mutant DC3000\emph{hrpA} (condition 2). The shaded area represents the 95\% credible region of the GP and the dashed line is the estimated mean of the model. }
\label{realdata_figure}
\end{figure}

To determine the biological utility of estimating perturbation times, we re-examined a large dataset recently published by \citet{Lewis:15} that captures the transcriptional reprogramming associated with defence and disease development in \emph{Arabidopsis thaliana} leaves inoculated with \emph{Pseudomonas syringae} pv. tomato DC3000 and the non-pathogenic DC3000hrpA mutant strain. The differences in gene expression between these two challenges is a result of the action of virulence factors delivered by the DC3000 strain into the plant cell, in this case predominately the collaborative activities of 28 bacterial effector proteins. Figure~\ref{realdata_figure} shows examples of an early and late perturbed gene identified by our method. A preliminary investigation of the perturbation times of differentially expressed genes revealed two peak times (Supplementary Figure \ref{Fig:Histtimes}), allowing genes to be assigned to one of three groups: early, intermediate and late perturbed genes. This initial characterisation was consistent with major phase changes in the infection process, and the onset of effector mediated transcriptional reprogramming: effectors are not delivered into plant cells until 90-120 minutes post inoculation \citep{grant2000rpm1}, and do not promote bacterial growth until $\sim 8$ hpi, when they have effectively disabled host defence processes. This general progression is reflected in GO and pathway analysis outlined in Supplementary Section \ref{section_s4.1} and \ref{section_s4.2}. 

\begin{figure}
\centering
\includegraphics[trim={-1cm 0.1cm 3cm 9cm},clip, width=\textwidth]{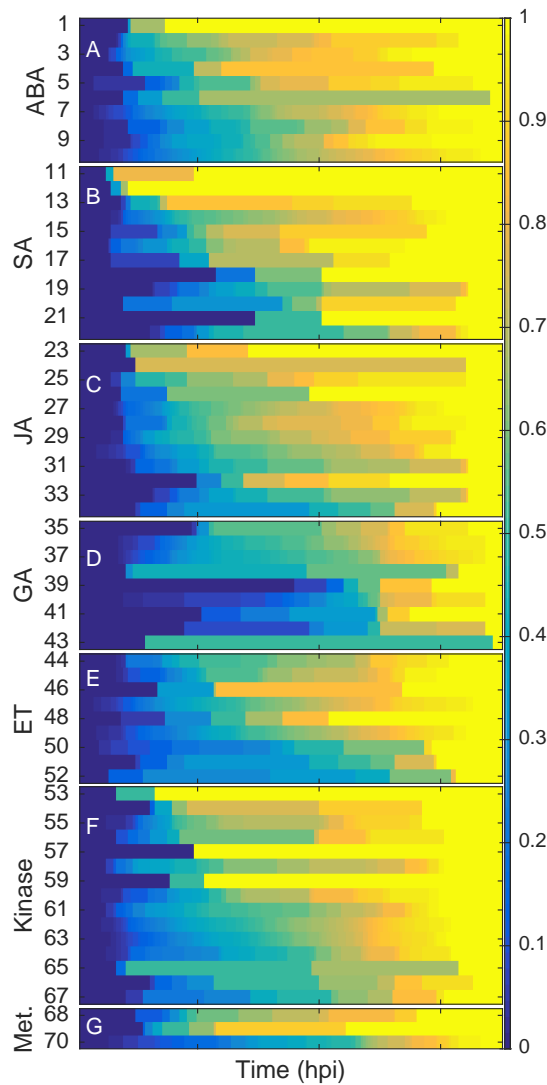}
\caption{The cumulative distribution of inferred perturbation times of gene sets associated with hormone, signalling and metabolism related gene ontology terms. See Supplementary Table \ref{Tab:11} for key.}
\label{GO_figure}
\end{figure}

The recent study by \citet{Lewis:15} provided a comprehensive overview of the transition from defence to disease. Thus we investigated if the calculation of perturbation times provided supporting evidence and additional novel insights not highlighted by \citet{Lewis:15}. To do so, genes were first grouped according to their GO or AraCyc Pathway annotation, and the cumulative perturbation time for each term calculated. The time at which more than $50\%$ of the genes associated with a particular term were perturbed could then be used to rank terms, allowing a high resolution understanding of the infection process. Heat maps showing the cumulative density function (CDF) of perturbation times for each term are shown in Supplementary Figures \ref{Fig:GO1}-\ref{Fig:GO12}. For clarity, we chose to focus predominately on the earliest processes perturbed by bacterial effectors as these are predicted to be processes integral to the suppression of innate immunity. As an initial proof of concept we focussed on the perturbation of hormone pathways, as modulation of these pathways are well known to be integral to pathogen virulence strategies (Fig. \ref{GO_figure}). 

First we looked at abscisic acid (ABA) pathways, as it has previously been shown that DC3000 rapidly induces de novo ABA biosynthesis and hijacks ABA signalling pathways to promote virulence \citep{de2007pseudomonas,de2009antagonism}. Fig. \ref{GO_figure} shows a strong link between various GOs associated with ABA processes and early perturbation, which is what is predicted in the literature and demonstrated by \citet{Lewis:15}. Amongst these early ABA signalling components induced were the classic ABA responsive TFs, RD26 and both ATAIB and AFP2 were induced around $2$ hpi. This prediction suggests that effectors are targeting ABA signaling very early in the infection process. Furthermore $>50\%$ of genes annotated with `regulation of abscisic acid biosynthetic process' were perturbed by 2.3 hpi, consistent with measurable increased in de novo ABA biosynthesis 6 hpi \citep{de2007pseudomonas}, with subsequent perturbation of `cellular response to abscisic acid stimulus' occurring by 3.5 hpi. Two genes showing perturbation at 4.1 hpi and annotated as ABA responsive, BLHL1 and TCP14, are predicted to be targeted by the DC3000 effector AvrPto in yeast two hybrid protein-protein interaction studies \citep{mukhtar2011independently}. Moreover a knockout of TCP14 results in enhanced disease resistance to DC3000, consistent with TCP14 being a virulence target of effectors \citep{wessling2014convergent}. Subsequently, a number of ABA related pathways appear to be further targeted later in the infection. Interestingly `negative regulation of abscisic acid-activated signaling pathway' was perturbed at 4.4 hpi suggesting this is an example of a failed host response \citep{Lewis:15}. Other notable perturbed ABA related ontologies included `abscisic acid transport' (4.9 hpi), `abscisic acid catabolic process' (5.1 hpi), `abscisic acid binding' (5.1 hpi), `abscisic acid-activated signaling pathway' (6.3 hpi), `abscisic acid biosynthetic process' (7.2 hpi) and `positive regulation of abscisic acid-activated signaling pathway' (7.2 hpi). Thus we can validate the importance of ABA in the infection process but, moreover, using our estimation of perturbation process we can see fine resolution of the increased impact of ABA biosynthesis and signaling on the infection process not evidenced by the previous analyses \citep{Lewis:15} as illustrated in Fig. \ref{GO_figure}A.

As expected, we also identified strong early perturbations in salicylic acid (SA, Fig. \ref{GO_figure}B) related ontologies, as these are key targets for effector mediated suppression \citep{debroy2004family}. For further validation, we looked at ontologies associated with the hormone jasmonic acid (JA, Fig. \ref{GO_figure}C). The JA ontologies show more delayed perturbation than ABA, particularly notably the ontologies associated with `response to jasmonic acid' (2.3 hpi), `jasmonic biosynthetic processes' (3.7 hpi) and `regulation of jasmonic acid mediate signaling pathways' (3.8 hpi). This is consistent with the recent study by \citet{de2015novel} using a specific targeted analysis of the same dataset which demonstrated that the JA contribution to DC3000 pathogenesis was preceded by a stronger ABA component. Thus both the ABA and JA analyses provide two examples that validate the utility of the perturbation estimation approach. Two other hormone signalling pathways, gibberellic acid (GA, Fig. \ref{GO_figure}D) and ethylene (ET, Fig. \ref{GO_figure}E), are predicted to play a minor role in establishment of virulence, with their contributions only occurring late in the infection process. 

We next identified two signalling and two primary metabolism pathways that are predicted to be important in the early conflict between plant defence and pathogen virulence: MAP kinase kinase (MAPKK) activity, regulation of protein kinase activity, NAD biosynthesis process and methionine biosynthesis (Figure \ref{GO_figure}F/G). 

MAMP signaling activates an early kinase phosphorylation cascade that initiates transcriptional activation \citep{zipfel2014plant}, however little is known about the transcriptional activation or kinases. Remarkably, 8 out of the 10 MAPKKs encoded by the Arabidopsis genome were perturbed early. Given that these MAPKKs are responsible for phosphophorylation of the 20 downstream MAPKs their respective roles are naturally extensive. However, MAPKKs are strongly implicated in biotic stress. Most notably, the DC3000 effector HopF2 can interact with Arabidopsis MKK5 and most likely other MAPKKs to inhibit MAPKs and PAMP-triggered immunity. This is probably through MAPKK inhibition via ADP-ribosylation as HopF2 delivery inhibited PAMP-induced MPK phosphorylation \citep{wang2010pseudomonas}. Functional evidence for a positive role of MKKs in defence comes from work in tobacco, where transient expression of AtMKK7/AtMKK9 and AtMKK4/AtMKK5 caused a hypersensitive response \citep{zhang2008arabidopsis}. However, the roles of MKKs are likely to be multifunctional and may be manipulated by effectors to promote virulence. The MAPKK, MKK1 was shown to negatively regulate immunity \citep{kong2012mekk1}. This may be through a dual role in activating ABA signalling as AtMKK1 as well as AtMKK2 and AtMKK3, could activate the ABA responsive RD29A promoter and MKK8 could activate the RD29B promoter \citep{hua2006activation}. Concomitant with perturbation of the MKK pathway was a significant early perturbation of a sets of genes associated with regulation of protein kinase activity. Strikingly, these genes belong to a class of evolutionarily conserved kinases functioning as metabolic sensors and are activated in response to declining energy levels. Their co-regulation is probably because they typically function as a heterotrimeric complex comprising two regulatory subunits, $\beta$ and $\gamma$, and an $\alpha$-catalytic subunit. Intriguingly, a recent study predicted that the two clade A type 2C protein phosphatases that are negative regulators of ABA signalling, ABI1 and PP2CA, negatively regulate the Snf1-related protein kinase1 and that PP2C inhibition by ABA results in SnRK1 activation \citep{rodrigues2013abi1}. Moreover, SnRK1 and ABA were shown to induce largely overlapping transcriptional responses, thus these data reveal a previously unknown link between ABA and energy signalling during DC3000 infection.

A pathway intimately linked to energy signalling and redox reactions is NAD biosynthesis, one of the most significantly perturbed pathways following effector delivery (Figure~\ref{GO_figure}G).  Although powdery mildew infection of barley leaves was reported to be associated with increased NAD content more than 40 years ago \citep{ryrie1969nicotinate} and recently the identification of the fin4 (flagellin insensitive 4) mutant as aspartate oxidase \citep{macho2012aspartate}, a precursor of the NANP biosynthetic pathway, the role of pyridines in plant defence has received little attention. NAD and NADP play crucial roles in pro-oxidant and antioxidant metabolism and have been linked to biotic stress responses, including production of nitric oxide and metabolism of reactive lipid derivatives \citep{crawford2005new,mano2005protection}. We highlight two possible, and contrasting, roles for rapid induction of NAD biosynthesis components by effectors. First, it has recently been shown that chloroplast ROS production is influenced by NADP:NADPH ratios and bacteria effector delivery rapidly suppresses a MAMP triggered chloroplast burst of hydrogen peroxide in an ABA dependent manner \citep{de2015chloroplasts}.  Secondly, poly(ADP-Ribose) polymerases (PARPs) is emerging as a key regulator of defence responses. PARPs are important NAD+ consuming enzymes induced by biotic stress, polymerising long poly(ADP-ribose) chains on target proteins including histones. \citep{adams2010disruption} reported a $40\%$ to $50\%$ decrease in NAD+ 12 hpi of DC3000 challenged leaves compared to a mock control and $\sim50\%$ increase in total cellular and nuclear poly(ADP-Rib) polymers \citep{adams2010disruption}. Consistent with these results, a knockout of PARP2, which is induced by MAMPs, restricts DC3000 growth \citep{song2015parp2} demonstrating that loss of poly(ADP-ribosyl)ation activity affects the capacity of Arabidopsis to limit DC3000 growth.

The second primary metabolism example we choose to highlight is the very rapid induction methionine biosynthesis pathway (Figure~\ref{GO_figure}G). Methionine is a sulphur amino acid involved in multiple cellular processes from being a protein constituent, to initiation of mRNA translation as well as functioning as a regulatory molecule in the form of S-adenosylmethionine (SAM). There are 13 unique genes associated with this ontology, and while it is outside the scope of this manuscript to explore these in detail it is worth noting that this includes DMR1 (Downy Mildew Resistance 1) \citep{van2009downy}, encoding homoserine kinase, which produces O-phospho-L-homoserine, a compound at the branching point of methionine and threonine biosynthesis. Mutations in dmr1 lead to elevated foliar homoserine and resistance to the biotrophic pathogens \emph{Hyaloperonospora arabidopsidis}, \emph{Oidium neolycopersici}, \emph{F. culmorum} and \emph{F. graminearum}, although the mechanism has yet to be identified \citep{van2009downy,huibers2013powdery,brewer2014mutations}.

Thus in summary, we have validated perturbation times against previous analyses, and provide four new examples derived from examining early perturbation times of biological pathways to identify novel signalling and, particularly, primary metabolic pathways that are implicated in the transition from defence to disease following infection with DC3000. These examples provide compelling leads for further investigation.



\section{Conclusion}

We have introduced a fully Bayesian approach to infer the initial point where two gene expression time profiles diverge using a novel GP regression approach. We model the data as noise-corrupted samples coming from a shared function prior to some ``perturbation time" after which it splits into two conditionally independent functions. The full posterior distribution of the perturbation point is obtained through a simple histogram approach, providing a straightforward method to infer the divergence time between two gene expression time profiles under different conditions. The proposed method is applied to a study of the timing of transcriptional changes in \emph{Arabidopsis thaliana} under a bacterial challenge with a wild-type and disarmed strain. Analysis of differences in the gene expression profiles between strains is shown to be informative about the immune response. 

Many transcriptional perturbation experiments are focused on a single perturbation. However, multiple perturbations occurring at different times or a single perturbation targeting many conditions will be needed to unmask complex gene regulatory strategies. An interesting future line of research would be the development of GP covariance structures to uncover the ordering of events under these more general scenarios.

\section*{Acknowledgement} MR and JY were supported by the EU FP7 project RADIANT (grant no. 305626 to MR), MRG and CAP by the BBSRC [BB/F005806/1 to MRG and CAP] and EPSRC [EP/I036575/1 to CAP] UK research councils. 

\bibliographystyle{natbib}
\bibliography{myreference}

\beginsupplement
%

\onecolumn
\Large
\centerline{Supplementary Files}
\normalsize
\section{Model Derivation}\label{section_s1}
Consider the case where two time profiles, $f({\bX})$ and $g({\bZ})$, evaluated at specified sets of time points $\bX$ and $\bZ$, respectively, cross at the point $x_p$ with
$f(x_p) = g(x_p) = u$ at the crossing point. We use the same Gaussian process (GP) prior for each function,
\begin{align*}
f({\bX}) &\sim \mathcal{GP}(\mu(x),K(x,x')), \\
g({\bZ}) &\sim \mathcal{GP}(\mu(z),K(z,z')).
\end{align*}
\noindent Imposing the constraint that the functions cross at $x_p$, $p(f|\bX,u)$ and $p(g|\bZ,u)$ are then,
\begin{align*}
p(f(\bX)|u) & \sim \mathcal{N}(\mu_\bX,C_\bX), \\
p(g(\bZ)|u) & \sim \mathcal{N}(\mu_\bZ,C_\bZ),
\end{align*}

\noindent with  
\begin{eqnarray*}
\mu_\bX & = &  \frac{K(\bX,x_p)u}{K(x_p,x_p)},  \\
C_\bX & = & K(\bX,\bX) - \frac{K(\bX,x_p)K(\bX,x_p)^\top}{K(x_p,x_p)},\\
\mu_\bZ  & = &  \frac{K(\bZ,x_p)u}{K(x_p,x_p)},  \\
C_\bZ & =  & K(\bZ,\bZ) - \frac{K(\bZ,x_p)K(\bZ,x_p)^\top}{K(x_p,x_p)}.
\end{eqnarray*}

In practice, the time profiles $f(\bX)$ and $g(\bZ)$ are typically measured at the same time points, so that $\bZ$ can be replaced by $\bX$. The joint probably distribution of f and g is then obtained by
integrating over the latent function $u = f(x_p)$ which has a Gaussian distribution with variance $k(x_p,x_p)= \alpha$, $u \sim {\cal N}(0,\alpha)$, so that:
\begin{eqnarray*}
p(f(\bX),g(\bX)) \!\!\!\! & = & \!\!\!\!  \int p(f|{\bX},u)p(g|\bX,u)p(u) du,\\
& \propto &  \int \exp\left[-\frac{1}{2}(f-\mu_\bX)C_\bX^{-1}(f-\mu_\bX)^T -  \frac{1}{2}(g-\mu_\bX)C_\bX^{-1}(g-\mu_\bX)^T -\frac{u^2}{2\alpha} \right] du, \\
& \propto & \exp\left( -\frac{f^TC_\bX^{-1}f + g^TC_\bX^{-1}g}{2} \right) \int \exp \left[ (f^TC_\bX^{-1}k_\bX + g^TC_\bX^{-1}k_\bX) \frac{u}{\alpha} - k_\bX^TC_\bX^{-1}k_\bX \frac{u^2}{\alpha^2}
-\frac{u^2}{2\alpha} \right] du, \\
&=& \exp\left( -\frac{f^TC_\bX^{-1}f + g^TC_\bX^{-1}g}{2} \right)\int \exp 
\left\{ -\frac{2k_\bX^TC_\bX^{-1}k_\bX + \alpha}{2\alpha^2} \right. \\ 
&& \cdot \left.\left[\left(u-\frac{\alpha(f^TC_\bX^{-1}k_\bX + g^TC_\bX^{-1}k_\bX)}{2k_\bX^TC_\bX^{-1}k_\bX + \alpha}\right)^2-\left(\frac{\alpha(f^TC_\bX^{-1}k_\bX + g^TC_\bX^{-1}k_\bX)}{2k_\bX^TC_\bX^{-1}k_\bX
+ \alpha}\right)^2 \right]\right\}du,\\
& \propto & \exp  \left( -\frac{f^TC_\bX^{-1}f + g^TC_\bX^{-1}g}{2} + \frac{ (f^T C_\bX^{-1} k_\bX + g^T C_\bX^{-1} k_\bX)^2 }{4k_\bX^T C_\bX^{-1}k_\bX + 2\alpha} \right), \\
& = & \exp \left[ -\frac{1}{2}f^T \left (C_\bX^{-1} - \frac{C_\bX^{-1}k_{\bX}k_{\bX}^T C_{\bX}^{-1}}{2k_{\bX}^T C_{\bX}^{-1}k_{\bX} + \alpha}\right)f - \frac{1}{2}g^T \left (C_\bX^{-1} -
\frac{C_{\bX}^{-1}k_{\bX}k_{\bX}^T C_{\bX}^{-1}}{2k_{\bX}^T C_{\bX}^{-1} k_{\bX} + \alpha } \right)g \right. \\ & & + \left. \frac{1}{2} g^T \left( \frac{C_{\bX}^{-1}k_{\bX}k_{\bX}^T C_{\bX}^{-1}}{2k_{\bX}^T
C_{\bX}^{-1} k_{\bX} + \alpha} \right) f + \frac{1}{2} f^T \left( \frac{C_{\bX}^{-1}k_{\bX}k_{\bX}^T C_{\bX}^{-1}}{2k_{\bX}^T C_{\bX}^{-1} k_{\bX} + \alpha} \right) g \right], \\
& = & \exp \left( -\frac{1}{2} \begin{pmatrix} f \\ g \end{pmatrix}^T \begin{pmatrix}  A & B \\ B & A \end{pmatrix} \begin{pmatrix} f \\ g \end{pmatrix} \right)
 =  \exp \left( -\frac{1}{2} \begin{pmatrix} f \\ g \end{pmatrix}^T \begin{pmatrix}  K_{ff}& K_{fg} \\ K_{fg}^T & K_{gg} \end{pmatrix}^{-1} \begin{pmatrix} f \\ g \end{pmatrix} \right),
\end{eqnarray*}

\noindent where $K$ and $k_{\bf X}$ are abbreviations for $K(\bf X, X)$ and $K({\bf X},x_p)$, respectively, and 
\begin{eqnarray*}
A & = & C_{\bX}^{-1} - \frac{C_{\bX}^{-1}k_{\bX}k_{\bX}^T C_{\bX}^{-1}}{\alpha + 2 k_{\bX}^T C_{\bX}^{-1}k_{\bX}},  \\
B & = & -\frac{C_{\bX}^{-1}k_{\bX}k_{\bX}^T C_{\bX}^{-1}}{\alpha + 2 k_{\bX}^T C_{\bX}^{-1}k_{\bX}}.
\end{eqnarray*}

Let us say $p_1=k_{\bX}^T K^{-1} k_{\bX}$  and $p_2=k_{\bX}^TC^{-1}k_{\bX}$,
\begin{eqnarray*}
p_2=k_{\bX}^TC_{\bX}^{-1}k_{\bX} &=& \frac{\alpha k_{\bX}^TK^{-1} k_{\bX}}{\alpha - k_{\bX}^T K^{-1}  k_{\bX}}=\frac{\alpha p_1}{\alpha-p_1}, \\
\end{eqnarray*}

\begin{eqnarray*}
B & = &  -\frac{C_{\bX}^{-1}k_{\bX}k_{\bX}^T C_{\bX}^{-1}}{\alpha + 2 k_{\bX}^T C_{\bX}^{-1}k_{\bX}} =- \frac{\alpha^2 K^{-1}k_{\bX}k_{\bX}^TK^{-1}} {(\alpha- p_{1})^2(\alpha+2p_2)} = -\frac{\alpha K^{-1}k_{\bX}k_{\bX}^TK^{-1}}{ (\alpha^2-p_1^2)},\\
A & = & C_{\bX}^{-1} - B = K^{-1} + \frac{K^{-1} k_{\bX} k_{\bX}^T K^{-1}}{\alpha -  p_{1}}  - \frac{\alpha K^{-1}k_{\bX}k_{\bX}^TK^{-1}} { (\alpha^2-p_1^2)}=K^{-1} +\frac{p_1K^{-1} k_{\bX} k_{\bX}^T K^{-1} }{ (\alpha^2-p_1^2)},\\
\end{eqnarray*}

\noindent so,
\begin{eqnarray*}
A^{-1} &=& K-\frac{\frac{p_1KK^{-1}k_{\bX}k_{\bX}^TK^{-1}K}{(\alpha^2-p_1^2)}}{1+\frac{p_1 k_{\bX}^TK^{-1}KK^{-1}k_{\bX}}{(\alpha ^2-p_1^2)}} = K - \frac{p_1k_{bX}k_{\bX}^T}{(\alpha^2-p_1^2)+p_1 k_{\bX}^TK^{-1}k_{\bX}} = K -\frac{p_1}{\alpha^2}k_{\bX}k_{\bX}^T,\\
\end{eqnarray*}

\noindent then,
\begin{eqnarray*}
K_{ff} &=& (A-BA^{-1}B)^{-1} =K,\\
K_{fg} &=& -(A-BA^{-1}B)^{-1}BA^{-1}=K\frac{\alpha K^{-1}k_{\bX}k_{\bX}^TK^{-1}}{ (\alpha^2-p_1^2)}\{K -\frac{p_1}{\alpha^2}k_{\bX}k_{\bX}^T\}=\frac{k_{\bX}k_{\bX}^T}{\alpha},\\
K_{gf} &=& -A^{-1}B(A-BA^{-1}B)^{-1}=\frac{k_{\bX}k_{\bX}^T}{\alpha}, \\
K_{gg} &=& (A-BA^{-1}B)^{-1}=K. \\
\end{eqnarray*}

\section{Covariance matrix of the perturbation model}\label{section_s2}

Without loss of generality, we consider the discrete functions $\{f(x_i):i=1:n\}$ and $\{g(x_j):j=1:n\}$, here $n$ is the length of the data. Assuming that the perturbation occurs at $x=x_p$, according to the
perturbation model assumption, the data $\{f(x_i):i=1:n\}$ and $\{g(x_j):j=1:p\}$ can be represented by the same GP while $\{g(x_j):j=p+1:n\}$ are perturbed and represented by another GP, therefore, the matrix
blocks $K_{ff}$, $K_{fg}$, $K_{gf}$ and $K_{gg}$ will be changed accordingly to $\bar{K}_{ff}$, $\bar{K}_{fg}$, $\bar{K}_{gf}$ and $\bar{K}_{gg}$,

\begin{eqnarray*}
\bar{K}_{ff}^{[1:n,1:n]} &  = & K_{ff}^{[1:n,1:n]}, \\
\bar{K}_{fg}^{[1:n,1:p]} &  = & K_{ff}^{[1:n,1:p]}, \\
\bar{K}_{fg}^{[1:n,p+1:n]} &  = & K_{fg}^{[1:n,p+1:n]}, \\
\bar{K}_{gg}^{[1:p,1:p]} & = & K_{ff}^{[1:p,1:p]}, \\
\bar{K}_{gg}^{[1:p,p+1:n]} &=& K_{fg}^{[1:p,p+1:n]}, \\
\bar{K}_{gg}^{[p+1:n,1:p]} &=& K_{gf}^{[p+1:n,1:p]}, \\
\bar{K}_{gg}^{[p+1:n,p+1:n]} &=& K_{gg}^{[p+1:n,p+1:n]}, \\
\bar{K}_{gf}^{[1:n,1:n]} &  = & [\bar{K}_{fg}^{[1:n,1:n]}]^T, 
\end{eqnarray*}

\noindent here $K_m^{[a,b]}$ represents the element at row $a$ and column $b$ of the matrix $K_m$. Taking into account the noisy measurements, a small noise variance $\sigma^2$ will be added to the diagonal of the
covariance matrix, and eventually the covariance matrix for $f$ and $g$, $\bar{K}_{final}$, becomes

\begin{eqnarray*}
\bar{K}_{final} &=& \begin{pmatrix} \bar{K}_{ff} & \bar{K}_{fg} \\  \bar{K}_{gf} & \bar{K}_{gg}  \end{pmatrix} +\sigma^2\mathit{I_{(2n \times 2n)}}.
\end{eqnarray*}

\noindent When replicates are encountered in the model, the matrix will have to be repeated, so in the case of two replicates, the covariance matrix $\bar{K}_{final}$ becomes 

\begin{eqnarray*}
\bar{K}_{final} &=& \begin{pmatrix} \bar{K}_{ff} & \bar{K}_{ff} & \bar{K}_{fg} & \bar{K}_{fg} \\ \bar{K}_{ff} & \bar{K}_{ff} & \bar{K}_{fg} & \bar{K}_{fg} \\ \bar{K}_{fg} & \bar{K}_{fg} & \bar{K}_{gg} & \bar{K}_{gg} \\ \bar{K}_{fg} & \bar{K}_{fg} & \bar{K}_{gg} & \bar{K}_{gg} \end{pmatrix}+\sigma^2\mathit{I_{(4n
\times 4n)}}.
\end{eqnarray*}

\section{Results on simulated data}\label{section_s3}

We generated data under a range of different scenarios to explore performance and robustness to deviations from the model. We generated expression profiles from three different covariance models, one matching the one used for inference and other two generating rougher profiles. We then add noise using three different noise models, one matching the Gaussian model used for inference and two from heavier-tailed distributions with the same set of hyperparameters. 

\begin{enumerate}
\item \emph{profile$_1$}: simulated data generated from the model $\mathcal{GP}_{\theta}(\mathbf{0},\hat{\Sigma}_{\theta})$ with $\hat{\Sigma}_{\theta}$ assuming a squared exponential covariance function. 
 \item \emph{profile$_2$}: simulated data generated from above Gaussian process model with the covariance function in the form of  a matern32 covariance function.
 \item \emph{profile$_3$}: simulated data generated from above Gaussian process model with the covariance function in the form of  a matern12 covariance function.
 \end{enumerate}

 Nine simulated dataset were induced with different kinds of i.i.d noise on top of \emph{profile$_1$}, \emph{profile$_2$} and \emph{profile$_3$}, respectively: Gaussian $\mathcal{N}(0,1.5)$, Student-t distributed with 3 $(T(3))$ and 6 $(T(6))$ degrees of freedom. The simulated data were sampled every hour from 0hrs till 18hrs. Perturbation times ${\mathbf x_p}$ were set in the range $\{0,1,\cdots, 17,18\}$ and 100 different sets of data were produced for each $x_p$ with each dataset having 2 replicates.    
 
\begin{figure*}
   \begin{minipage}[c]{0.5\textwidth}
  \includegraphics[width=\textwidth]{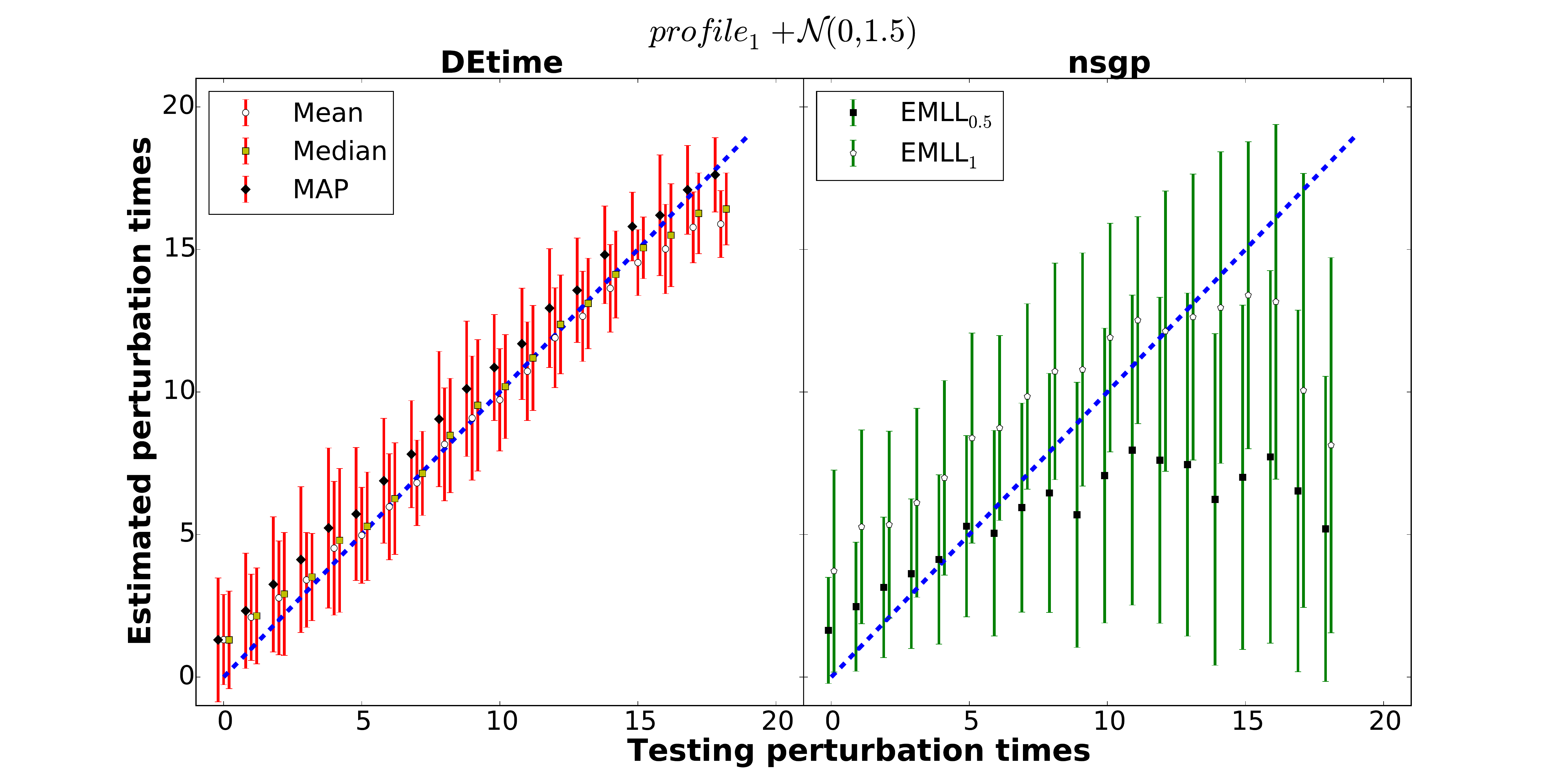}
  \end{minipage}
 \begin{minipage}[c]{0.5\textwidth}
  \includegraphics[width=\textwidth]{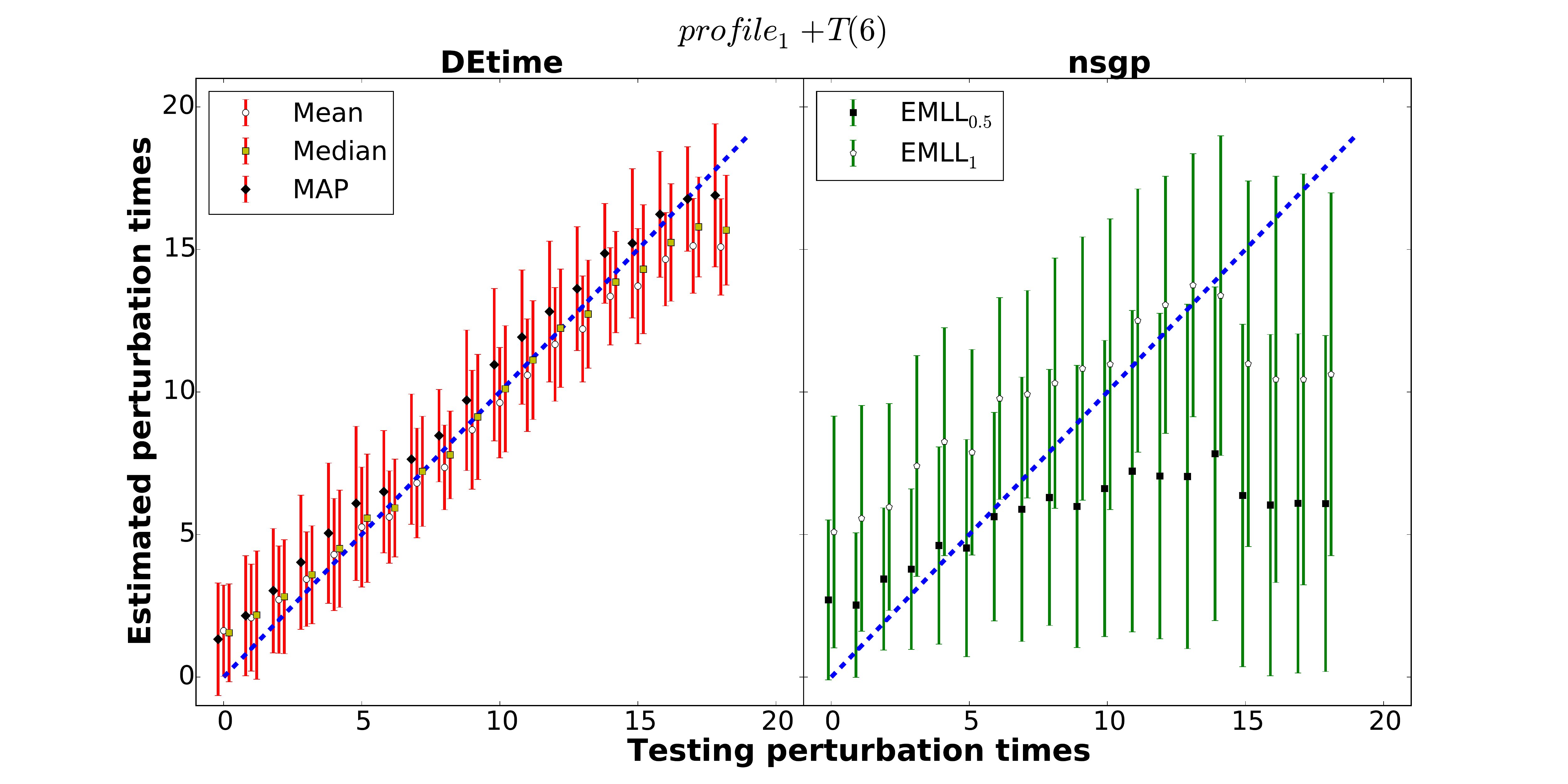}
 \end{minipage}
 \begin{minipage}[c]{0.5\textwidth}
  \includegraphics[width=\textwidth]{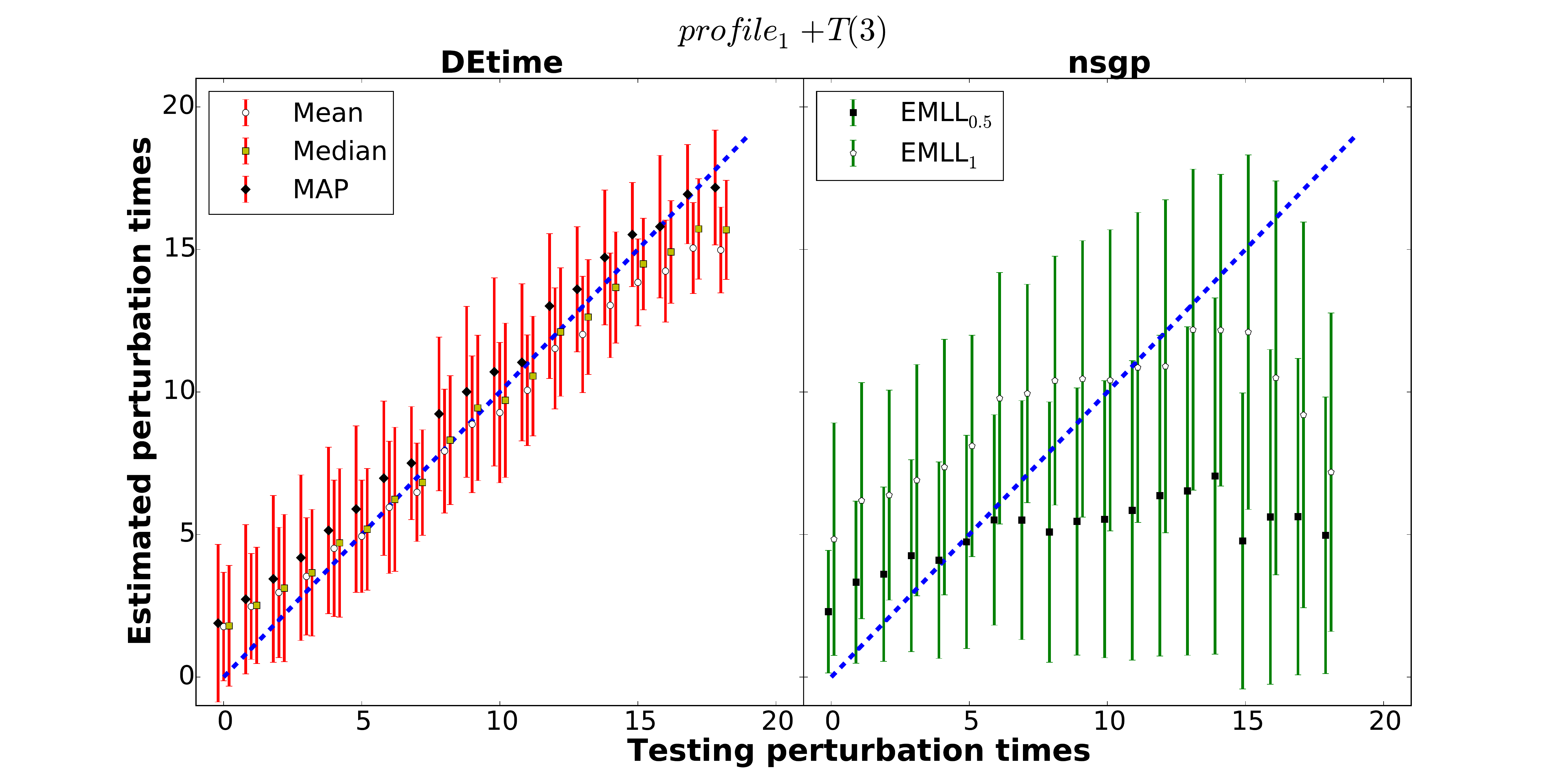}
 \end{minipage}
  \begin{minipage}[c]{0.5\textwidth}
  \includegraphics[width=\textwidth]{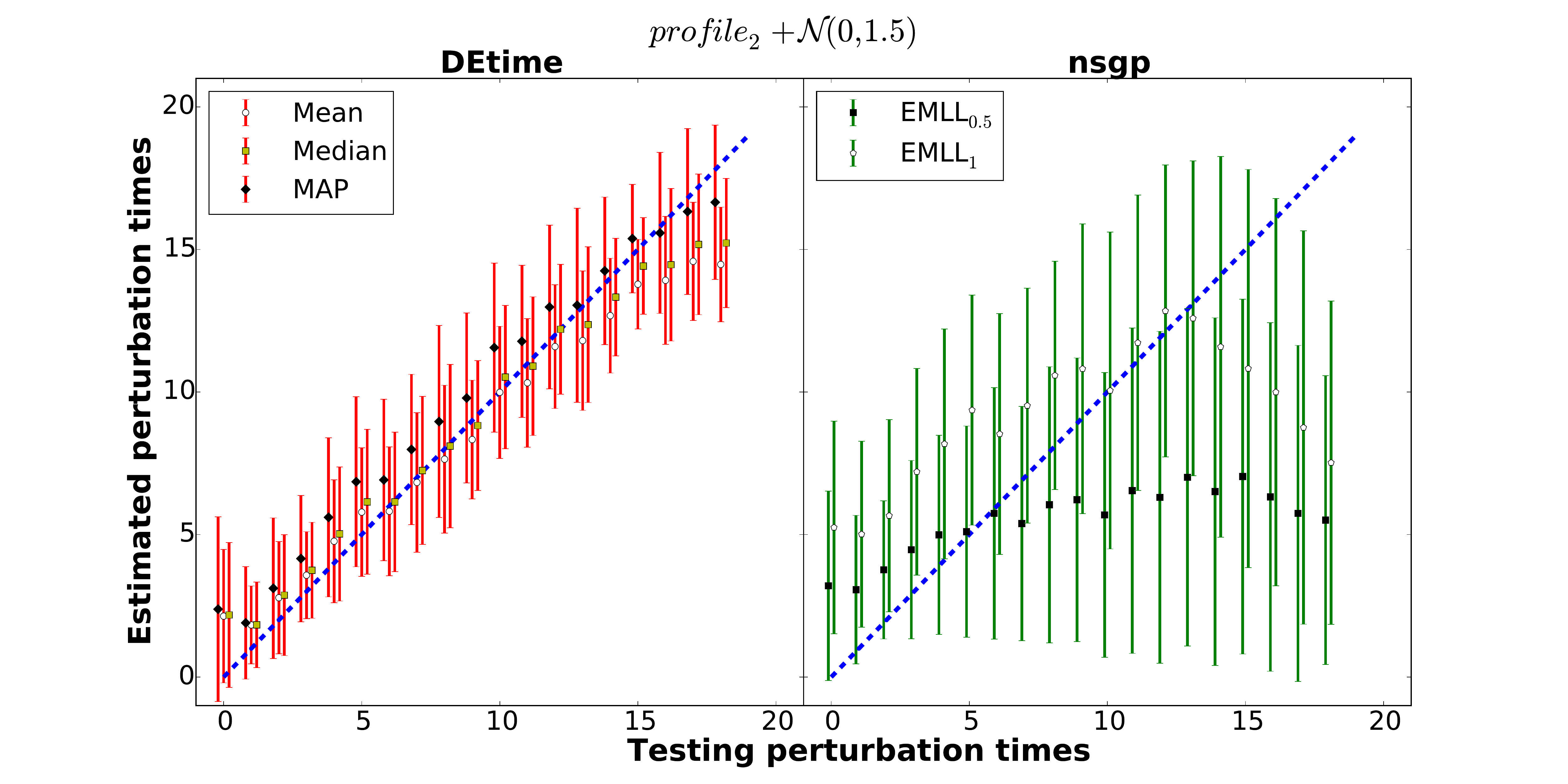}
  \end{minipage}
 \begin{minipage}[c]{0.5\textwidth}
  \includegraphics[width=\textwidth]{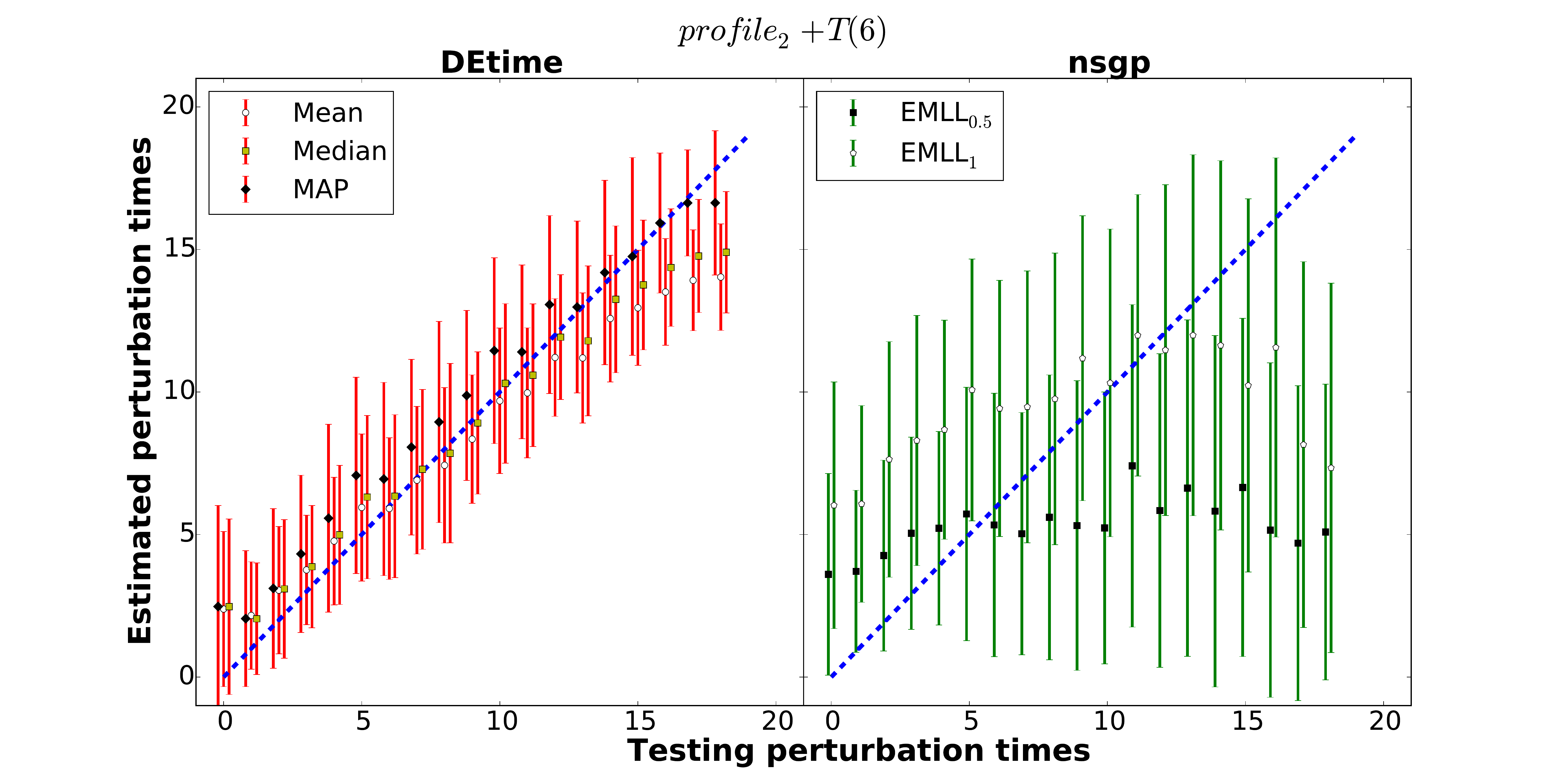}
 \end{minipage}
 \begin{minipage}[c]{0.5\textwidth}
  \includegraphics[width=\textwidth]{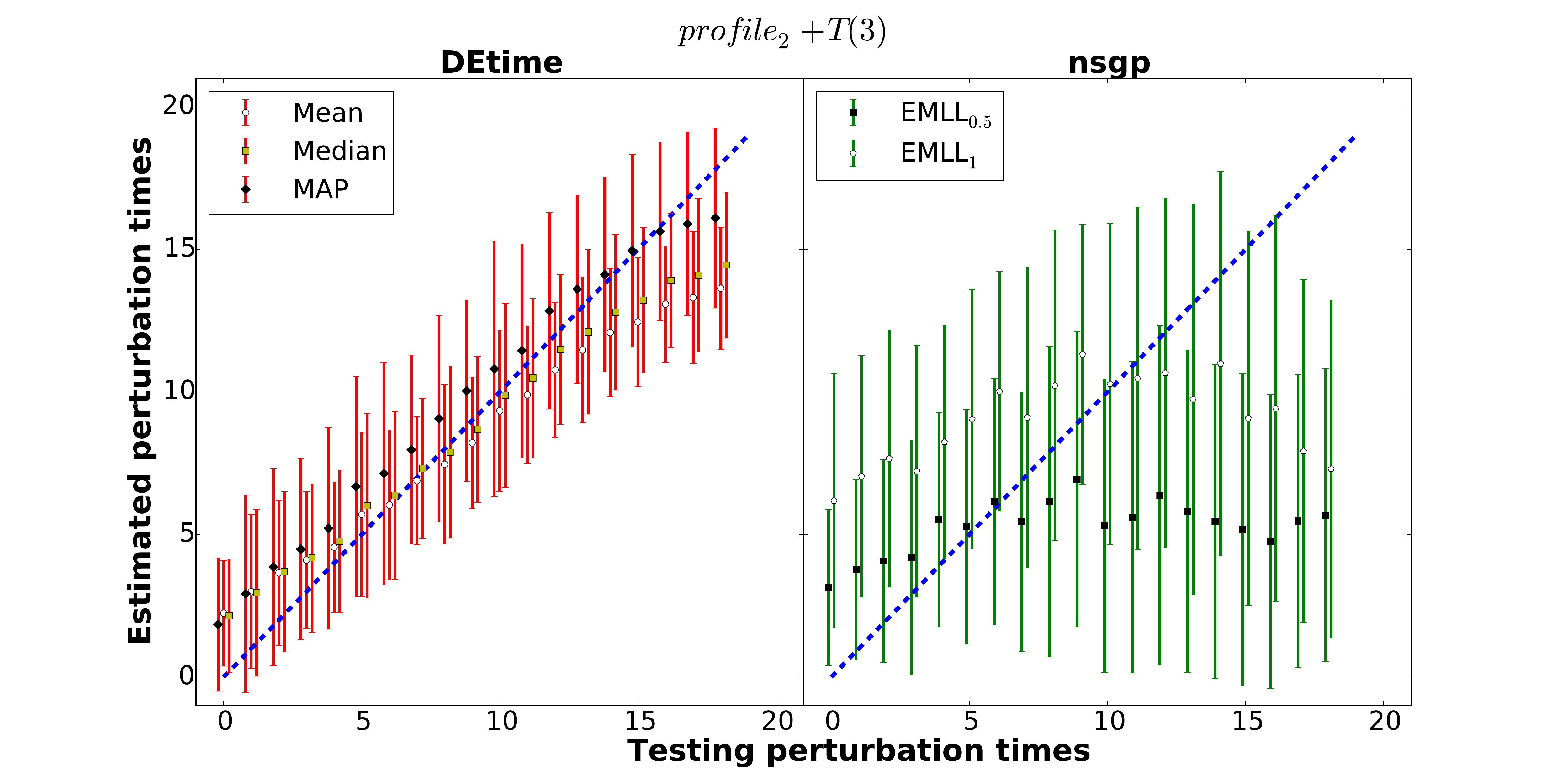}
 \end{minipage}
   \begin{minipage}[c]{0.5\textwidth}
  \includegraphics[width=\textwidth]{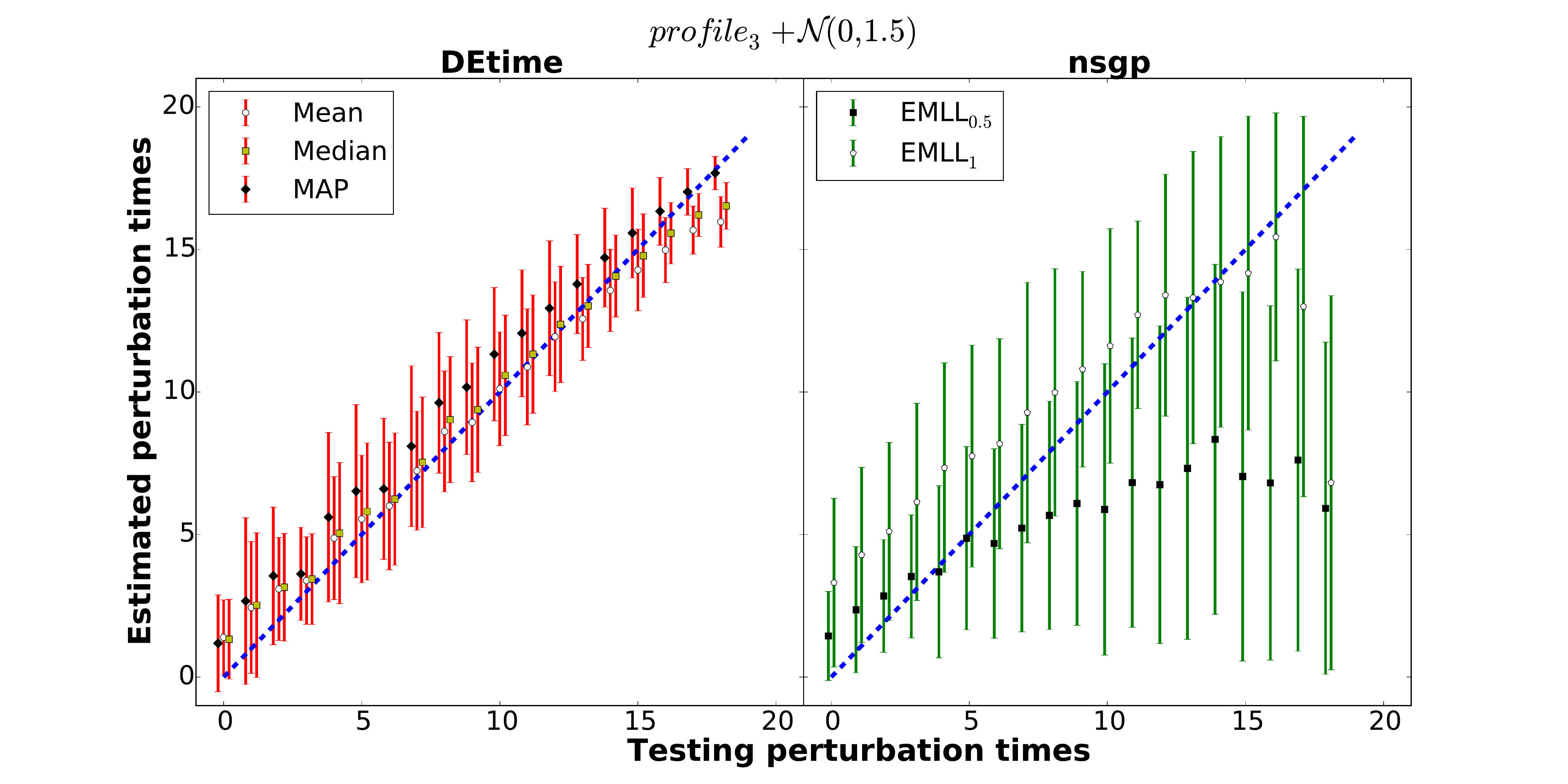}
  \end{minipage}
 \begin{minipage}[c]{0.5\textwidth}
  \includegraphics[width=\textwidth]{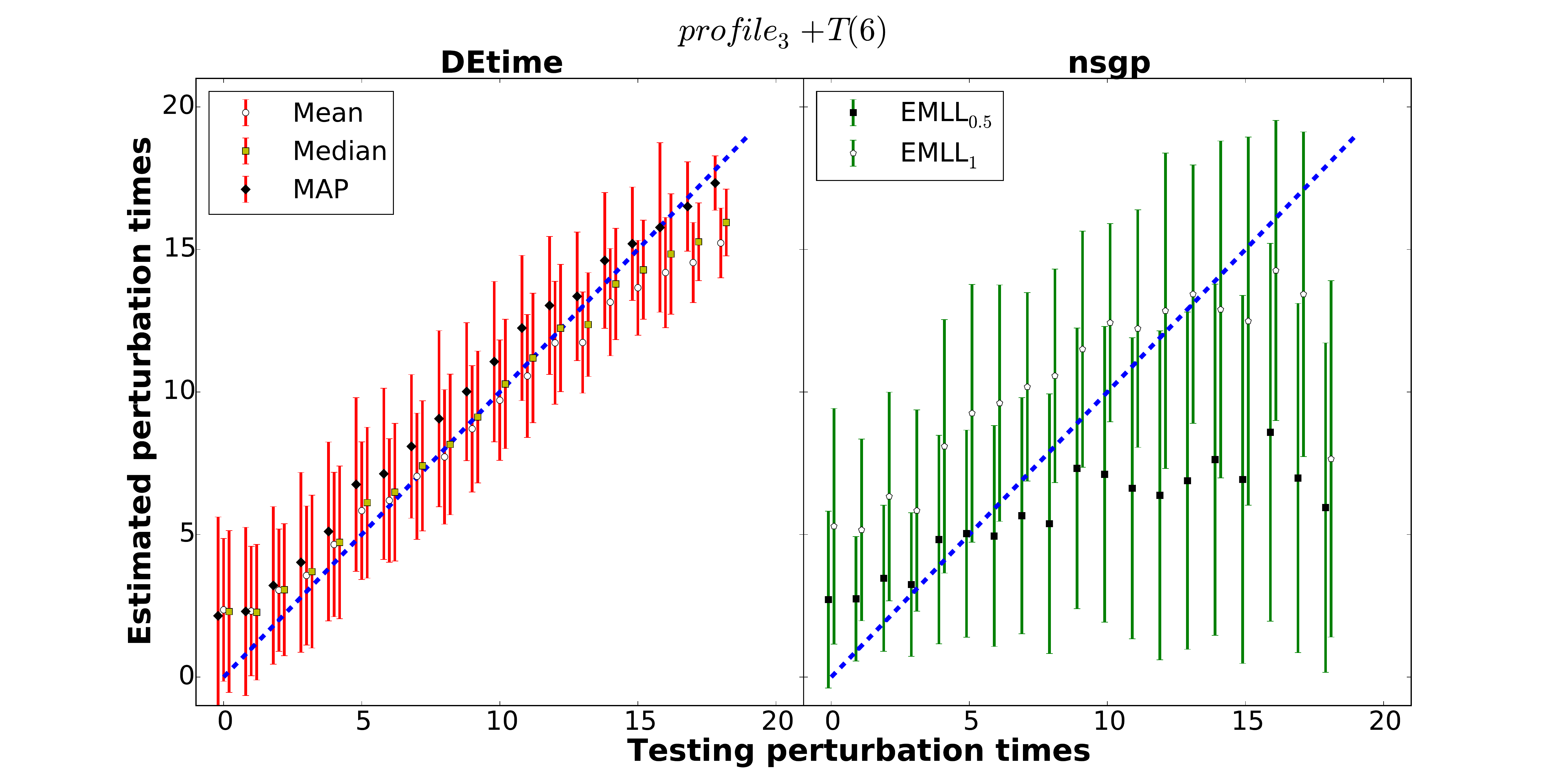}
 \end{minipage}
 \begin{minipage}[c]{0.5\textwidth}
  \includegraphics[width=\textwidth]{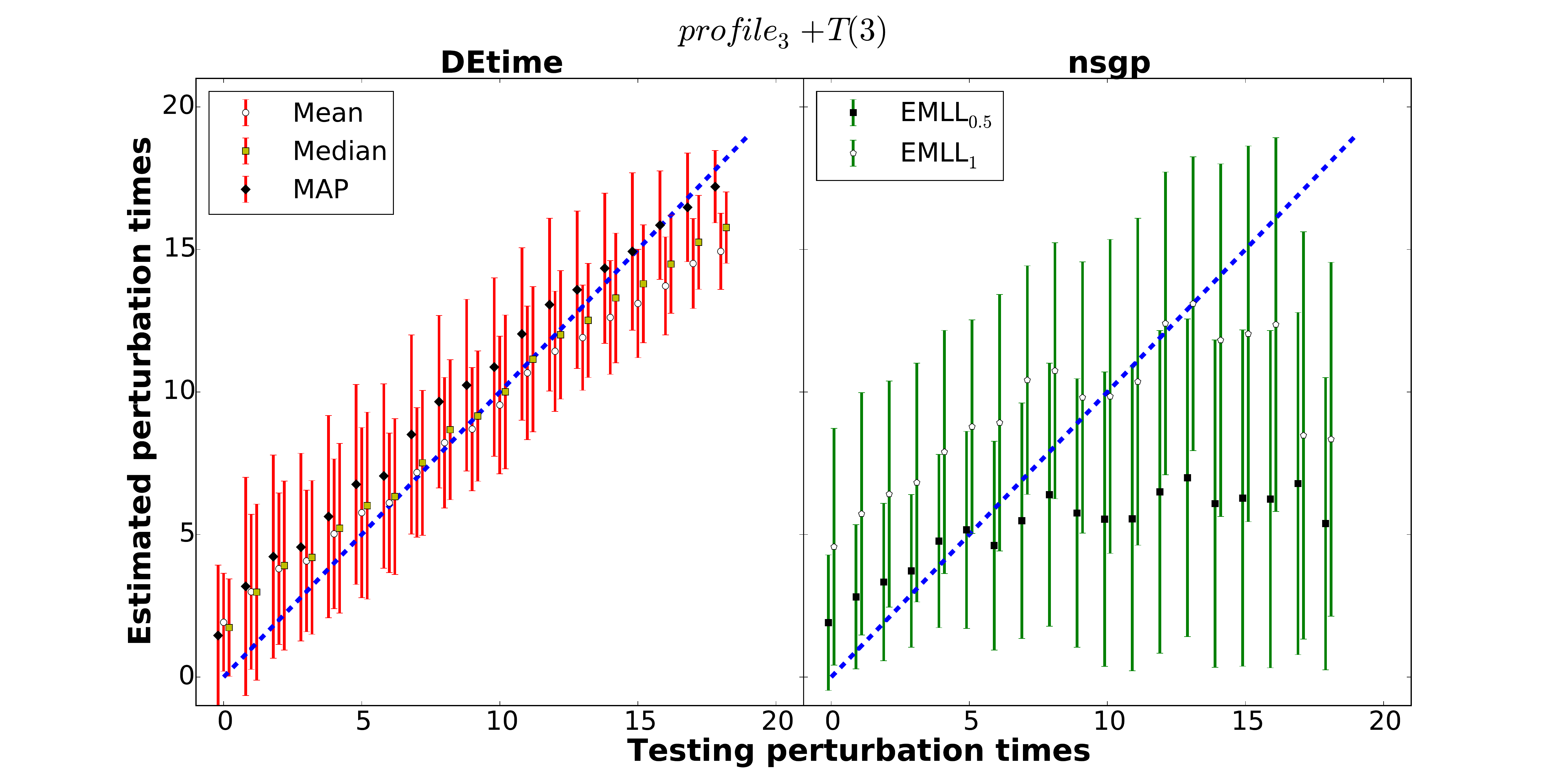}
 \end{minipage}
 \caption{Estimated mean, median, MAP (from DEtime package), EMLL$_{0.5}$, EMLL$_{1.0}$ (from nsgp package) and 5-95 percentile coverage of the posterior distributions on tested perturbation times of 100 dataset each generated over the nine types of simulated data with hyperparameters $\alpha=30.0, l=8.0$.}
 \label{mean_median_MAP}
 \end{figure*}

Here we compare our approach to the most recently published package of this type, developed by \citet{Heinonen2014} implemented in the nsgp R-package. The nsgp package infers the differentially expressed time periods and uses four likelihood ratios: marginal log-likelihood ratio (MLL), expected marginal log-likelihood ratio (EMLL), the posterior concentration (PC) and the noisy posterior concentration (NPC) to quantify these regions. We adopt thresholds of 0.5 and 1.0 to define the initial perturbation points, respectively. The mean, median and mode of the posterior distribution of the inferred perturbation points from our method are also computed.  In order to test the robustness of our proposed model, we fixed the lengthscale parameter ($l=8.0$) and varied the amplitude parameter ($\alpha=30.0,20.0,10.0,1.5$) in different kernel profiles. The performance of ranking $x_p$ using each method is measured by Spearman's rank correlation coefficient with the known ground truth and the mean and standard deviation of the rank correlation coefficients across 100 dataset are illustrated in Table \ref{table_s1}. It is clear that the proposed algorithm provides robust estimations to the perturbation points under reasonable noise conditions. EMLL with threshold of 1.0 provides the best results among those from nsgp package, however, its performance is much worse than those from the DEtime package. The mean, meadian and MAP results from DEtime package and EMLL$_{0.5}$ and EMLL$_{1.0}$ from nsgp package with $\alpha=30.0$ are illustrated in Fig. \ref{mean_median_MAP}.

\begin{table*}\scriptsize
\setlength{\tabcolsep}{2pt}
\renewcommand{\arraystretch}{1.1}
\caption{Comparison of the means and stds of the Spearman's rank correlation coefficients of the mean, median and MAP estimation of the perturbation times from our DEtime package and different likelihood ratios with various thresholds from the nsgp package under various $\alpha$. $M_n$ represents the $M$ ratio with a threshold of $n$. }\label{table_s1}
\begin{tabular}{c| c| c| c| c| c| c| c| c| c| c| c }
\hline
  R Package & \multicolumn{3}{c}{DEtime} & \multicolumn{8}{|c}{ nsgp}\\

\hline

\multicolumn{11}{c}{$\alpha=30.0$}\\
\hline
 Data & Mean & Median & MAP & MLL$_{0.5}$& EMLL$_{0.5}$ & PC$_{0.5}$ & NPC$_{0.5}$ & MLL$_{1.0}$& EMLL$_{1.0}$ & PC$_{1.0}$ & NPC$_{1.0}$ \\ 
\hline
\emph{profile$_1$}$+\mathcal{N}(1.5)$ & 0.94$\pm$0.04& 0.94$\pm$0.04& 0.93$\pm$0.05 & -0.02$\pm$0.23 & 0.26$\pm$0.22 &0.36$\pm$0.22 &0.26$\pm$0.23 & -0.02$\pm$0.23 & 0.67$\pm$0.16 & 0.29$\pm$0.22 &0.48$\pm$0.21  \\
\emph{profile$_2$}$+\mathcal{N}(1.5)$ & 0.90$\pm$0.06& 0.90$\pm$0.06& 0.88$\pm$0.08 & -0.05$\pm$0.23 & 0.17$\pm$0.23 &0.21$\pm$0.25 &0.24$\pm$0.22 & -0.06$\pm$0.23 & 0.57$\pm$0.20 & 0.18$\pm$0.24 &0.42$\pm$0.21  \\
\emph{profile$_3$}$+\mathcal{N}(1.5)$ & 0.93$\pm$0.04& 0.93$\pm$0.04& 0.92$\pm$0.05 & -0.04$\pm$0.26 & 0.31$\pm$0.22 &0.21$\pm$0.25 &0.28$\pm$0.25 & -0.05$\pm$0.26 & 0.75$\pm$0.16 & 0.20$\pm$0.23 &0.47$\pm$0.23  \\
\emph{profile$_1$}$+T(6)$ & 0.93$\pm$0.05& 0.92$\pm$0.06& 0.91$\pm$0.07 & 0.01$\pm$0.24 & 0.22$\pm$0.24 &0.24$\pm$0.27 &0.23$\pm$0.24 & 0.02$\pm$0.23 & 0.59$\pm$0.18 & 0.17$\pm$0.23 &0.40$\pm$0.23  \\
\emph{profile$_2$}$+T(6)$ & 0.87$\pm$0.07& 0.86$\pm$0.07& 0.83$\pm$0.09 & 0.04$\pm$0.23 & 0.27$\pm$0.25 &0.08$\pm$0.24 &0.15$\pm$0.22 & 0.03$\pm$0.23 & 0.42$\pm$0.24 & 0.02$\pm$0.24 &0.28$\pm$0.24  \\
\emph{profile$_3$}$+T(6)$ & 0.89$\pm$0.06& 0.89$\pm$0.06& 0.87$\pm$0.08 & 0.01$\pm$0.26 & 0.24$\pm$0.24 &0.13$\pm$0.25 &0.26$\pm$0.26 & -0.00$\pm$0.27 & 0.64$\pm$0.20 & 0.07$\pm$0.25 &0.41$\pm$0.23  \\
\emph{profile$_1$}$+T(3)$ & 0.91$\pm$0.05& 0.90$\pm$0.06& 0.89$\pm$0.06 & -0.02$\pm$0.24 & 0.14$\pm$0.22 &0.19$\pm$0.22 &0.20$\pm$0.21 & -0.03$\pm$0.24 & 0.48$\pm$0.21 & 0.15$\pm$0.23 &0.32$\pm$0.21  \\
\emph{profile$_2$}$+T(3)$ & 0.83$\pm$0.09& 0.83$\pm$0.10& 0.80$\pm$0.11 & -0.03$\pm$0.26 & 0.08$\pm$0.23 &0.12$\pm$0.23 &0.16$\pm$0.21 & -0.04$\pm$0.26 & 0.36$\pm$0.23 & 0.05$\pm$0.22 &0.24$\pm$0.24  \\
\emph{profile$_3$}$+T(3)$ & 0.87$\pm$0.07& 0.87$\pm$0.07& 0.84$\pm$0.09 & -0.01$\pm$0.25 & 0.20$\pm$0.21 &0.09$\pm$0.23 &0.20$\pm$0.18 & 0.00$\pm$0.25 & 0.54$\pm$0.22 & 0.09$\pm$0.23 &0.33$\pm$0.23  \\

\hline

\multicolumn{11}{c}{$\alpha=20.0$}\\
\hline
 Data & Mean & Median & MAP & MLL$_{0.5}$& EMLL$_{0.5}$ & PC$_{0.5}$ & NPC$_{0.5}$ & MLL$_{1.0}$& EMLL$_{1.0}$ & PC$_{1.0}$ & NPC$_{1.0}$ \\ 
\hline
\emph{profile$_1$}$+\mathcal{N}(1.5)$ & 0.93$\pm$0.05& 0.93$\pm$0.05& 0.91$\pm$0.07 & -0.02$\pm$0.23 & 0.20$\pm$0.22 & 0.29$\pm$0.20 &0.22$\pm$0.18 & 0.01$\pm$0.23 & 0.62$\pm$0.18 & 0.26$\pm$0.23 &0.47$\pm$0.20  \\
\emph{profile$_2$}$+\mathcal{N}(1.5)$ & 0.88$\pm$0.08& 0.88$\pm$0.10& 0.86$\pm$0.10 & 0.01$\pm$0.23 & 0.16$\pm$0.23 &0.17$\pm$0.23 & 0.25$\pm$0.24 & -0.01$\pm$0.25 & 0.46$\pm$0.24 &0.11$\pm$0.23 &0.34$\pm$0.22  \\
\emph{profile$_3$}$+\mathcal{N}(1.5)$ & 0.91$\pm$0.05& 0.91$\pm$0.06& 0.89$\pm$0.06 & -0.05$\pm$0.25 & 0.34$\pm$0.20 &0.18$\pm$0.24 &0.31$\pm$0.22 & -0.05$\pm$0.24 & 0.66$\pm$0.18 & 0.12$\pm$0.24 &0.46$\pm$0.23  \\
\emph{profile$_1$}$+T(6)$ & 0.89$\pm$0.07& 0.89$\pm$0.08& 0.86$\pm$0.09 & -0.04$\pm$0.22 & 0.12$\pm$0.22 & 0.18$\pm$0.21 &0.17$\pm$0.25 & -0.03$\pm$0.22 & 0.43$\pm$0.20 & 0.12$\pm$0.21 &0.34$\pm$0.21  \\
\emph{profile$_2$}$+T(6)$ & 0.82$\pm$0.11& 0.82$\pm$0.12& 0.78$\pm$0.13 & -0.04$\pm$0.25 & 0.08$\pm$0.20 &0.07$\pm$0.23 &0.14$\pm$0.21 & -0.03$\pm$0.25 & 0.35$\pm$0.22 & 0.05$\pm$0.22 &0.24$\pm$0.23  \\
\emph{profile$_3$}$+T(6)$ & 0.87$\pm$0.06& 0.87$\pm$0.07& 0.84$\pm$0.08 & -0.04$\pm$0.25 & 0.22$\pm$0.21 &0.07$\pm$0.23 &0.22$\pm$0.23 & -0.04$\pm$0.25 & 0.51$\pm$0.21 & 0.09$\pm$0.21 &0.35$\pm$0.22 \\
\emph{profile$_1$}$+T(3)$ & 0.87$\pm$0.07& 0.87$\pm$0.07& 0.85$\pm$0.08 & 0.03$\pm$0.22 & 0.13$\pm$0.20 &0.15$\pm$0.22 &0.18$\pm$0.18 & 0.03$\pm$0.22 & 0.46$\pm$0.21 & 0.09$\pm$0.23 &0.31$\pm$0.21  \\
\emph{profile$_2$}$+T(3)$ & 0.77$\pm$0.10& 0.76$\pm$0.11& 0.71$\pm$0.15 & -0.01$\pm$0.25 & 0.04$\pm$0.22 &0.06$\pm$0.25 &0.13$\pm$0.26 & -0.01$\pm$0.25 & 0.31$\pm$0.24 & 0.07$\pm$0.21 &0.21$\pm$0.25  \\
\emph{profile$_3$}$+T(3)$ & 0.85$\pm$0.07& 0.85$\pm$0.08& 0.83$\pm$0.09 & -0.02$\pm$0.25 & 0.12$\pm$0.21 & 0.05$\pm$0.24 &0.16$\pm$0.21 & -0.00$\pm$0.25 & 0.46$\pm$0.24 & 0.03$\pm$0.23 &0.23$\pm$0.24 \\

\hline
\multicolumn{11}{c}{$\alpha=10.0$}\\
\hline
 Data & Mean & Median & MAP & MLL$_{0.5}$& EMLL$_{0.5}$ & PC$_{0.5}$ & NPC$_{0.5}$ & MLL$_{1.0}$& EMLL$_{1.0}$ & PC$_{1.0}$ & NPC$_{1.0}$ \\ 
\hline
\emph{profile$_1$}$+\mathcal{N}(1.5)$ & 0.89$\pm$0.06& 0.88$\pm$0.07& 0.85$\pm$0.10 & -0.02$\pm$0.25 & 0.18$\pm$0.22 & 0.18$\pm$0.25 &0.23$\pm$0.22 & -0.03$\pm$0.24 & 0.45$\pm$0.24 & 0.13$\pm$0.23 &0.34$\pm$0.23  \\
\emph{profile$_2$}$+\mathcal{N}(1.5)$ & 0.81$\pm$0.11& 0.80$\pm$0.12& 0.76$\pm$0.13 & -0.01$\pm$0.22 & 0.09$\pm$0.22 &0.07$\pm$0.22 & 0.19$\pm$0.20 & -0.01$\pm$0.23 & 0.34$\pm$0.24 &0.04$\pm$0.22 &0.23$\pm$0.21  \\
\emph{profile$_3$}$+\mathcal{N}(1.5)$ & 0.88$\pm$0.06& 0.88$\pm$0.06& 0.85$\pm$0.09 & -0.03$\pm$0.20 & 0.20$\pm$0.21 &0.03$\pm$0.25 &0.25$\pm$0.21 & -0.03$\pm$0.21 & 0.55$\pm$0.19 & 0.03$\pm$0.23 &0.32$\pm$0.21  \\
\emph{profile$_1$}$+T(6)$ & 0.83$\pm$0.08& 0.83$\pm$0.09& 0.80$\pm$0.11 & -0.03$\pm$0.25 & 0.16$\pm$0.24 & 0.11$\pm$0.25 &0.20$\pm$0.24 & -0.02$\pm$0.25 & 0.36$\pm$0.23 & 0.05$\pm$0.24 &0.26$\pm$0.22  \\
\emph{profile$_2$}$+T(6)$ & 0.75$\pm$0.12& 0.73$\pm$0.12& 0.69$\pm$0.14 & -0.02$\pm$0.23 & 0.10$\pm$0.21 &0.03$\pm$0.25 &0.15$\pm$0.22 & -0.01$\pm$0.24 & 0.27$\pm$0.21 & -0.01$\pm$0.23 &0.16$\pm$0.22  \\
\emph{profile$_3$}$+T(6)$ & 0.82$\pm$0.10& 0.82$\pm$0.11& 0.77$\pm$0.14 & -0.02$\pm$0.23 & 0.14$\pm$0.23 &0.03$\pm$0.24 &0.19$\pm$0.24 & -0.01$\pm$0.23 & 0.37$\pm$0.20 & 0.03$\pm$0.23 &0.23$\pm$0.22 \\
\emph{profile$_1$}$+T(3)$ & 0.82$\pm$0.09& 0.81$\pm$0.11& 0.77$\pm$0.11 & 0.04$\pm$0.22 & 0.07$\pm$0.22 &0.05$\pm$0.25 &0.10$\pm$0.24 & 0.04$\pm$0.23 & 0.30$\pm$0.21 & 0.04$\pm$0.25 &0.17$\pm$0.23  \\
\emph{profile$_2$}$+T(3)$ & 0.74$\pm$0.12& 0.73$\pm$0.12& 0.69$\pm$0.14 & -0.02$\pm$0.23 & 0.11$\pm$0.21 &0.02$\pm$0.24 &0.15$\pm$0.22 & -0.02$\pm$0.24 & 0.27$\pm$0.21 & -0.01$\pm$0.23 &0.16$\pm$0.22  \\
\emph{profile$_3$}$+T(3)$ & 0.75$\pm$0.12& 0.75$\pm$0.13& 0.70$\pm$0.15 & 0.01$\pm$0.22 & 0.08$\pm$0.24 &-0.00$\pm$0.24 &0.12$\pm$0.23 & 0.00$\pm$0.22 & 0.29$\pm$0.24 & 0.03$\pm$0.24 &0.15$\pm$0.25 \\

\hline
\multicolumn{11}{c}{$\alpha=1.5$}\\
\hline
 Data & Mean & Median & MAP & MLL$_{0.5}$& EMLL$_{0.5}$ & PC$_{0.5}$ & NPC$_{0.5}$ & MLL$_{1.0}$& EMLL$_{1.0}$ & PC$_{1.0}$ & NPC$_{1.0}$ \\ 
\hline
\emph{profile$_1$}$+\mathcal{N}(1.5)$ & 0.69$\pm$0.14& 0.68$\pm$0.15& 0.62$\pm$0.17 & 0.01$\pm$0.24 & 0.05$\pm$0.22 &-0.01$\pm$0.20 &0.10$\pm$0.23 & -0.01$\pm$0.23 & 0.18$\pm$0.24 & 0.01$\pm$0.20 &0.10$\pm$0.20  \\
\emph{profile$_2$}$+\mathcal{N}(1.5)$ & 0.60$\pm$0.18& 0.58$\pm$0.18& 0.52$\pm$0.20 & 0.02$\pm$0.23 & -0.03$\pm$0.21 &-0.05$\pm$0.21 &0.03$\pm$0.21 & 0.01$\pm$0.22 & 0.14$\pm$0.24 & -0.04$\pm$0.21 &0.02$\pm$0.23  \\
\emph{profile$_3$}$+\mathcal{N}(1.5)$ & 0.60$\pm$0.17& 0.59$\pm$0.18& 0.51$\pm$0.22 & -0.05$\pm$0.23 & 0.04$\pm$0.21 &0.02$\pm$0.24 &0.08$\pm$0.23 & -0.06$\pm$0.23 & 0.20$\pm$0.23 & 0.02$\pm$0.23 &0.13$\pm$0.24  \\
\emph{profile$_1$}$+T(6)$ & 0.59$\pm$0.16& 0.57$\pm$0.16& 0.51$\pm$0.18 & -0.02$\pm$0.25 & 0.06$\pm$0.24 &-0.01$\pm$0.25 &0.09$\pm$0.21 & 0.00$\pm$0.25 & 0.15$\pm$0.22 & 0.03$\pm$0.28 &0.07$\pm$0.23  \\
\emph{profile$_2$}$+T(6)$ & 0.49$\pm$0.18& 0.47$\pm$0.19& 0.41$\pm$0.21 & 0.01$\pm$0.23 & 0.01$\pm$0.23 &-0.04$\pm$0.23 &0.03$\pm$0.24 & 0.02$\pm$0.23 & 0.11$\pm$0.21 & -0.02$\pm$0.23 &-0.04$\pm$0.23  \\
\emph{profile$_3$}$+T(6)$ & 0.49$\pm$0.19& 0.49$\pm$0.21& 0.41$\pm$0.24 & 0.03$\pm$0.23 & 0.05$\pm$0.23 &-0.05$\pm$0.25 &0.05$\pm$0.23 & 0.02$\pm$0.23 & 0.11$\pm$0.25 & -0.03$\pm$0.26 &0.06$\pm$0.26 \\
\emph{profile$_1$}$+T(3)$ & 0.52$\pm$0.15& 0.52$\pm$0.17& 0.44$\pm$0.19 & 0.00$\pm$0.23 & 0.00$\pm$0.22 &-0.03$\pm$0.22 &0.06$\pm$0.21 & 0.01$\pm$0.22 & 0.08$\pm$0.24 & -0.02$\pm$0.21 &0.04$\pm$0.22  \\
\emph{profile$_2$}$+T(3)$ & 0.40$\pm$0.19& 0.38$\pm$0.20& 0.30$\pm$0.20 & 0.02$\pm$0.22 & 0.00$\pm$0.22 &-0.04$\pm$0.25 &0.03$\pm$0.22 & 0.03$\pm$0.22 & 0.03$\pm$0.22 & -0.02$\pm$0.23 &0.01$\pm$0.25  \\
\emph{profile$_3$}$+T(3)$ & 0.41$\pm$0.21& 0.41$\pm$0.20& 0.35$\pm$0.20 & -0.03$\pm$0.25 & 0.07$\pm$0.21 &-0.02$\pm$0.26 &0.06$\pm$0.24 & -0.02$\pm$0.25 & 0.10$\pm$0.24 & -0.02$\pm$0.26 &0.07$\pm$0.26 \\

\hline
\end{tabular}
\end{table*}

\section{GO and GSEA analysis results}

\subsection{GO and GSEA analysis results for differentially expressed genes}\label{section_s4.1}

We used the method to study \emph{Arabidopsis thaliana} genes following inoculation with the hemibiotrophic bacteria \emph{Pseudomonas syringae} \citep{Lewis:15}. This dataset includes time series from two conditions: (i) infection of Arabidopsis with virulent \emph{Pseudomonas syringage} pv. tomato DC3000, which leads to disease development (condition 1); and (ii) infection of Arabidopsis with the disarmed strain DC3000\emph{hrpA} (condition 2). Each of the three time series comprised $13$ time points at times $t = [0,2,3,4,6,7,8,10,11,12,14,16,17.5]$ hours post inoculation (hpi). Each treatment/challenge was undertaken on leaf $8$ and comprised $4$ biological replicates. The data are deposited at Gene Expression Omnibus under the accession number GSE56094. 

A key difference between time series in condition 1 and condition 2 is that the DC3000\emph{hrpA} mutant is compromised in production of a major component of the Type Three Secretion System and cannot deliver bacterial effectors into the plant cell. DC3000\emph{hrpA} challenge triggers basal defence through activation of innate immune receptors in response to microbe/pathogen-associated molecular pattern (M/PAMP). Thus DC3000\emph{hrpA} (condition 2) reports induced basal defences. By contrast, in condition 1, DC3000 delivers $\sim 28$ ``effector'' proteins into the plant cell and these effectors collectively suppress innate immunity and reconfigure plant metabolism for bacterial sustenance \citep{cunnac2009pseudomonas}. Comparisons between condition 1 and condition 2 will reveal important information about how basal immune responses are suppressed or subverted by the pathogen whilst later time points should reveal the metabolic reprogramming to provide nutrient to the apoplastically localised bacteria.

\begin{figure}
\centering
\includegraphics[width=0.5\textwidth]{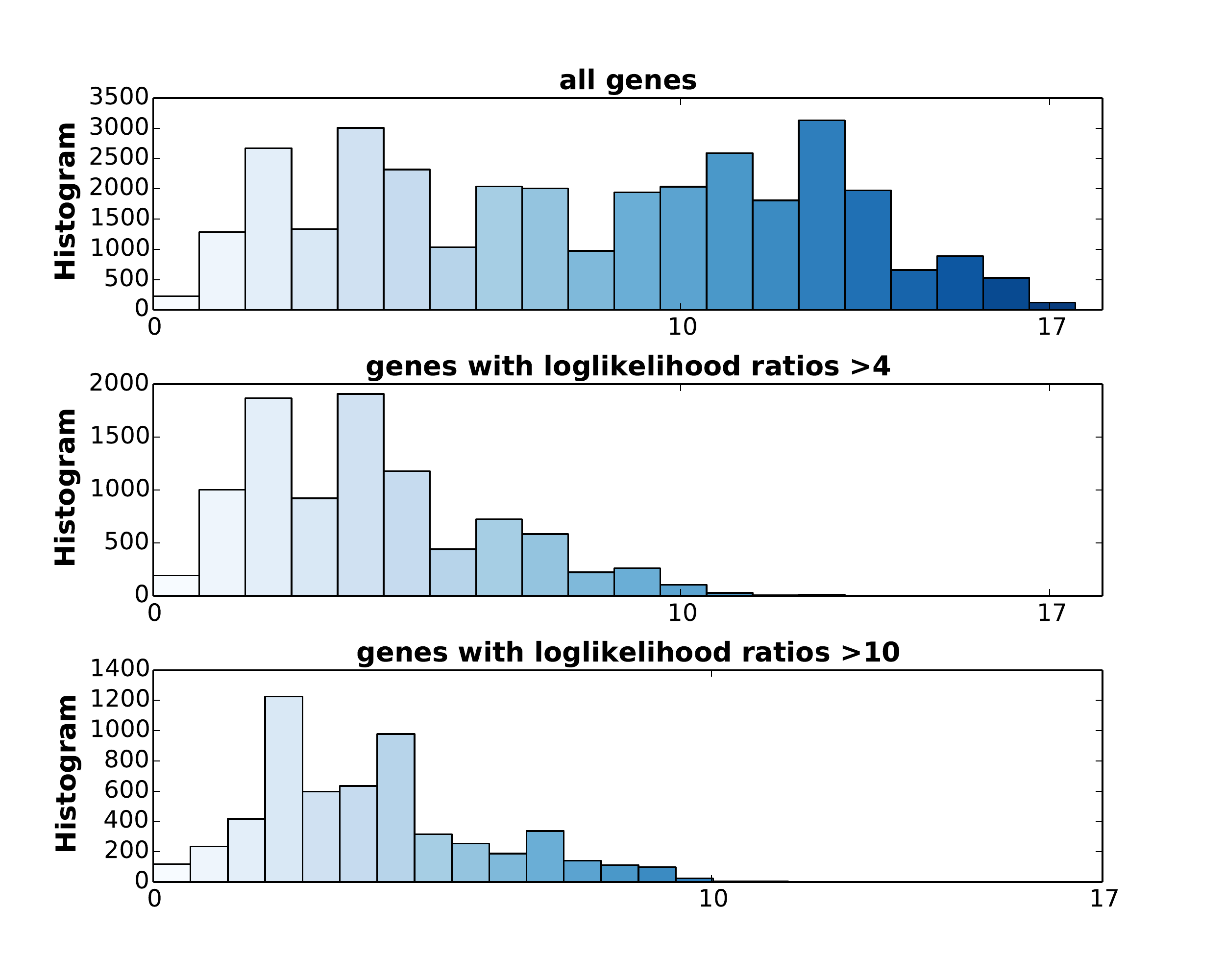}
 \caption{Histograms of estimated perturbation times for all the genes (upper panel), genes with loglikelihood ratios over 4 (middle) and genes with loglikelihood ratios over 10 (lower panel), respectively. \label{Fig:Histtimes}}
 \end{figure}

A histogram of the perturbation times for all differentially expressed (DE) genes between DC3000 (condition 1) and DC3000\emph{hrpA} (condition 2), using a log-likelihood ratio filter $r> 4$ and $r>10$, is shown in Fig. \ref{Fig:Histtimes} (middle) and Fig. \ref{Fig:Histtimes} (bottom) respectively. Two peaks in perturbation times were identified, the first between $t = 0$ and $t \leq 2.5$ hours post inoculation (hpi) representing genes that become DE early in the time series, with a second peak between $t > 2.5$ and $t \leq 7$ hpi. A third set of genes, whose perturbation time was between $t > 7$ hpi was taken as representative of late genes. Characterisation of genes as early, middle and late responsive are consistent with the general progression of bacterial infection, including delivery of effectors and onset of effector mediated transcriptional reprogramming as effectors are not delivered into plant cells until 90-120 minutes post inoculation \citep{grant2000rpm1}, the failed immune response of the plant, and finally subversion of plant metabolism to provide nutrient to the apoplastically localised bacteria 

Previous studies by \citet{tao2003quantitative} measured infection at 3, 6 and 9 hpi, and consequently, missed the earliest responses. Additionally, this paper did not compare virulent with the disarmed DC3000\emph{hrpA} strain, but instead compare DC3000 with strains carrying the avrRpt2 or avrB effectors, and could therefore not disentangle basal immune response from bacterial subversion in the same way. Whilst other studies by \citet{thilmony2006genome} measured gene expression in both DC3000 and DC3000\emph{hrpA} at 7, 10, and 24 hpi, again missing the earliest responses. Furthermore, the initial inoculum concentrations for each of the three time points were different, thus pairwise comparisons were not indicative of the natural temporal progression of pseudomonas infection.

GO analysis of genes perturbed between DC3000 and DC3000\emph{hrpA} are summarised in Tables \ref{Tab:1} (for genes with loglikelihood ratios $>4$) and \ref{Tab:2} (for genes with likelihood ratios $>10$), respectively. GO analysis was run using BINGO \citep{maere2005bingo} with differentially expressed genes grouped into three categories: early genes that were perturbed between $t=0$ and $t  < 2.5$ hpi; genes that were perturbed midway through the time series ($t \geq 2.5$ and $t\leq 7$ hpi);  and late genes (perturbed after $t=7$ hpi). The tables details significant GO terms following Benjamini and Hochberg FDR correction at $p<0.01$. At both thresholds similar GO terms are evident. 

Genes that were perturbed early on in the time series, between $t = 0$ and $t < 2.5$ hpi, were enriched for kinase-related terms that included kinase activity (GO ID 16301) and protein kinase activity (4672). Early genes also included terms such as signalling (23052), receptor signalling protein activity (5057) and signalling pathway (23033). This group of genes was also enriched for Receptor kinase-like protein family (PFAM). Other terms included some related to changes in molecular activity, including phosphorylation (16310), transferase activity (16740) and post-translational protein modification (43687). 

Perhaps most significant were terms that suggest differences in the basal immune response already beginning to arise, including detection of biotic stimulus (9595), response to other organisms (51707), immune system process (2376), response to biotic stimulus (9607), defence response (6952), response to stress (6950). These observations appear consistent with the inability of the mutant strain DC3000\emph{hrpA} to form a functional type III secretion system thus deliver effector proteins to suppress plant immunity \citep{roine1997hrp}. Consequently the DC3000\emph{hrpA} time series is expected to capture information about pathogen-associated molecular pattern (PAMP) induced basal immune responses, whilst the DC3000 strain, which is able to deliver effector proteins via the type III secretion system, should capture information about how effector proteins suppress basal immune responses and enable bacterial proliferation

Furthermore, BONZAI (BON1), a calcium-dependent phospholipid binding and its interacting partner BAP2 (BON ASSOCIATION PROTEIN 2) are perturbed around 1.5 hpi and 3.9 hpi respectively. Multiple plant disease resistance genes are suppressed by BON1 \citep{li2009multiple} consistent with its induction as a pathogen strategy to suppress \emph{R} genes. In agreement with this, two TIR-NBS-LRR plant disease resistance proteins were suppressed, as was a LRR-receptor kinase (LRR-RK) and receptor kinase protein. While a simple interpretation of these data are that this is a consequence of effector suppression of basal defence, the analysis also revealed some counter-intuitive responses. Unexpectedly, three TIR-NBS-LRR genes were induced.  More remarkable, PAD4, a key regulator of SA defence responses and NPR3, a salicylic acid receptor \citep{fu2012npr3} were perturbed around 1.5 hpi. Similarly genes encoding a set MAPKKKs, upstream activators of MAP kinase signalling cascades, 3, 7, 15, 18, 19, and 20 were also induced 1.7, 2.9, 2.1, 2.4, 2.1, 2.4 and 3.9 hpi respectively. MAPKK9 was recently shown to activate MAPK3 and MAPK6 \citep{liu2014novel}, two core MAPKs previously shown to have major roles in plant defence responses. We interpret these unexpected results to be a component an early plant defence response to mitigate effector virulence activities. 

Genes perturbed between $t \geq 2.5$ and $t \leq 7$ hpi were enriched for generic ``response terms'' including defence response (6952), defence response to bacterium (42742), response to bacterium (9617), response to stress (6950), response to other organism (51707), multi-organism process (51704), response to osmotic stress (6970), response to chemical stimulus (42221), response to abiotic stimulus (9628) adn response to salicylic acid (9751). Collectively these terms highlight ontologies expected for pathways engaged in the battle between the PAMP triggered immune responses and the pathogens virulence strategy designed to suppress PAMP immunity. Ontologies including response to abiotic stimuli reflect the cross-talk between biotic and abiotic stress responses. As highlighted above, DC3000 induces abscisic acid (ABA) to compromise host defence \citep{de2007pseudomonas}. ABA is also induced in plant responses to drought. In contrast to the first early response genes, no clear examples of modules of genes that are experimentally associated with plant defence are evident. This may therefore represent a transitional phase in the virulence strategy where mechanisms to suppress basal defences have been activated and the pathogen is targeting host metabolism to provide bacterial nutrition, or may simply reflect previously unexpected complexity in the transcriptional reprogramming by effectors. Two genes worth noting are the induction of the auxin receptor TIR1 (3.7 hpi) and its cognate interacting partner SKP1 INTERACTING PARTNER 1 (3 hpi). Like ABA, Pseudomonas activates auxin signalling to promote bacterial multiplication \citep{cui2013pseudomonas}.

GO terms enriched after $t = 7$ hpi (late genes) were predominately related to photosynthesis and include chloroplast (9507), thylakoid (9579), photosynthetic membrane (34357) and light-harvesting complex (30076). Comparison of the distribution of perturbation times for genes with the GO term Chloroplast suggest a striking difference compared to the distribution for all differentially expressed genes. Similarly, these genes were enriched for photosynthesis related pathways (KEGG) and Chloroplast gene families (PFAM). These terms appear consistent with a central role of chloroplasts in the production of precursors of salicylic acid, jasmonic acid and other key hormone components \citep{nomura2012chloroplast} as well as energy generation and primary metabolism. Modulation of host hormones is a key virulence strategy deployed by many plant pathogens \citep{grant2002hormone,robert2011hormone}. Indeed recent studies suggest that \emph{P. syringae} effectors can localise to chloroplasts \citep{jelenska2007aj,li2014distinct}. Again, like the intermediate gene list, there were no obvious components that could be readily associated with plant defence responses, probably reflecting the previous lack of experimentation in this area.  The most notable feature was the suppression of AUXIN RESISTANT 4 (AXR4) at 7.9 hpi, which we have previously shown to be required for systemic immune signalling \citep{truman2010arabidopsis} and suppression of CYCLIC NUCLEOTIDE-GATED CHANNEL 12 (CNGC12) at 8.3 hpi. CNGC12 contributes to the activation of multiple pathogen resistance pathways \citep{yoshioka2006chimeric}. Thus, suppression of these genes would be predicted to contribute to systemic and local enhanced susceptibility respectively. 

Interestingly, a number of the genes identified as being differentially expressed between DC3000 and DC3000\emph{hrpA} have additionally been identified in yeast two-hybrid screens as targets of a variety of effector proteins from a range of pathogens (\citealp{mukhtar2011independently}; see Table \ref{Tab:5}). These include a class of effector proteins of \emph{P. syringae} pathovars: Hop effector proteins \citep{lindeberg2005unified}. HopBF1 was found to target two adjacent AAA-ATPase genes, AT5G19990 and AT5G20000, likely involved in protein degradation (26S Proteasome). Interfering or subverting host proteasome systems appears to be a common strategy for plant pathogens  \citep{banfield2015perturbation,dudler2013manipulation}. Additionally, of the remaining 6 genes, it is interesting to note that one of the genes AT3G46370, is a LRR-RK, which are often associated with plant defence responses. Indeed AvrPto is a member of the HopAB1 effector family targets tomato Prf, one of the earliest plant disease resistance proteins to be identified \citep{salmeron1994tomato}. An emerging paradigm is that effectors target kinase pathways to block innate immunity including AvrPto \citep{xiang2008pseudomonas}, and AvrPphB \citep{zhang2010receptor}, HopAI1 \citep{zhang2012disruption}. Currently little is know about HopAB1, but observation of AT3G46370 suggests this kinase gene is initially turned on but is subsequently repressed in DC3000, implying an important role in plant immunity.

\vspace{.2in}
\noindent To further augment the GO analysis, a plant Gene Set Enrichment Analysis was run for the same sets of genes using PlantGSEA \citep{yi2013plantgsea}. Results are summarised in Tables \ref{Tab:3} and \ref{Tab:4}. Early genes were enriched for Receptor kinase-like protein family (PFAM, \citealp{bateman2004pfam}), complementing the kinase and signalling GO terms. Whilst no pathways or gene families appeared to overrepresented in genes perturbed midway through the experiment, this set of genes were overrepresented for confirmed and suspected targeted of AtbHLH15, suggesting enrichment for a common transcriptional program in this group. Late perturbed genes were enriched for the KEGG pathways \citep{kanehisa2000kegg,kanehisa2014data} relating to photosynthesis, with additional evidence of enrichment for chloroplast gene families (PFAM), again complementing the GO enrichment analysis. 

\subsection{Alternative filtering}\label{section_s4.2}
\begin{figure}
\centering
\includegraphics[width=0.5\textwidth]{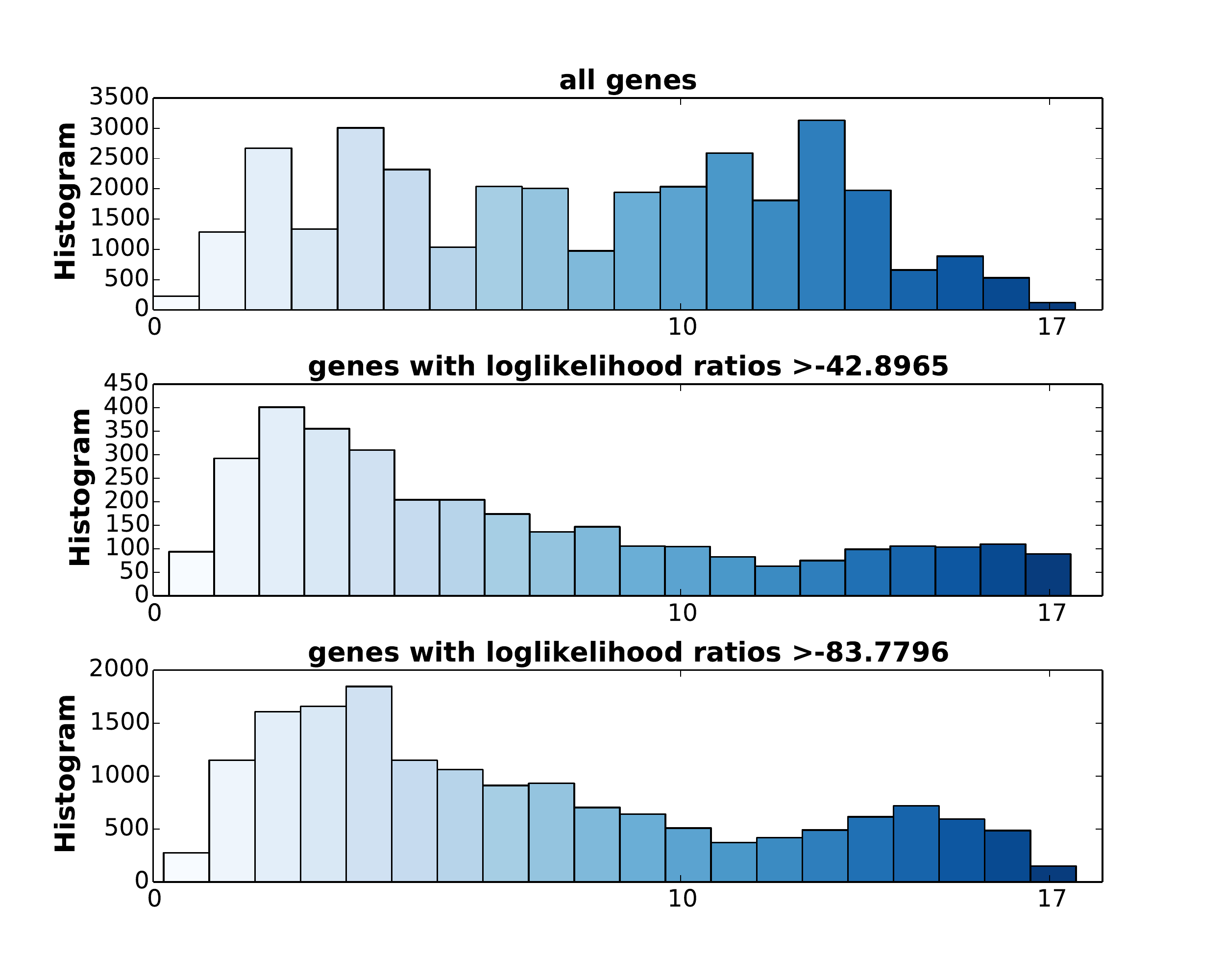}
 \caption{Histograms of estimated perturbation times for all the genes (upper panel), genes with loglikelihood ratios over -42.8965 (middle) and genes with loglikelihood ratios over -83.7796 (lower panel), respectively.}
 \label{histogram_newfiltering}
 \end{figure}

\noindent In order to identify if our approach could be applied without first identifying differentially expressed genes, we instead filtered according to how dynamic the genes were \citep{kalaitzis2011simple}. In particular for each profile we could fit a GP model (dynamic data) versus a flat model (i.e., a noise model) and calculate the corresponding log likelihood ratio for the gene being dynamic within that perturbation, with the sum of the log likelihoods over both the DC3000 and DC3000\emph{hrpA} perturbations taken to give a final log likelihood. Genes were filtered to remove those with low levels of dynamics using thresholds of $-83.7796$ and $-42.8965$, which filtered out $50\%$ and $10\%$ of genes respectively.  Histograms of estimated perturbation times for all the genes, genes with loglikelihood ratios over -42.8965 and genes with loglikelihood ratios over -83.7796 are illustrated in Fig. \ref{histogram_newfiltering}. It is obvious that the new filtering method keeps a good number of non-DE genes, and might not be necessary when the focus is on DE genes only.

Genes were subsequently grouped into three groups: those whose perturbation time was $t<2.5$, those whose perturbation time was $2.5 \leq t \leq 7$ and those perturbed $7 < t < 10$. In Tables \ref{Tab:6} and \ref{Tab:7} we indicate GO enrichment for genes filtered according to how dynamic their profiles were, whilst Tables \ref{Tab:8} and \ref{Tab:9} indicate GSEA. There is some evidence that early perturbed genes were again enriched for GO terms relating to signalling, and the PFAM Receptor kinase-like protein, although these terms were not apparent at the higher threshold. At the lower thresholds the presence of GO terms such as immune system process and defense response appear to show consistency with our earlier analysis based upon differentially expressed genes, with the higher threshold showing enrichment of other ``response to ...'' type terms including, crucially, response to abscisic acid stimulus. Genes that were perturbed midway through the time series again show consistent enrichment for a variety of ``response to ..'' type terms, including response to bacterium, biotic stimulus, abiotic stimulus, and chemical stimulus. Late genes appear to be generally, and consistently enriched for GO terms and KEGG pathways related to chloroplasts and photosynthesis, as well as enrichment of related PFAM protein families. The general overlap of various terms, both at different likelihood thresholds, and for the different methods used for filtering, suggest that our approach has identified a consistent and genuine picture of the dynamics of plant-pathogen interactions. 

\subsection{Distribution of Perturbation Times of Gene Ontologies and Pathways}

\begin{figure*}
\centering
\includegraphics[trim={0 3cm 0 2cm},clip,width=0.9\textwidth]{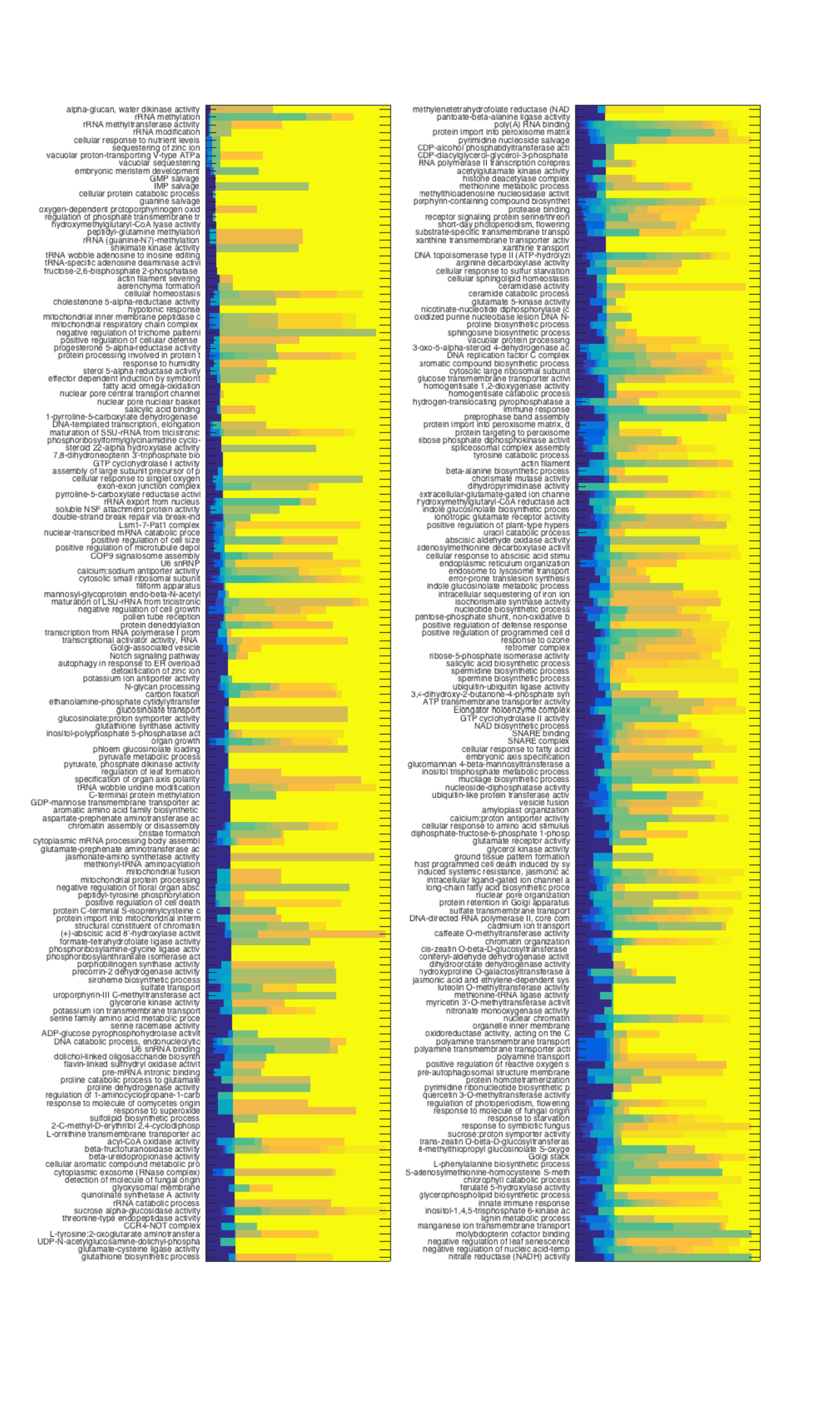}
\caption{Temporal ordering of significantly perturbed GO terms. }
\label{Fig:GO1}
\end{figure*}

\begin{figure*}
\centering
\includegraphics[trim={0 3cm 0 2cm},clip,width=0.9\textwidth]{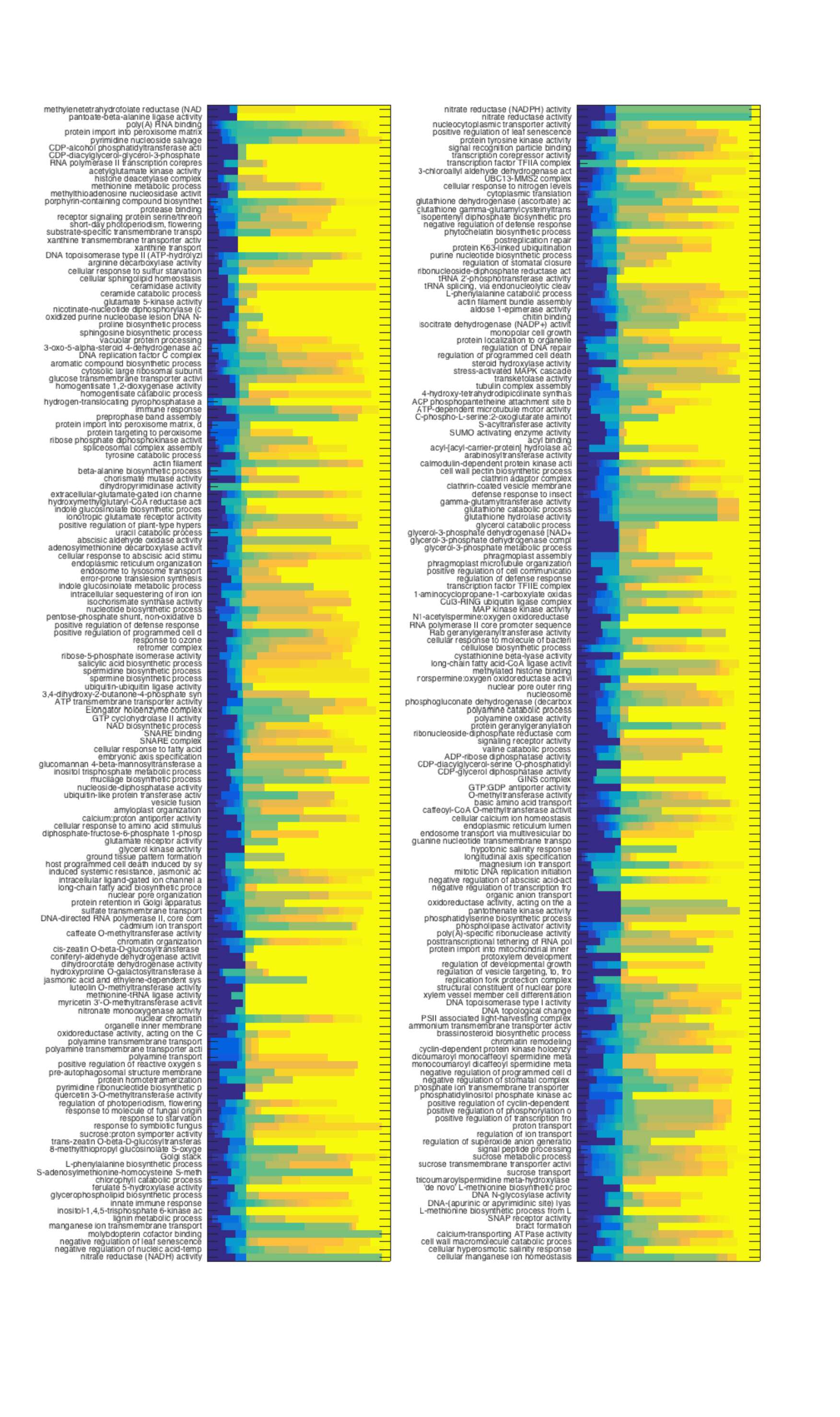}
\caption{Temporal ordering of significantly perturbed GO terms. }
\label{Fig:GO2}
\end{figure*}

\begin{figure*}
\centering
\includegraphics[trim={0 3cm 0 2cm},clip,width=0.9\textwidth]{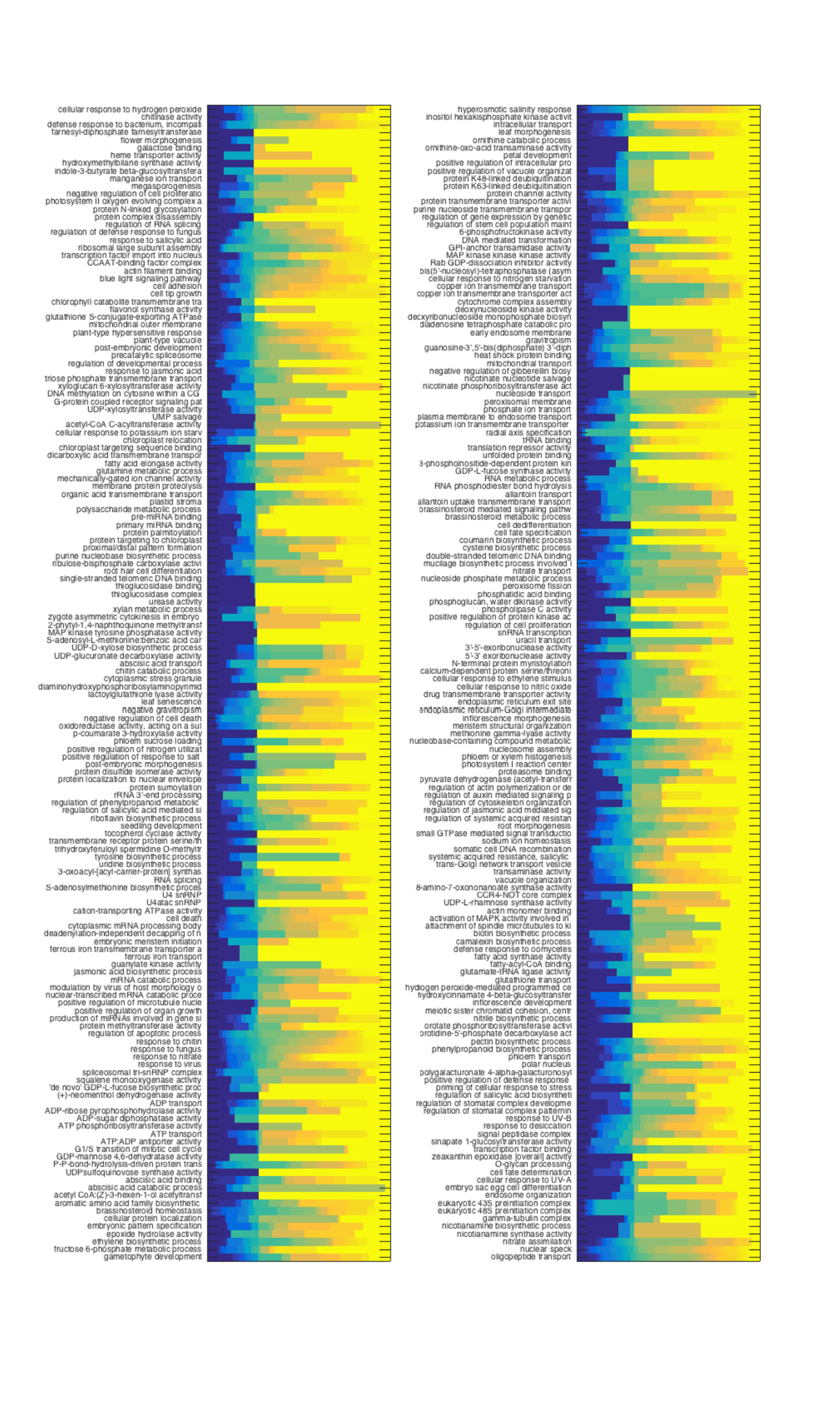}
\caption{Temporal ordering of significantly perturbed GO terms. }
\label{Fig:GO3}
\end{figure*}

\begin{figure*}
\centering
\includegraphics[trim={0 3cm 0 2cm},clip,width=0.9\textwidth]{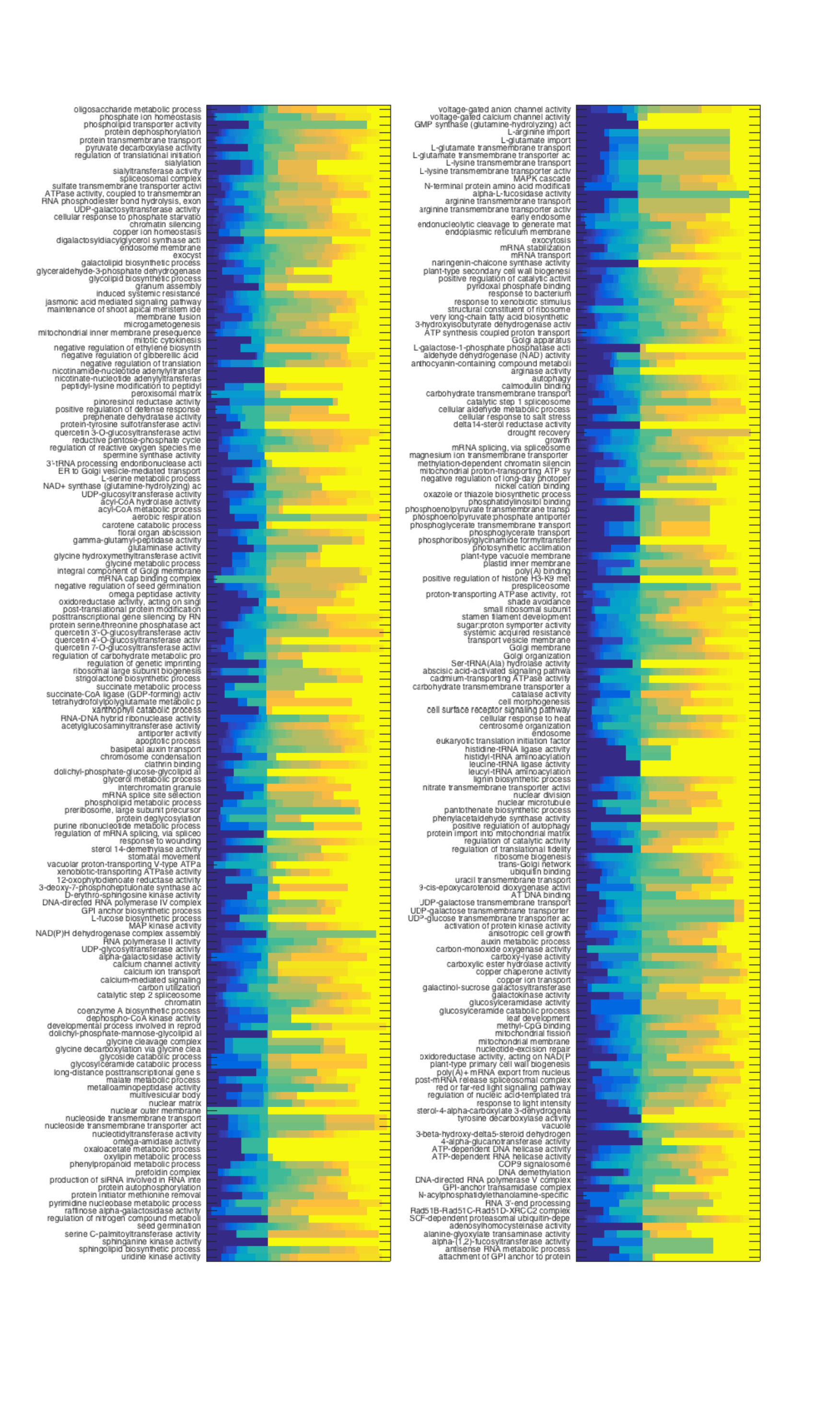}
\caption{Temporal ordering of significantly perturbed GO terms. }
\label{Fig:GO4}
\end{figure*}

\begin{figure*}
\centering
\includegraphics[trim={0 3cm 0 2cm},clip,width=0.9\textwidth]{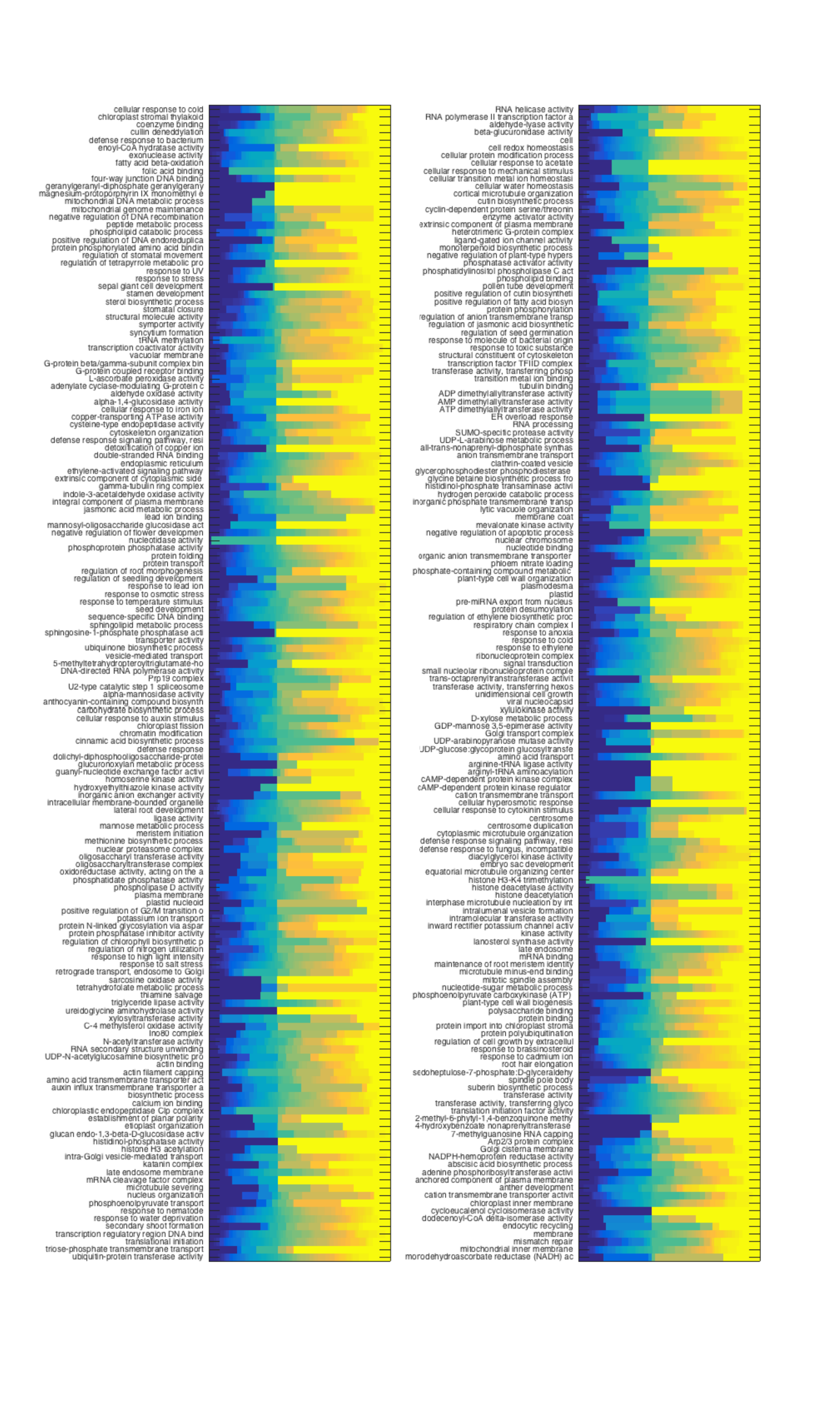}
\caption{Temporal ordering of significantly perturbed GO terms. }
\label{Fig:GO5}
\end{figure*}

\begin{figure*}
\centering
\includegraphics[trim={0 3cm 0 2cm},clip,width=0.9\textwidth]{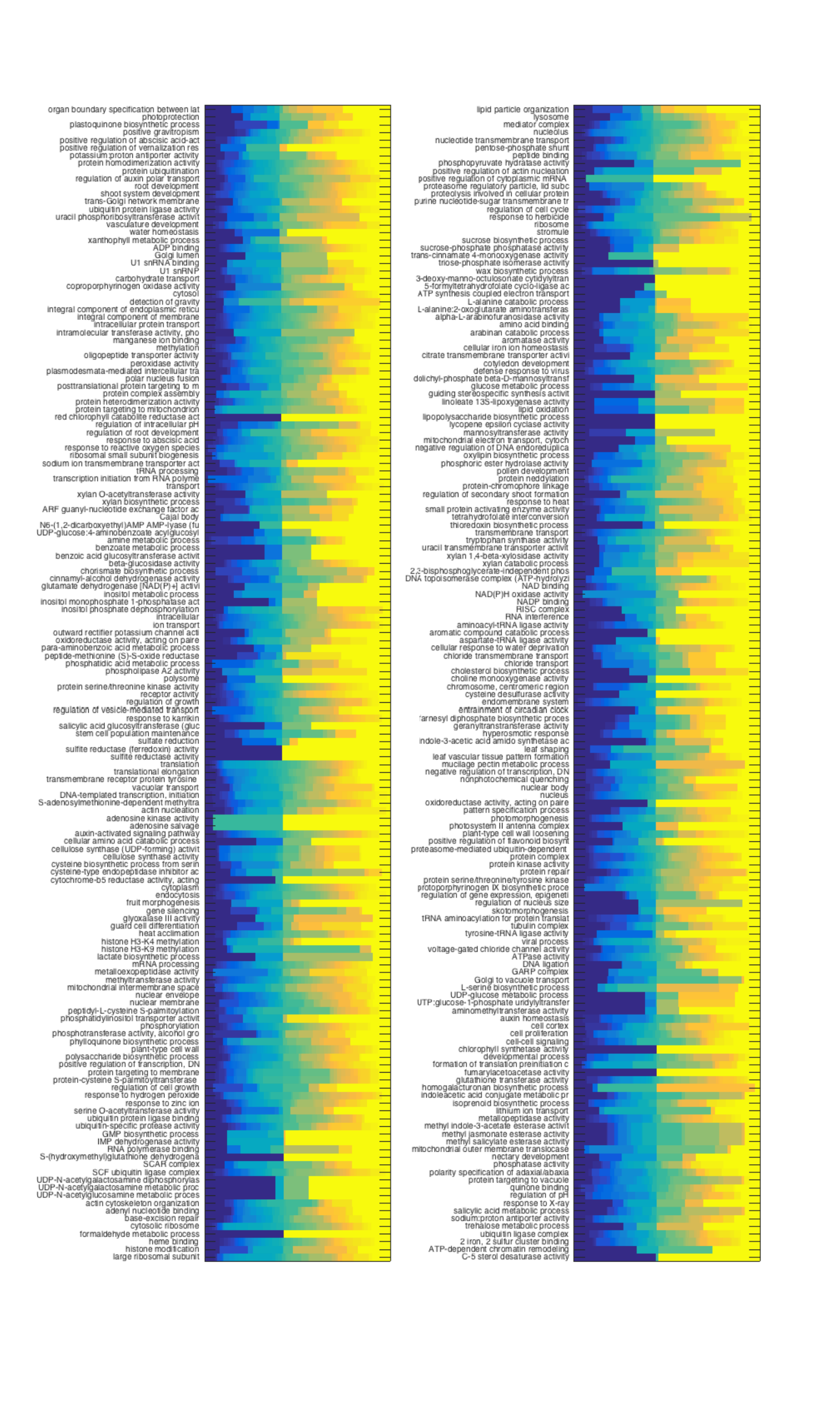}
\caption{Temporal ordering of significantly perturbed GO terms. }
\label{Fig:GO6}
\end{figure*}

\begin{figure*}
\centering
\includegraphics[trim={0 3cm 0 2cm},clip,width=0.9\textwidth]{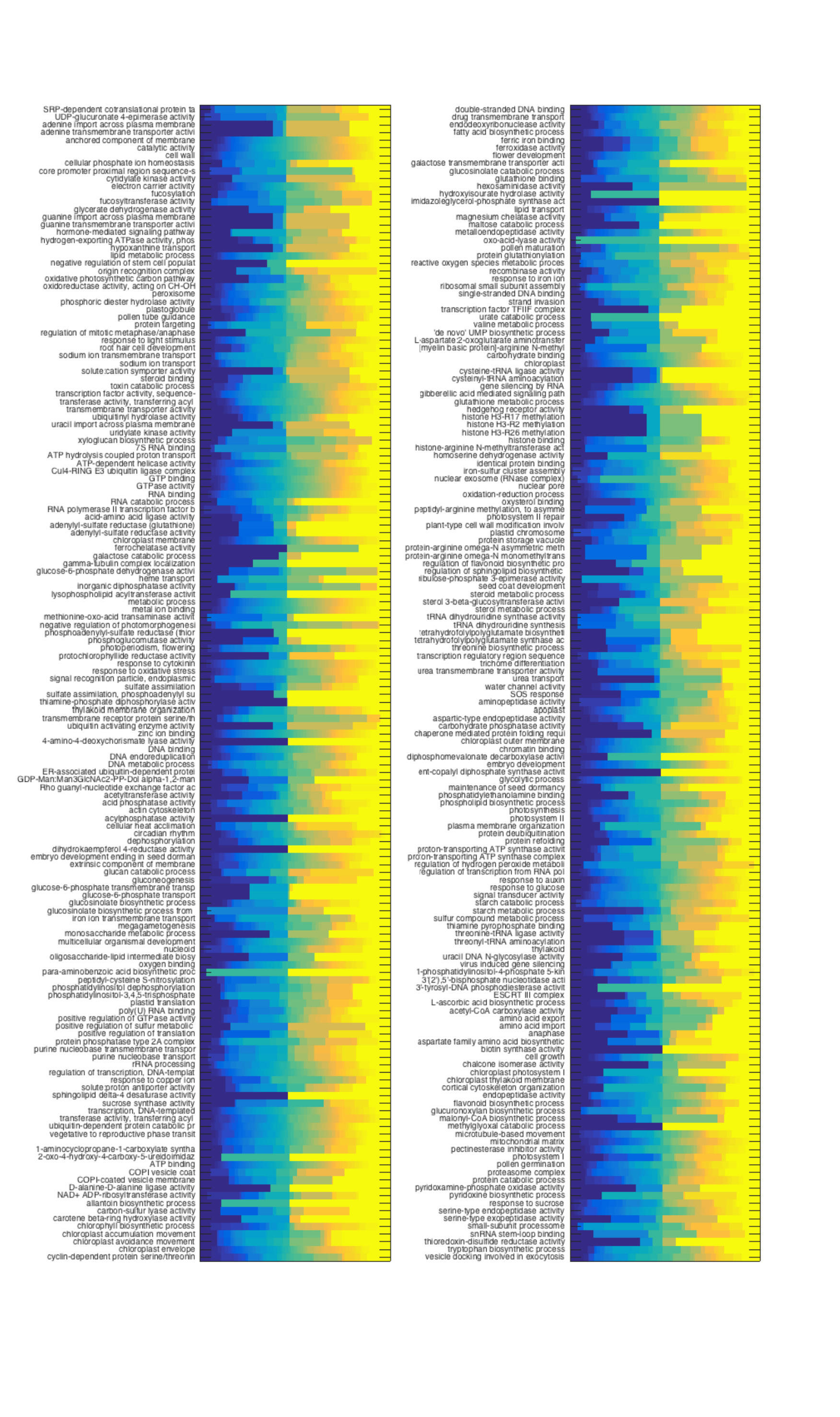}
\caption{Temporal ordering of significantly perturbed GO terms. }
\label{Fig:GO7}
\end{figure*}

\begin{figure*}
\centering
\includegraphics[trim={0 3cm 0 2cm},clip,width=0.9\textwidth]{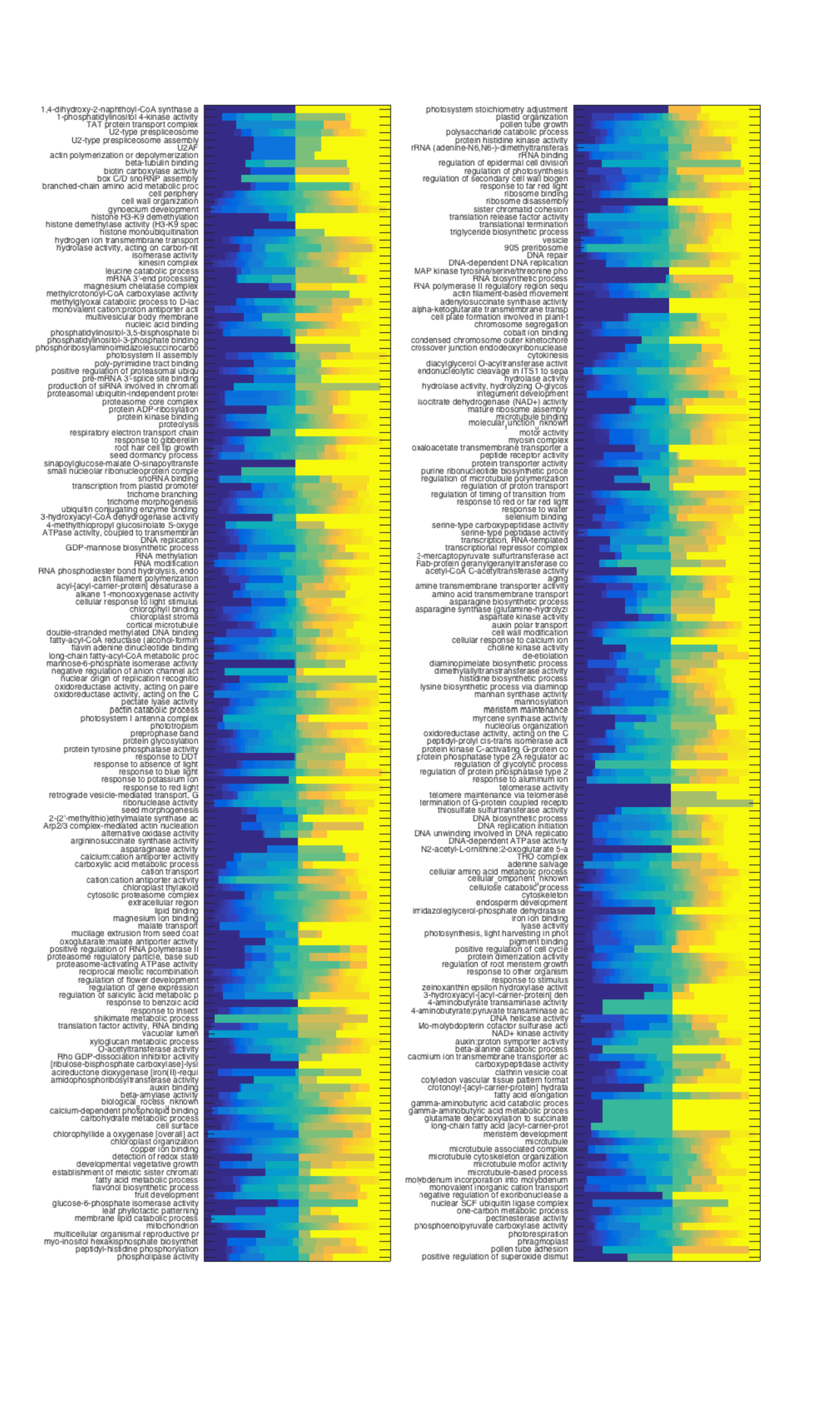}
\caption{Temporal ordering of significantly perturbed GO terms. }
\label{Fig:GO8}
\end{figure*}

\begin{figure*}
\centering
\includegraphics[trim={0 3cm 0 2cm},clip,width=0.9\textwidth]{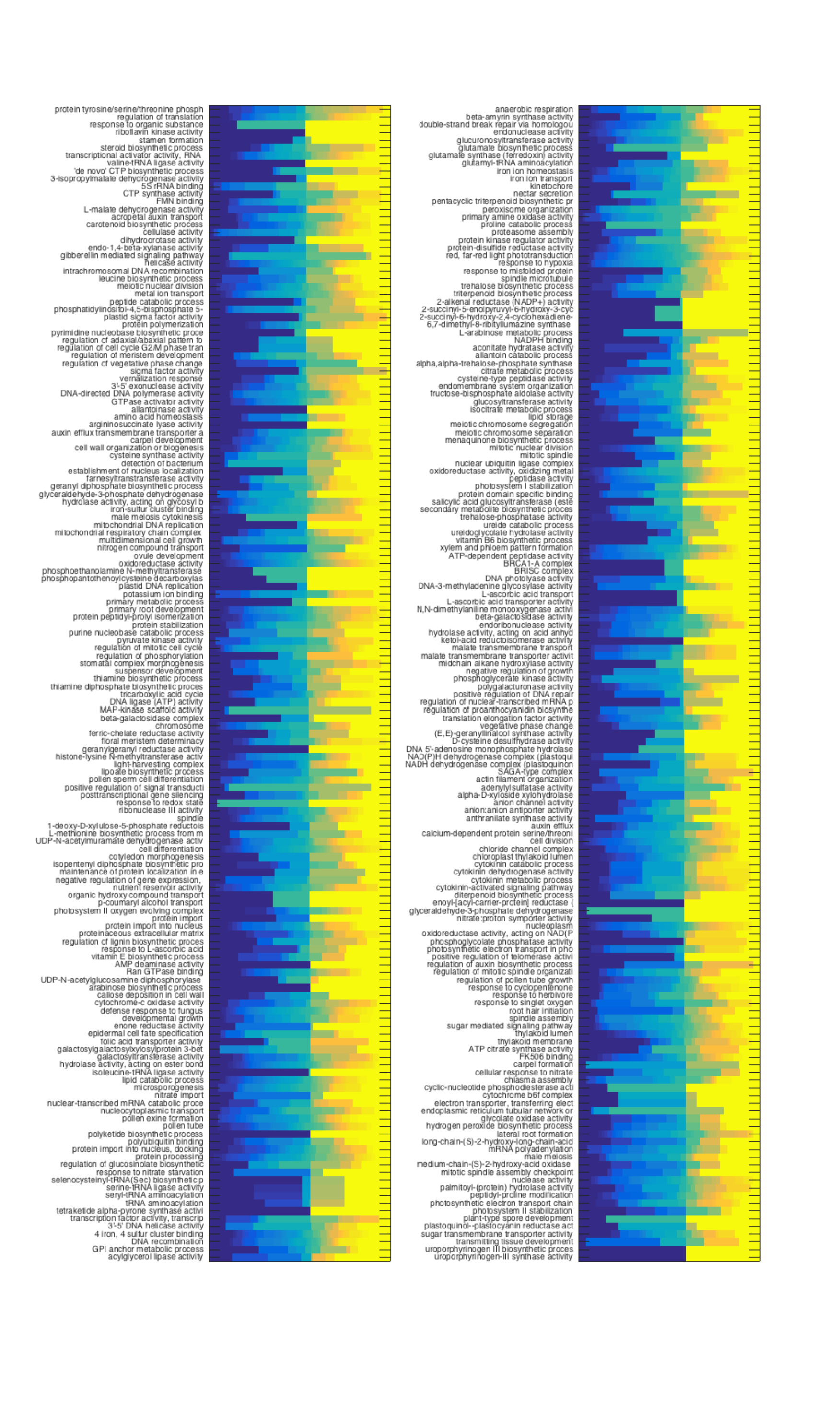}
\caption{Temporal ordering of significantly perturbed GO terms. }
\label{Fig:GO9}
\end{figure*}

\begin{figure*}
\centering
\includegraphics[trim={0 3cm 0 2cm},clip,width=0.9\textwidth]{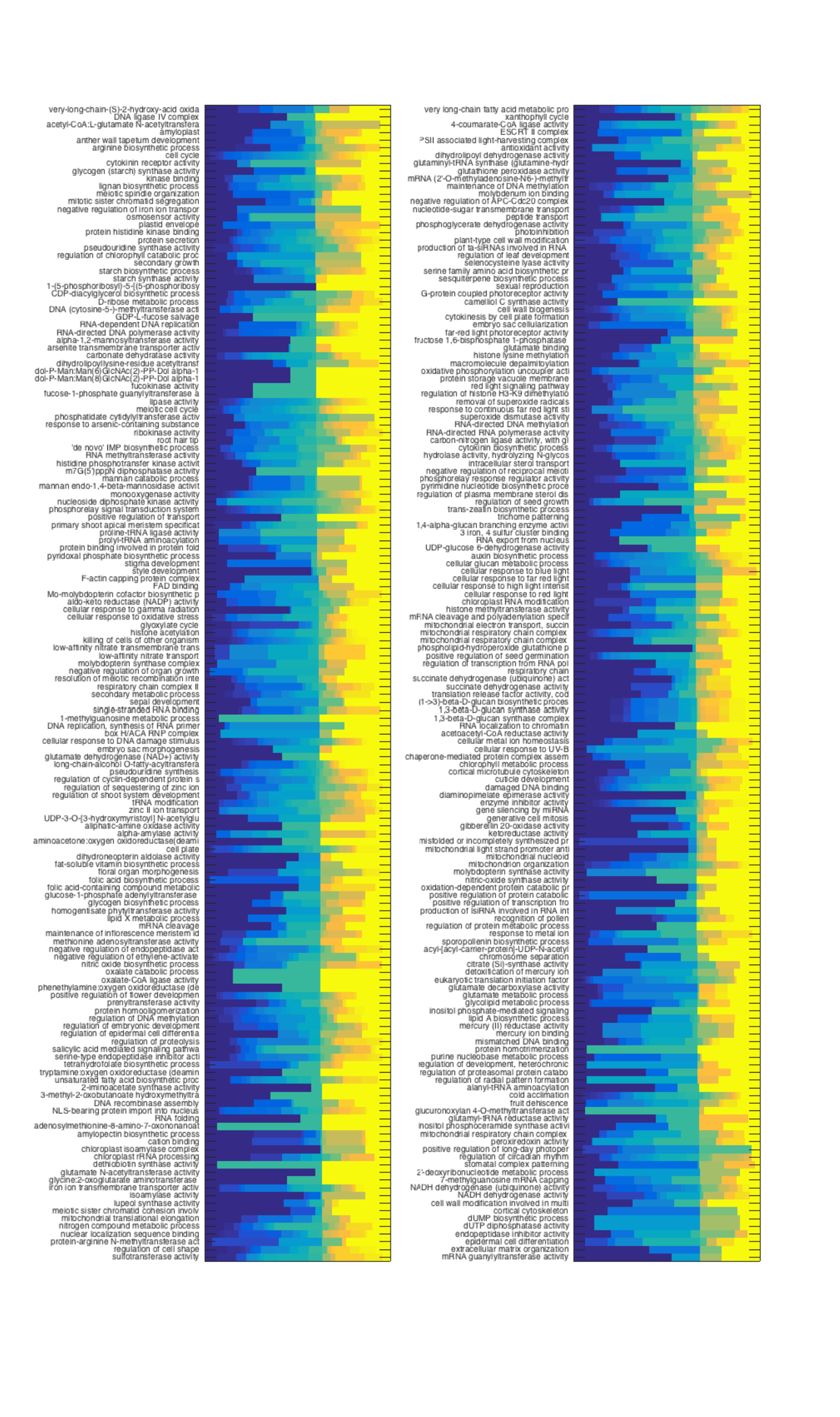}
\caption{Temporal ordering of significantly perturbed GO terms. }
\label{Fig:GO10}
\end{figure*}

\begin{figure*}
\centering
\includegraphics[trim={0 3cm 0 2cm},clip,width=0.9\textwidth]{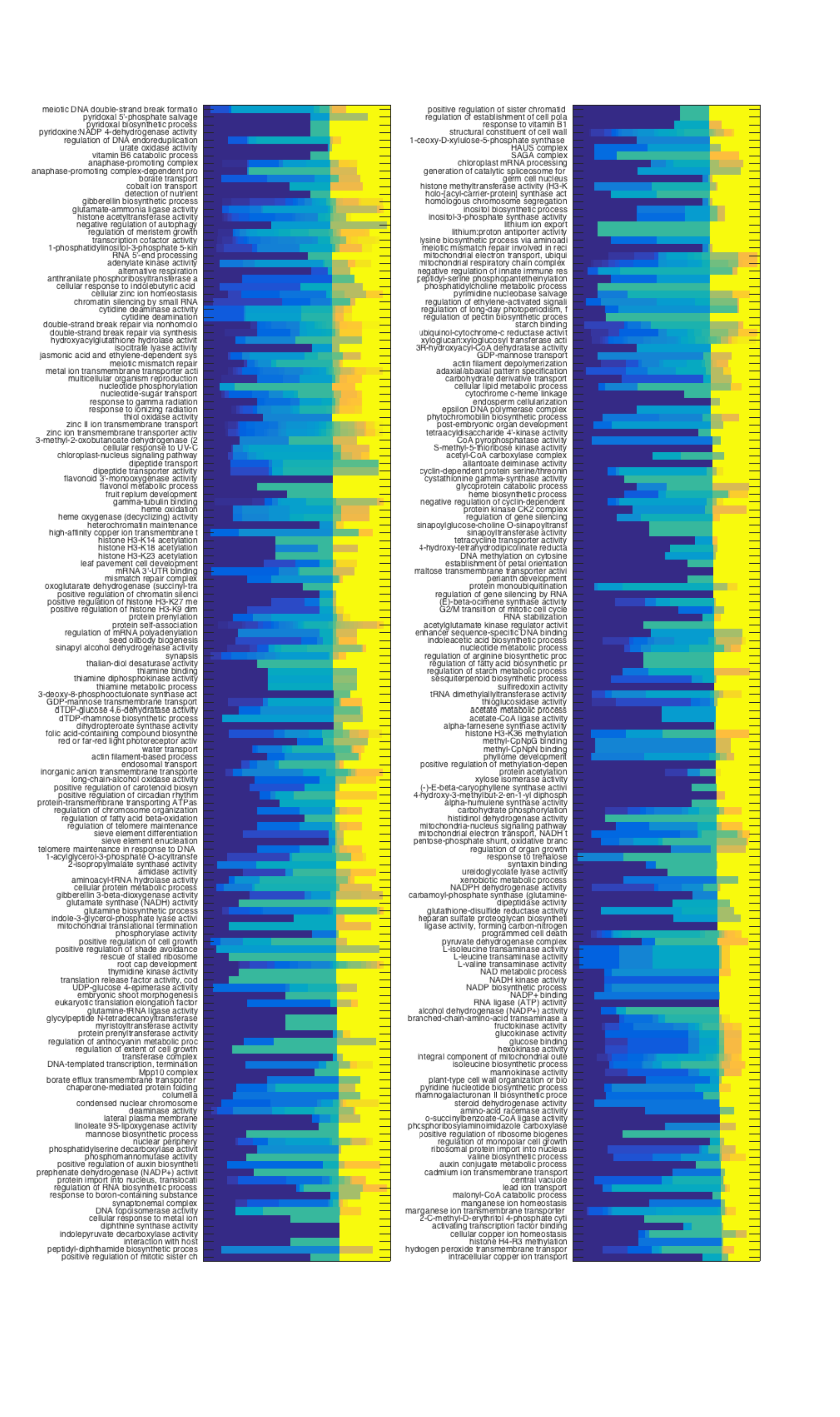}
\caption{Temporal ordering of significantly perturbed GO terms. }
\label{Fig:GO11}
\end{figure*}

\begin{figure*}
\centering
\includegraphics[trim={0 3cm 0 2cm},clip,width=0.9\textwidth]{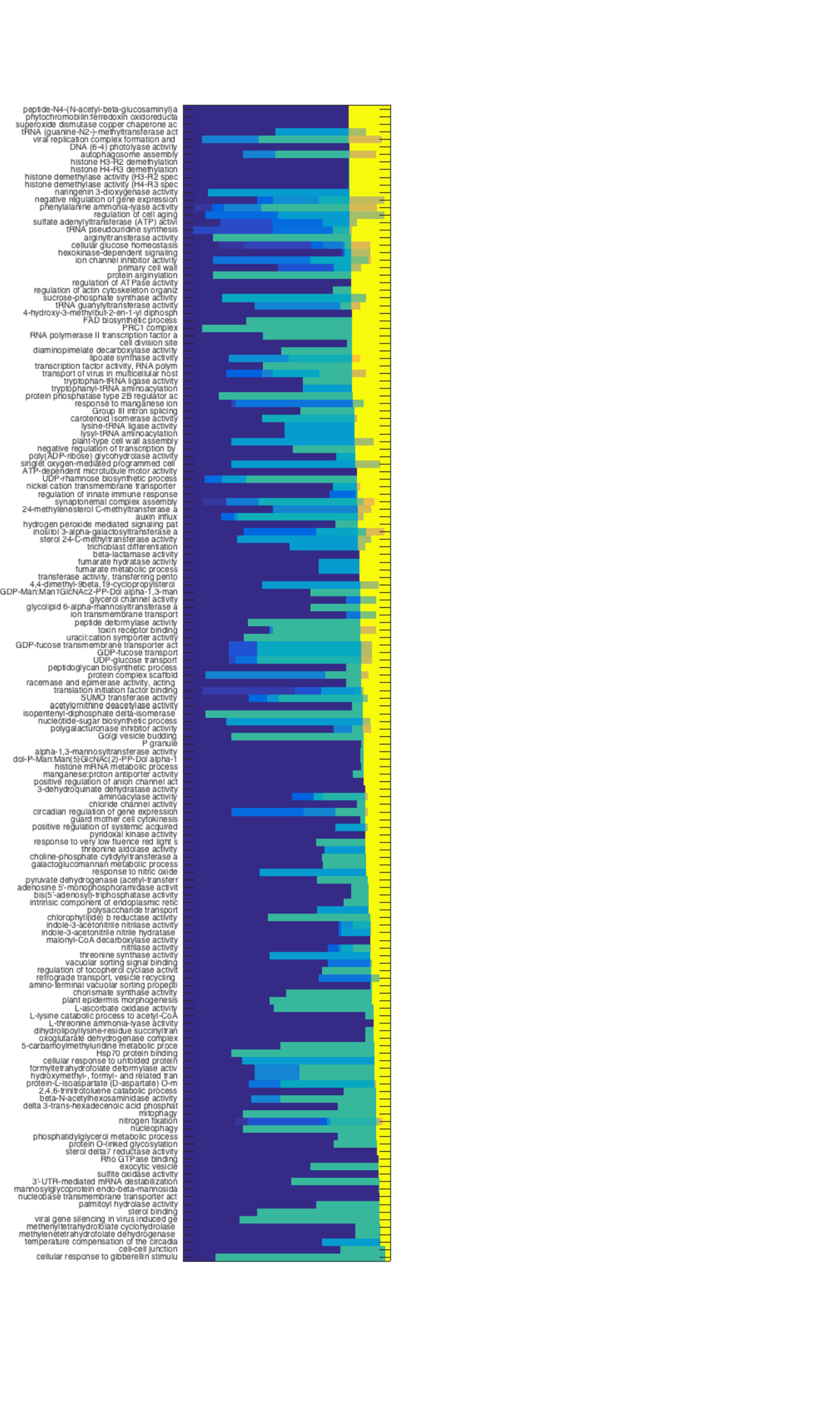}
\caption{Temporal ordering of significantly perturbed GO terms. }
\label{Fig:GO12}
\end{figure*}

\begin{figure*}
\centering
\includegraphics[width=\textwidth]{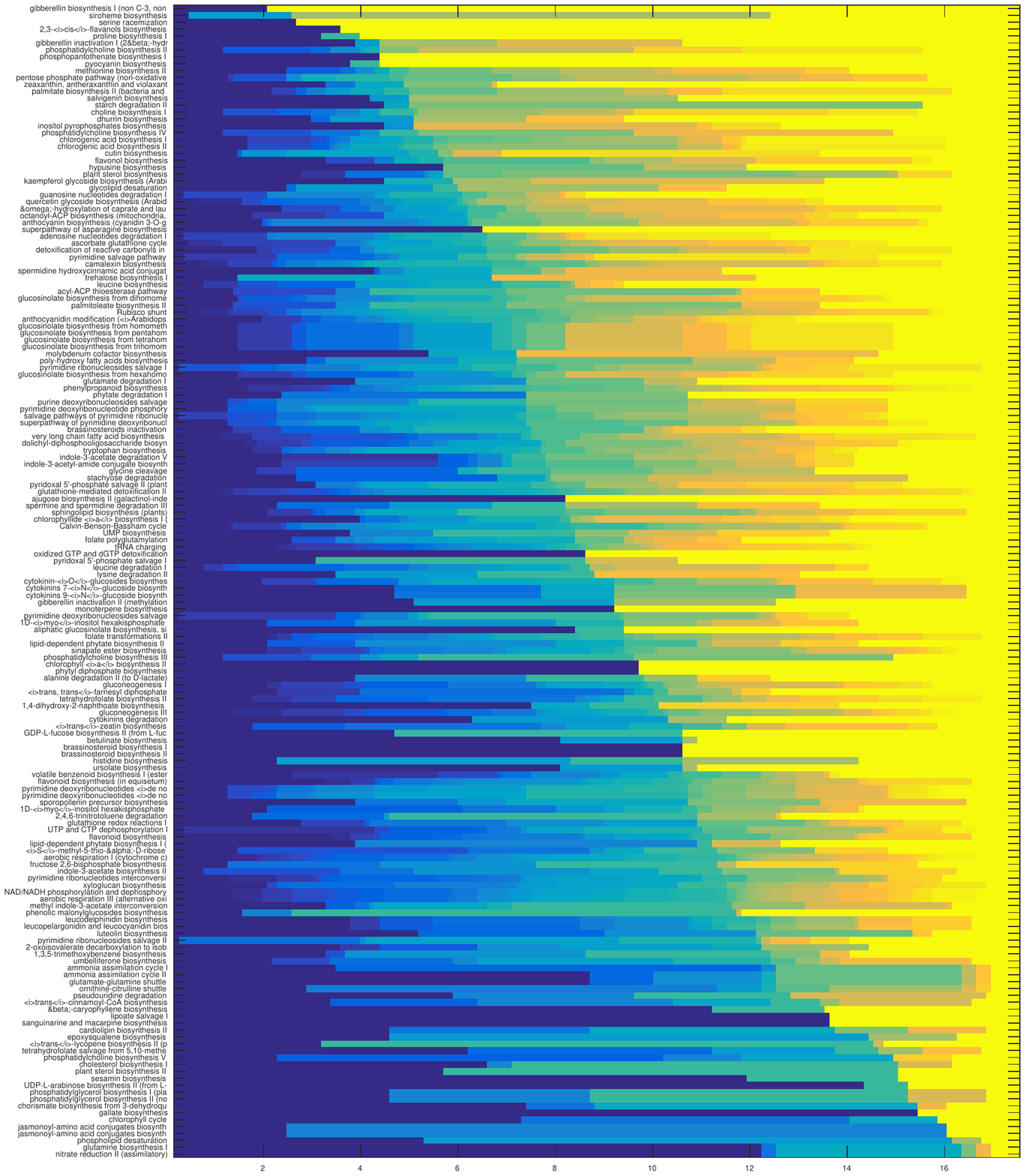}
\caption{Temporal ordering of significantly perturbed Aracyc terms. }
\label{Fig:Pathway}
\end{figure*}

Whilst the grouping of genes into early, middle and late genes well reflects the phases of bacterial infection, the calculation of perturbation times allows for the calculation of the distribution of particular Gene Ontology or Pathway terms. In turn this should allow the various terms to be ordered, producing a high resolution picture of the differences in processes that occur in Arabidopsis when infected with virulent DC3000 versus the disarmed mutant DC3000\emph{hrpA}. To do so we calculated the perturbation times for all genes associated with a particular GO/Pathway term, and compared this distribution against that for all genes. Statistical significance was calculated using empirically corrected Kolmogorov-Smirnov (KS) tests and Kullback-Leibler (KL) divergence. The cumulative distribution function of all GO terms that are significantly perturbed using either KS or KL ($p<0.05$) are shown in heatmap form in Figures \ref{Fig:GO1} - \ref{Fig:GO12} , with significantly perturbed Pathways terms indicated in Fig. \ref{Fig:Pathway}. For clarity, terms are ordered by the time at which $>50\%$ of associated genes are perturbed.

\vspace{.2in}


\section{Reference table of GO and AraCyc pathway terms}

\noindent Finally in Table \ref{Tab:11} we show the reference key to the Fig. 5 in the main text for GO and AraCyc pathway terms. 

\begin{longtable}{lllll}
\caption{GO analysis of genes perturbed between DC3000 and DC3000\emph{hrpA} following identification of differentially expressed genes (log likelihood ratio $>4$)\label{Tab:1}}\\
\toprule
PT	&	GO-ID	&	corr p-value	&	Description	\\
\midrule
$0 \leq t < 2.5$	&	3735 &	5.16E-15	&	structural constituent of ribosome	\\
$0 \leq t < 2.5$	&	5622	&	3.12E-11	&	intracellular	\\
$0 \leq t < 2.5$	&	5198	&	3.15E-11	&	structural molecule activity	\\
$0 \leq t < 2.5$	&	44424	&	3.56E-11	&	intracellular part	\\
$0 \leq t < 2.5$	&	5515	&	1.00E-09	&	protein binding	\\
$0 \leq t < 2.5$	&	43229	&	1.35E-09	&	intracellular organelle	\\
\end{longtable}

\begin{longtable}{lllll}
\caption{GO analysis of genes perturbed between DC3000 and DC3000\emph{hrpA} following identification of differentially expressed genes (log likelihood ratio $>10$)\label{Tab:2}}\\
\toprule
PT	&	GO-ID	&	corr p-value	&	Description	\\
\midrule
$0 \leq t < 2.5$	&	5515	&	2.14E-09	&	protein binding	\\
$0 \leq t < 2.5$	&	43687	&	2.22E-08	&	post-translational protein modification	\\
$0 \leq t < 2.5$	&	6796	&	4.21E-08	&	phosphate metabolic process	\\
$0 \leq t < 2.5$	&	6793	&	4.21E-08	&	phosphorus metabolic process	\\
$0 \leq t < 2.5$	&	6468	&	1.15E-07	&	protein amino acid phosphorylation	\\
$0 \leq t < 2.5$	&	6464	&	2.01E-07	&	protein modification process	\\
\bottomrule
\end{longtable}

\begin{longtable}{lllll}
\caption{Genes that were perturbed between DC3000 and DC3000\emph{hrpA}, and were also targeted by an effector from a range of pathogens in a yeast two-hybrid screen \citep{mukhtar2011independently}. \label{Tab:5}}\\
\toprule
Gene	&	Effector	&	Effector type	&	Likelihood ratio	&	Perturbation time	\\
\midrule
AT5G19990	&	HopBF1\_Pla 107	&	Hop	&	11.8721721	&	1.02857142	\\
AT4G38800	&	HopAF1\_Pph 1448A	&	Hop	&	16.7558441	&	2.05714285	\\
AT5G43700	&	HopH1\_Psy B728A	&	Hop	&	8.40857712	&	2.05714285	\\
AT3G46370	&	HopAB1\_Pph 1448A	&	Hop	&	10.6853893	&	2.05714285	\\
AT3G48550	&	HopR1-C-TERM\_Pto DC3000	&	Hop	&	10.4462306	&	3.08571428	\\
AT3G11720	&	HopM1\_Pph 1448A	&	Hop	&	11.8544659	&	4.11428571	\\
\end{longtable}

\begin{longtable}{lllll}
\caption{Plant Gene Set Enrichment Analysis (GSEA) for genes differentially expressed between DC3000 and DC3000\emph{hrpA} at a log likelihood ratio threshold $>4$\label{Tab:3}}\\
\toprule
PT	&	corr p-value	&	Description	\\
\midrule
$0 \leq t < 2.5$	&	1.10E-08	&	Ribosome	\\
$0 \leq t < 2.5$	&	2.75E-06	&	Receptor kinase-like protein family	\\
$0 \leq t < 2.5$	&	4.48E-03	&	Confirmed target genes of transcription factor: HY5	\\
$0 \leq t < 2.5$	&	4.48E-03	&	Confirmed and Unconfirmed target genes of transcription factor: HY5	\\
	&		&		\\
$2.5 \leq t \leq 7$	&	2.29E-05	&	Confirmed and Unconfirmed target genes of transcription factor: AtbHLH15	\\
$2.5 \leq t \leq 7$	&	1.60E-03	&	Confirmed target genes of transcription factor: AtbHLH15	\\
	&		&		\\
$t > 7$	&	1.98E-07	&	Metabolic pathways	\\
$t > 7$	&	7.80E-05	&	Photosynthesis - antenna proteins	\\
$t > 7$	&	7.93E-04	&	Biosynthesis of alkaloids derived from histidine and purine	\\
$t > 7$	&	0.0136	&	Chloroplast and Mitochondria gene families ,Chlorophyll a/b-binding protein family	\\
$t > 7$	&	0.0242	&	Ubiquitin mediated proteolysis	\\
$t > 7$	&	0.0296	&	Valine, leucine and isoleucine biosynthesis	\\
$t > 7$	&	0.0356	&	Photosynthesis	\\
$t > 7$	&	0.0464	&	chlorophyllide a biosynthesis I	\\
\bottomrule
\end{longtable}

\begin{longtable}{lll}
\caption{Plant Gene Set Enrichment Analysis (GSEA) for genes differentially expressed between DC3000 and DC3000\emph{hrpA} at a log likelihood ratio threshold of $>10$ \label{Tab:4}}\\
\toprule
PT		&	corr p-value	&	Description	\\
\midrule
$0 \leq t < 2.5$	&	7.71E-09	&	Receptor kinase-like protein family	\\
$0 \leq t < 2.5$	&	0.0293	&	Confirmed target genes of transcription factor: HY5	\\
$0 \leq t < 2.5$	&	0.0293	&	Confirmed and Unconfirmed target genes of transcription factor: HY5	\\
$0 \leq t < 2.5$	&	0.0431	&	Confirmed and Unconfirmed target genes of transcription factor: AtbHLH15	\\
	&		&		\\
$2.5 \leq t \leq 7$	&	3.64E-03	&	Confirmed and Unconfirmed target genes of transcription factor: AtbHLH15	\\
$2.5 \leq t \leq 7$	&	3.64E-03	&	Confirmed target genes of transcription factor: AtbHLH15	\\
	&		&		\\
$t > 7$	&	3.05E-03	&	Biosynthesis of alkaloids derived from histidine and purine	\\
$t > 7$	&	0.0227	&	Ubiquitin mediated proteolysis	\\
$t > 7$	&	0.0248	&	Glucosinolate biosynthesis	\\
$t > 7$	&	0.0264	&	Metabolic pathways	\\
\bottomrule
\end{longtable}

\begin{longtable}{llll}
\caption{GO analysis of genes perturbed between DC3000 and DC3000\emph{hrpA} following identification of dynamically expressed genes with second most number of genes (log likelihood ratio of $>-83.7796$)\label{Tab:6}}\\
\toprule
PT	&	GO-ID	&	corr p-value	&	Description	\\
\midrule
$0 \leq t < 2.5$	&	9987	&	2.91E-12	&	cellular process	\\
$0 \leq t < 2.5$	&	44267	&	2.99E-10	&	cellular protein metabolic process	\\
$0 \leq t < 2.5$	&	44260	&	1.39E-09	&	cellular macromolecule metabolic process	\\
$0 \leq t < 2.5$	&	44237	&	3.64E-07	&	cellular metabolic process	\\
$0 \leq t < 2.5$	&	19538	&	3.64E-07	&	protein metabolic process	\\
$0 \leq t < 2.5$	&	43170	&	9.96E-07	&	macromolecule metabolic process	\\
\bottomrule
\end{longtable}

\begin{longtable}{llll}
\caption{GO analysis of genes perturbed between DC3000 and DC3000\emph{hrpA} following identification of dynamically expressed genes with fewest number of genes (log likelihood ratio $>-42.8965$)\label{Tab:7}}\\
\toprule
PT	&	GO-ID	&	corr p-value	&	Description	\\
\midrule
$0 \leq t < 2.5$	&	44267	&	9.27E-09	&	cellular protein metabolic process	\\
$0 \leq t < 2.5$	&	6412	&	2.72E-08	&	translation	\\
$0 \leq t < 2.5$	&	10467	&	1.02E-07	&	gene expression	\\
$0 \leq t < 2.5$	&	44260	&	1.49E-07	&	cellular macromolecule metabolic process	\\
$0 \leq t < 2.5$	&	19538	&	3.22E-07	&	protein metabolic process	\\
$0 \leq t < 2.5$	&	9987	&	1.42E-06	&	cellular process	\\
\bottomrule
\end{longtable}

\begin{longtable}{lll}
\caption{Plant Gene Set Enrichment Analysis (GSEA) for genes dynamically expressed with a log likelihood ratio $>-83.7796$ \label{Tab:8}}\\
\toprule
PT		&	corr p-value	&	Description	\\
\midrule
$0 \leq t < 2.5$	&	2.79E-32	&	Ribosome	\\
$0 \leq t < 2.5$	&	4.79E-03	&	Receptor kinase-like protein family	\\
$0 \leq t < 2.5$	&	0.0343	&	Spliceosome	\\
	&		&		\\
$2.5 \leq t \leq 7$	&	9.46E-05	&	Confirmed and Unconfirmed target genes of transcription factor: AtbHLH15	\\
$2.5 \leq t \leq 7$	&	6.19E-03	&	Metabolic pathways	\\
	&		&		\\
$7 > t \leq 10$	&	4.56E-15	&	Metabolic pathways	\\
$7 > t \leq 10$	&	1.51E-05	&	Biosynthesis of alkaloids derived from histidine and purine	\\
$7 > t \leq 10$	&	9.45E-05	&	Biosynthesis of alkaloids derived from shikimate pathway	\\
$7 > t \leq 10$	&	1.54E-04	&	Biosynthesis of alkaloids derived from ornithine, lysine and nicotinic acid	\\
$7 > t \leq 10$	&	7.87E-04	&	Aminoacyl-tRNA biosynthesis	\\
$7 > t \leq 10$	&	1.41E-03	&	Biosynthesis of plant hormones	\\
$7 > t \leq 10$	&	1.41E-03	&	Biosynthesis of phenylpropanoids	\\
$7 > t \leq 10$	&	1.41E-03	&	Valine, leucine and isoleucine biosynthesis	\\
$7 > t \leq 10$	&	2.32E-03	&	Carbon fixation in photosynthetic organisms	\\
$7 > t \leq 10$	&	2.32E-03	&	Photosynthesis - antenna proteins	\\
$7 > t \leq 10$	&	0.0115	&	Biosynthesis of alkaloids derived from terpenoid and polyketide	\\
$7 > t \leq 10$	&	0.0189	&	Biosynthesis of terpenoids and steroids	\\
$7 > t \leq 10$	&	0.0217	&	Phenylalanine, tyrosine and tryptophan biosynthesis	\\
$7 > t \leq 10$	&	0.0217	&	Glycolysis / Gluconeogenesis	\\
$7 > t \leq 10$	&	0.026	&	Citrate cycle (TCA cycle)	\\
$7 > t \leq 10$	&	0.0398	&	Pentose phosphate pathway	\\
$7 > t \leq 10$	&	0.0398	&	Proteasome	\\
$7 > t \leq 10$	&	0.0431	&	Starch and sucrose metabolism	\\
\bottomrule
\end{longtable}

\begin{longtable}{lll}
\caption{Plant Gene Set Enrichment Analysis (GSEA) for genes dynamically expressed with a log likelihood ratio $>-42.8965$ \label{Tab:9}}\\
\toprule
PT		&	corr p-value	&	Description	\\
\midrule
$0 \leq t < 2.5$	&	1.64E-28	&	Ribosome	\\
$0 \leq t < 2.5$	&	0.0131	&	Cytoplasmic ribosomal protein gene family ,L11	\\
	&		&		\\
$2.5 \leq t \leq 7$	&	1.34E-03	&	Confirmed and Unconfirmed target genes of transcription factor: AtbHLH15	\\
$2.5 \leq t \leq 7$	&	1.42E-03	&	Confirmed target genes of transcription factor: AtbHLH15	\\
$2.5 \leq t \leq 7$	&	0.0344	&	Metabolic pathways	\\
\bottomrule
\end{longtable}

\begin{longtable}{lll}
\caption{Reference table of GO and AraCyc Pathway Terms used in Figure 5 in the main text)\label{Tab:11}}\\
\toprule
Index in figure	&	Ontology		\\
\midrule
1	&	regulation of abscisic acid biosynthetic process	\\
2	&	cellular response to abscisic acid stimulus	\\
3	&	negative regulation of abscisic acid-activated signaling pathway	\\
4	&	abscisic acid transport	\\
5	&	abscisic acid binding	\\
6	&	abscisic acid catabolic process	\\
\end{longtable}

\newpage

\end{document}